%% file: bitto_fruehwirth_final.tex
\documentclass[12pt,a4]{article}

\usepackage{natbib}
\usepackage{color,graphicx}
\usepackage{amsthm, amsfonts}
\usepackage{amsmath, amsthm, amssymb}
\usepackage{makeidx}
\usepackage{csquotes}
\usepackage{array}
\usepackage{graphicx}
 \usepackage[onehalfspacing]{setspace}
\usepackage{amsmath, amsthm, amsfonts}
\usepackage{amssymb}
\usepackage{rotating}
\usepackage{color}
\usepackage[]{algorithm2e}
\usepackage{pbox}
\usepackage{longtable}
\usepackage{subcaption}
\usepackage{hyperref} % For hyperlinks in the PDF
\usepackage{listings}
\usepackage{lscape}
\usepackage{rotate}

\newtheorem{alg}{Algorithm}

\usepackage[usenames,dvipsnames]{xcolor}

\title{Achieving Shrinkage in a Time-Varying Parameter Model Framework}
\author{Angela Bitto and Sylvia Fr\"uhwirth-Schnatter\footnote{Institute for Statistics and Mathematics, 
Department of Finance, Accounting and Statistics, WU Vienna University of Economics and Business, Vienna, Austria; email:  \href{mailto:angela.bitto@wu.ac.at}{angela.bitto@wu.ac.at} and \href{mailto:sfruehwi@wu.ac.at}{sfruehwi@wu.ac.at}.}}
\date{\today}
\begin{document}
%\doublespacing

\onehalfspacing

%\centerline{\large}

% \thispagestyle{fancy} % All pages have headers and footers

%----------------------------------------------------------------------------------------
%	ABSTRACT
%----------------------------------------------------------------------------------------
\input{newcommands}

\maketitle % Insert title

\begin{abstract}

\noindent {\small Shrinkage for time-varying parameter (TVP)  models is investigated within a Bayesian framework, with the aim to automatically reduce time-varying parameters to static ones, if the model is overfitting. This is achieved through placing the double gamma shrinkage prior on the process variances. An efficient Markov chain Monte Carlo  scheme is developed, exploiting boosting based on the ancillarity-sufficiency interweaving strategy. The method is applicable both to TVP models for univariate as well as multivariate time series. Applications include a TVP generalized Phillips curve for EU area inflation modelling and a multivariate TVP Cholesky stochastic volatility model for joint modelling of the returns from the DAX-30 index.}
 \end{abstract}


\onehalfspacing
\section{Introduction}

%\marginpar{\Sylvia{Clearly state contribution in the Introduction}}
Time-varying parameter (TVP) models are widely used in time series analysis to
deal with processes which gradually change over time and provide an interesting alternative to models that allow multiple change points as considered, for instance,  in \citet{gew-jia:inf}.  A variety of  interesting econometric applications of  TVP models  appeared in recent years; for example, \citet{pri:tim}  used time-varying structural VAR models in a monetary policy application, \citet{dan-hal:pre} used TVP models for equity return prediction and \citet{bel-etal:hie-tv} used a TVP model to model EU-area inflation.

A huge  advantage of TVP models is their flexibility in capturing gradual  changes. However, the risk of overfitting increases
with a growing number of coefficients, as many of  them might in reality be constant over the entire observation period.
 This will be exemplified in the present paper for  a TVP Cholesky stochastic volatility (SV)  model \citep{lop-etal:par} for a  time series of returns from the DAX-30 index, where out of 406 potentially  time-varying  coefficients only a small fraction actually changes over time.
 Allowing static coefficients to  be time-varying leads to a considerable loss of statistical efficiency compared to a model, where   coefficients  are constant apriori. %

% \citet{nak-wes:bay-ana} present a Bayesian analysis of latent threshold dynamics models.
% \citet{kal-gri:tim} use a dynamic regression model based on continuous shrinkage priors to forecast equity premium.
% To shrink   process variances  toward zero, \citet{fru-wag:sto} use spike-and-slab priors  and  \citet{bel-etal:hie-tv}  consider Bayesian Lasso type priors.

Identifying fixed coefficients in a TVP model   amounts to a {\em variance selection} problem, involving a decision whether the variances of the shocks driving the dynamics of  a  time-varying  parameter  are equal to zero.  Variance selection in latent variable models is known to be a non-regular problem
within the framework of classical statistical hypothesis testing \citep{har:for}. The introduction of   shrinkage priors for  variances within a Bayesian  framework has proven to be an attractive alternative both for
random effects models \citep{fru-tue:bay,fru-wag:bay} as well as state space models  \citep{fru:eff,fru-wag:sto,nak-wes:bay-ana,bel-etal:hie-tv,kal-gri:tim}.  For TVP models,  shrinkage priors  can  automatically reduce time-varying coefficients to static ones, if the model is overfitting.

% See \citet{fru-wag:bay}for a comparison of  various shrinkage priors for random intercept models.
 % For state space models,  this goal was  achieved in  by formulating  appropriate shrinkage priors for  the process variances of the shocks driving  the dynamics of the various parameters. , see e.g.   \citet{fru-wag:sto}, \citet{nak-wes:bay-ana}, \citet{bel-etal:hie-tv},  and \citet{kal-gri:tim}.

 The literature on variance selection in TVP models is still rather slender, despite this pioneering work,
compared to the vast  literature on {\em variable selection} using shrinkage priors to shrink coefficients toward zero  in a  common regression
 framework.
This class includes  mixture priors such as  spike-and-slab priors  which
assign  positive  probability to zero values \citep{mit-bea:bay} and stochastic search variable selection (SSVS) priors \citep{geo-mcc:var}
as well as  continuous shrinkage priors  with a pronounced spike at zero,  well-known  examples being the  Bayesian Lasso  prior \citep{par-cas:bay},
 %Lasso  \citep{tib:reg} which can be derived as the Bayesian posterior mode estimator under independent double-exponential priors  \citep{par-cas:bay},
  the normal-gamma   prior \citep{gri-bro:inf,car-dou:spa} %, the Bayesian elastic net \citep{li-lin:bay},
and the horseshoe prior \citep{car-etal:hor}, among many others;
%the normal-exponential gamma prior \citep{gri-bro:bay}, the generalized beta mixtures \citep{arm-etal:gen_bet}, the double Pareto prior \citep{arm-etal:gen_dou} and the exponential power prior \citep{pol-etal:bay_bri};
see  \citet{fah-etal:bay} and \citet{pol-sco:shr}  for a  review.

One of the main contributions of \citet{fru-wag:sto} has been to  recast  the {\em variance selection} problem for state space models as a
 {\em variable selection} problem in the so-called non-centered parametrization of the state space model.  This established
 the possibility  to extend   shrinkage priors from  standard  regression analysis  to this more general framework to define a \lq\lq sparse\rq\rq\
 state space model. To this aim,  \citet{fru-wag:sto} employed spike-and-slab priors, %  and,  in a similar vein,
 whereas \citet{bel-etal:hie-tv} relied on the  Bayesian Lasso prior for variance selection in TVP models.   However, other shrinkage priors  might be useful  and overcome limitations of these priors, such as computational issues for the spike-and-slab prior  and the risk  of   overshrinking  coefficients  for the Bayesian Lasso prior.

 The present paper makes several contributions in the context of sparse state space models.  We  develop a new  continuous shrinkage prior for  process variances   by introducing the  normal-gamma prior  in the non-centered parametrization. This leads to a gamma-gamma (called double gamma) prior for the process variances,  which has many attractive properties compared to the popular   inverted gamma prior \citep{pet-etal:dyn}. % \citep{gri-bro:inf,car-dou:spa}.
  We show that the double gamma  prior
is more   flexible than the Bayesian Lasso  prior  (which is a special case of the double gamma)
and  yields posterior distributions  with a  pronounced spike at zero for coefficients which are not time-varying, while at the same time overshrinkage is  avoided for  time-varying coefficients.  A second shrinkage prior allows to shrink  static coefficients
to coefficients which are not significant over the entire observation period.   As a result, we are  able  to discriminate between time-varying coefficients,  coefficients which are significant, but static and insignificant coefficients. We compare   different prior settings %, including the common inverted gamma prior for the process variances,
using  log predictive density scores \citep{gew-ami:com} and discuss an accurate approximation of  the one-step ahead predictive density.
 % to evaluate one-step ahead predictive densities. and highlight the advantages of using a Kalman mixture approximation

Based on these priors, we define a very general class of sparse TVP models,   both for  univariate and multivariate times series,  and allow for  homoscedastic error variances as well as  error variances following a stochastic volatility (SV) model \citep{jac-etal:bayJBES}.
The later model  has proven to be useful in various applications, because  neglecting time-varying volatilities might lead to overstating the role of time-varying coefficients in explaining structural changes in the dynamics of macroeconomic variables, as exemplified  by \citet{sims2001evolving} and \citet{nak:tim}.

Finally, we develop a new Markov chain Monte Carlo  (MCMC)  scheme for Bayesian inference in  sparse TVP models.  Using
the scale-mixture representation of the normal-gamma prior  allows us to implement  full conditional Gibbs sampling,
thus avoiding Metropolis-Hastings steps which are often used to implement MCMC methods for non-Gaussian state space models,
see  e.g. \citet{gew-tan:mar}.  To improve MCMC performance, we exploit the  ancillarity-sufficiency interweaving strategy of \citet{yu-men:cen}.

The rest of the paper is structured as follows. Section~\ref{sec_mod_spe} discusses our novel  shrinkage method in the context of  sparse TVP models.  In Section~\ref{sec:mcmc}, we present the
MCMC scheme. Section~\ref{sec:forecasting} discusses  evaluation of  various priors  using log predictive density scores.   In Section~\ref{sec:mult_TVP},  we extend  our method to a multivariate framework.
Section~\ref {sec:sim_demo1}  presents a simulated data example and  Section~\ref{sec:appl} exemplifies our approach  through  EU area inflation modelling based on the generalized Phillips curve as well as  estimating a time-varying covariance matrix  based on a TVP Cholesky SV model for a multivariate time series of returns of the DAX-30 index. Section~\ref{sec:con} concludes.

 %\section{Model specification} \label{sec_mod_spe}
 \section{Sparse time-varying parameter models} \label{sec_mod_spe}

\subsection{Bayesian inference for time-varying parameter  models}
\label{sec:model}

Starting point  is the well known state space model, %(dynamic linear trend model),
which has been studied in many fields, see e.g.~\citet{wes-har:bay} % and \citet{dur-koo:tim}
for a comprehensive review.
For the ease of exposition, we consider in this section a univariate time series $y_t$, observed for $T$ time points $t=1, \ldots, T$,
whereas multivariate time series are discussed in Section~\ref{sec:mult_TVP}.
In a state space model, the distribution of  $y_t$ is driven by a latent  $d$-dimensional state vector $\betav_t$ which
we are unable to observe.
The time-varying parameter (TVP) model is a special case of  a state space model and can be regarded as a regression model with
time-varying regression coefficients $\betav_{t}$ following a random walk:
\begin{eqnarray}
\label{eq:TVP2}
\betav_{t} &=& \betav_{t-1}+\boldsymbol{\omega}_{t}, \quad \boldsymbol{\omega}_{t} \sim \Normult{d}{\bfz,\Qm}, \\
\label{eq:TVP}
y_{t} &=& \xm_{t}\betav_t +\varepsilon_{t},\quad \varepsilon_{t} \sim  \Normal{0,\sigma^2_{t}},
\end{eqnarray}
where  $\xm_{t}=(x_{t1}, x_{t2}, \dots,  x_{td})$   is a $d$-dimensional row vector,
containing the regressors of the model, one of them being a constant (e.g.~$x_{t1} \equiv 1$). %\footnote{
To avoid any scaling issues, we assume that all covariates except  the intercept are standardized such that for each $j$ the average of $x_{tj}$ over $t$ is equal to  zero and the sample variance is equal to 1.
The unknown initial value
$\betav_{0}$  is assumed to follow a normal prior distribution,
\begin{equation}
\label{equbeta0}
\betav_{0}| \betav,\Qm \sim \Normult{d}{\betav, \Pm_{0} \Qm},
\end{equation}
with  $\betav=(\beta_1, \ldots, \beta_d)'$  being  unknown  fixed  regression coefficients
%in the time-varying regression model~(\ref{eq:TVP})
and $\Pm_{0}=\Diag{P_{0,11}, \ldots, P_{0,dd}} $ being  a diagonal matrix.
Furthermore,  $\betav_{0}$ is independent of the innovations $(\varepsilon_t)$ and
$(\boldsymbol{\omega}_t)$, which are independent Gaussian white noise processes.

We assume  that  $\Qm=\Diag{\theta_{1},\dots,\theta_{d}}$  is a diagonal matrix,
 hence
 each element $\beta_{jt}$ of $\betav_t =(\beta_{1t},\dots,\beta_{dt})^{\prime}$ follows a random walk for $j=1,\dots,d$:
\begin{equation}  \label{eq:state}
\beta_{jt}=\beta_{j,t-1}+\omega_{jt}, \quad \omega_{jt} \sim \Normal{0,\theta_{j}},
\end{equation}
with  initial value $\beta_{j0}|\beta_j,\theta_{j},P_{0,jj} \sim\Normal{\beta_j,\theta_{j}P_{0,jj}}$. Hence, $\theta_j$ is the process variance governing the dynamics of  the time-varying coefficient $\beta_{jt}$.\footnote{\citet{eis-etal:sto} discuss an extension
where the  covariance matrix $\Qm$ in the state equation (\ref{eq:TVP2}) is a full matrix instead of a diagonal matrix.}

Concerning the error variances in the observation equation (\ref{eq:TVP}), we consider the homoscedastic case ($\sigma^2_t \equiv \sigma^2$ for all $t=1,\dots,T$)  as well as a more flexible model specification, where $\sigma^2_{t}$ is time-dependent.
To capture heteroscedasticity, we use a  stochastic volatility (SV) specification as in \citet{jac-etal:bayJBES}  %for the error variance
  where $ \sigma^2_t =\e^{h_t}$ and the log volatility $h_t$ follows an AR(1) process:
  %\footnote{Alternative specifications are possible such as switching  state space models \citep{fru:ful}.}
\begin{eqnarray} \label{svht}
 h_t | h_{t-1}, \mu, \phi, \sigma_\eta^2 \sim \Normal{\mu+\phi(h_{t-1}-\mu),\sigma^2_\eta}.
\end{eqnarray}
In this setup, the latent volatility process $\bold h=(h_0,\dots,h_T)$ is not observed and the initial state $h_0$
is assumed to follow the stationary distribution of the autoregressive process,
i.e. $h_0 | \mu, \phi, \sigma_\eta^2  \sim \Normal{\mu,  \sigma^2_\eta /(1-\phi^2)}$.
%\begin{eqnarray*} \label{svhtinit}
%h_0 | \mu, \phi, \sigma_\eta \sim \Normal{\mu, \frac{\sigma^2_\eta }{(1-\phi^2)}}.
%\end{eqnarray*}

% \subsection{Bayesian inference} \label{sec:prioral}

We perform Bayesian inference for the TVP model
 based on a new family of  shrinkage priors for the unknown model parameters  $\betav=(\beta_1, \ldots,\beta_d)'$ and $\thetav=(\theta_1,\dots,\theta_d)'$
 to be introduced in Section~\ref{sec:shrinkage}.
A shrinkage prior  for the process variance $\theta_j$  allows to  pull  the $j$th time-varying regression coefficient $\{\beta_{j0},  \beta_{j1}, \ldots, \beta_{jT}\}$  toward the fixed regression coefficient $\beta_j$, if the model is overfitting and the effect of the $j$th covariate $x_{tj}$ is, in fact,  not changing over time. This requires the definition of priors on the process variances $\theta_j$ that are able to shrink $\theta_j$ toward the boundary  value 0. At the same time, these priors are flexible enough to avoid overshrinking for regression coefficients % $\beta_{j0},  \beta_{j1}, \ldots, \beta_{jT}$
that are, actually,  changing over time $t$ and  are characterized by a non-zero  process variance   $\theta_{j} \neq 0$.

Concerning the remaining priors, we  assume that the scaling factor   $P_{0,jj}$ in the initial distribution  $\beta_{j0}|\beta_j,\theta_{j}, P_{0,jj} \sim\Normal{\beta_j,\theta_{j}P_{0,jj}}$
 is  unknown, following  the prior $P_{0,jj} \sim \Gammainv{\nu_P,(\nu_P-1)c_P}$ with
hyperparameters    $c_P=1$ and $\nu_P=20$, implying that no prior moments exist. We employ commonly used priors for the parameters of the  error distribution  in the  equation (\ref{eq:TVP2}), namely   a hierarchical prior for the homoscedastic  case,
  \begin{eqnarray} \label{priorsigma}
 \sigma^2|C_0 \sim \Gammainv{c_0,C_0}, \qquad  C_0 \sim \Gammad{g_0,G_0},
 \end{eqnarray}
 with hyperparameters $c_0$, $g_0$, and $G_0$. In our practical applications,   $c_0 = 2.5$,
 $g_0 = 5$, and $G_0 = g_0/\E{(\sigma^2)}(c_0-1)$, with $\E{(\sigma^2)}$ being a prior guess of $\sigma^2$.

In the SV framework  (\ref{svht}),  unknown  parameters  are the level $\mu$, % \in \mathbb{R}$,
the persistence $\phi$, % \in (-1,1)$,
 and the  volatility of  volatility $\sigma_\eta^2$. % \in \mathbb{R}^+$.
 The  priors are chosen  as in \citet{kas-fru:anc}, assuming prior independence, i.e.
$p(\mu,\phi,\sigma_\eta^2)$ = $p(\mu)p(\phi)p(\sigma_\eta^2)$, with $\mu \sim \Normal{b_{\mu}, B_{\mu}}$, {$(\phi+1)/2 \sim \Betadis{a_0, b_0}$}, and  $\sigma_\eta^2 \sim %B_\sigma\times \chi^2_1=
\Gammad{\frac{1}{2},\frac{1}{2B_\sigma}}$,
with hyperparameters %$b_\mu$, $B_\mu$, $a_0$, $b_0$, and $B_\sigma $.
% In our practical applications,
$b_\mu = 0$, $B_\mu = 100$, $a_0 = 20$, $b_0 = 1.5$, and $B_\sigma = 1$.

An important building block of  our approach is a non-centered parametrization of the TVP model
in  the vein of \citet{fru-wag:sto}. First, we define $d$ independent random walk processes $\tilde\beta_{jt}, j=1, \ldots, d,$ with
standard  normal independent increments, i.e.
\begin{eqnarray}\label{eq:solve}
\tilde\beta_{jt}=\tilde\beta_{j,t-1}+\tilde\omega_{jt},\quad \tilde\omega_{jt}  \sim \Normal{0,1},
\end{eqnarray}
and initial value  $\tilde\beta_{j0}|P_{0,jj} \sim \Normal{0,{P}_{0,jj}}$.
Using the transformation
\begin{eqnarray}  \label{betatrans}
\beta_{jt} = \beta_j + \sqrt{\theta_j} \tilde \beta_{jt}, \qquad  t=0,\ldots,T,
\end{eqnarray}
we rewrite the state space model (\ref{eq:TVP}) and  (\ref{eq:state})
 by  combining  the $d$ state equations  for  $\tilde\beta_{jt}$   given in (\ref{eq:solve})
   with following observation equation:
\begin{eqnarray} \label{eq:obs}
 y_{t}=\xm_{t}\betav+\xm_{t}\text{Diag}(\sqrt{\theta_{1}},\dots,\sqrt{\theta_{d}}) \tilde \betav_{t}+\varepsilon_{t}.
\end{eqnarray}
The resulting state space model with state vector $\tilde \betav_{t}=(\tilde\beta_{1t},\dots,\tilde\beta_{dt})'$ is an alternative   parametrization of the TVP model, where the observation equation
 (\ref{eq:obs}) contains all unknown  parameters, i.e. the fixed regression coefficients $\beta_1, \dots, \beta_d$, %$\betav$; $\betav=(\beta_1, \dots, \beta_d)'$
as well as  the   (square roots of the) unknown  process variances  $\theta_{1},\dots,\theta_{d}$, whereas the state equations (\ref{eq:solve}) are  independent of any parameter.   Such a parameterization is called non-centered   in the spirit of \citet{pap-etal:gen}, whereas the original parametrization (\ref{eq:TVP}) and  (\ref{eq:state}) is called centered. Note that the initial state in the non-centered parametrization follows $\tilde \betav_0|\Pm_{0} \sim \Normult{d}{0,\Pm_{0}}$ with  $\Pm_{0}=\Diag{  P_{0,11}, \ldots, P_{0,dd}}$.

%Note that the initial value in the non-centered parameterization is assumed to be random (i.e.~$\tilde \betav_0 \sim \Normult{d}{0,\Pm_{0}}$) rather than zero (i.e.~$\tilde \betav_0 =\boldmath{0}$) as in earlier work \citep{fru-wag:sto,bel-etal:hie-tv}. We found that this additional randomness avoids overshrinking of the time-varying parameters $\betav_t$ toward $\betav$ for the first few time points. \marginpar{\comment{ADD Angela.}}

%\subsection{Shrinking process variances through single, double and triple gamma priors}  \label{sec:shrinkage}
\subsection{Shrinking process variances through the double gamma prior}  \label{sec:shrinkage}

 A popular prior choice  for the process variance $\theta_j$ is
 the inverted gamma  distribution, which is the conjugate prior for $\theta_j$ in the centered parameterization (\ref{eq:state}), see e.g.~\citet{pet-etal:dyn}:
\begin{eqnarray} \label{gaminv}
 \theta_j  \sim  \Gammainv{s_0,S_0}.
 \end{eqnarray}
However,   as shown by \citet{fru-wag:sto},  this prior fails to introduce shrinkage as it is bounded  away  from zero.   \citet{fru:eff} introduced a shrinkage prior for the process variance in a univariate TVP model (that is  $d=1$)  through the scale parameter   in the non-centered parametrization (\ref{eq:obs})  and  \citet{fru-wag:sto} extended  this idea to state space models with $d>1$.
 The scale parameter  $\sqrt{\theta}_ j \in \mathbb{R}$  is defined  as the positive and  the  negative root  of  ${\theta}_ j$ and
is allowed  to  take on positive and negative values.
Since the conjugate prior for $\sqrt{\theta}_j$ in the
non-centered parameterization (\ref{eq:obs}) is the normal distribution,
  $ \sqrt{\theta}_j$ is  assumed to be Gaussian with zero mean and scale parameters $\xi^{2}_j$:
 \begin{eqnarray} \label{normal}
  \sqrt{\theta}_j|\xi^{2}_j  \sim \Normal{0,\xi^{2}_j} \quad \Leftrightarrow  \quad \theta_j |\xi^{2}_j  \sim  \Gammad{\frac{1}{2},\frac{1}{2\xi^{2}_j}}.
 \end{eqnarray}
Shrinking $\theta_{j}$ toward the boundary value  is achieved  by shrinking  $  \sqrt{\theta}_j $   toward 0 (which is an interior point of the parameter space in the non-centered parametrization). For a sparse state space model,  prior (\ref{normal}) substitutes the inverted gamma prior (\ref{gaminv}) by a gamma prior.\footnote{We use  the parametrization  of the $\Gammad{\alpha,\beta}$ distribution with pdf given by $f(y) = \beta^\alpha y^{\alpha-1}e^{-\beta y}/\Gamma(\alpha)$.}  % namely a $\chi^2_1$-distribution.

 \begin{figure}[t!]%[h]
\centering
 \includegraphics[width=\textwidth,height=7cm]{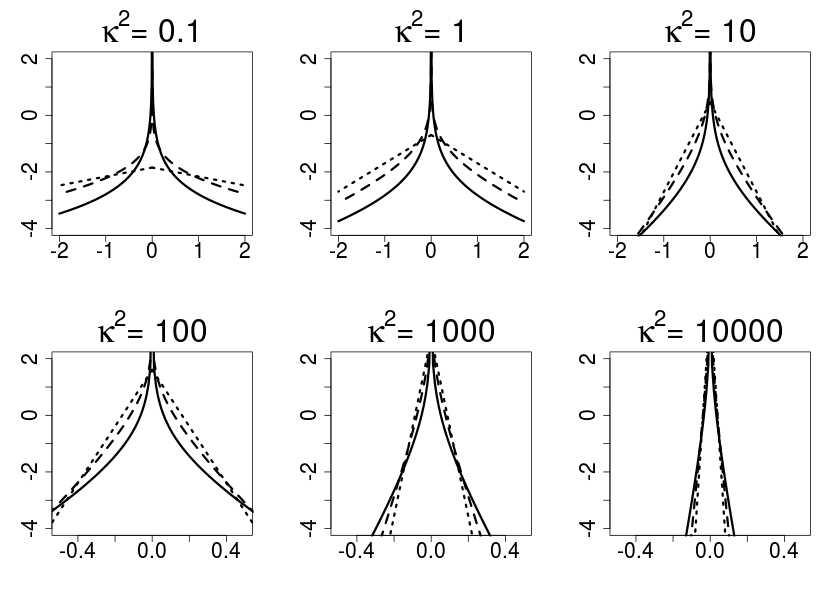}
% {./plots_for_diss/plots_12042016/log_density_normal_gamma_prior_13042016}
  \caption{Log  $p(\sqrt{\theta}_j|a^\xi,\kappa^2)$  of the double gamma prior for different values of $\kappa^2$
  and $a^\xi=0.1$ (solid line), $a^\xi=1/3$ (dashed line) and $a^\xi=1$ (dotted line).}
 \label{log_density_normal_gamma_prior}
\end{figure}

To discriminate between static  and time-varying components, \citet{fru-wag:sto} introduced  spike-and-slab priors, where  $\xi^{2}_j=0$
with positive prior probability
and $\xi^{2}_j$ is  fixed, otherwise  (e.g. $\xi^{2}_j=10$).
Instead of using spike-and-slab priors,  \citet{bel-etal:hie-tv} extended prior (\ref{normal}) by  adding  two levels of hierarchy to define a hierarchical
Bayesian Lasso   prior, where $ \xi^{2}_j$ follows an exponential distribution.
%Second, they allow this  population distribution to depend on a  random hyperparameter.

In the present paper, we introduce  a more general family of shrinkage priors derived from the normal-gamma prior,  introduced  by \citet{gri-bro:inf} for variable selection in standard regression models and applied in \citet{car-dou:spa} to multivariate regression models. The main idea is to use the
normal-gamma prior as a prior for $\sqrt\theta_{j}$  in the non-centered state space model, extending (\ref{normal}).
The normal-gamma prior is a scale mixture of normal distributions %  \citep{and-mal:sca},
with following  hierarchical representation:
 \begin{eqnarray} \label{equNGtheta}
 \displaystyle  %\pm
 \sqrt\theta_{j}| \xi^2_j &\sim& \Normal{0,\xi^2_j}, \qquad \xi_j^2|a^\xi,\kappa^2  \sim  \Gammad{a^\xi,a^\xi \kappa^2/2}.
\end{eqnarray}
In terms of the process variances $\theta_{j}$, (\ref{equNGtheta}) implies  that  $\theta_j$ follows a \lq\lq double gamma\rq\rq\  prior:
\begin{eqnarray} \label{normaltwo}
\theta_j|\xi^{2}_j  \sim \Gammad{\frac{1}{2},\frac{1}{2\xi^{2}_j}},  \qquad \xi_j^2|a^\xi,\kappa^2  \sim  \Gammad{a^\xi,a^\xi \kappa^2/2}.
 \end{eqnarray}
 For  $a^\xi =1$,   $\xi_j^2|a^\xi,\kappa^2$   reduces to an exponential distribution and the Bayesian Lasso prior  considered by
 \citet{bel-etal:hie-tv} results as a special case of the double gamma prior.

%The hierarchical representations (\ref{equNGtheta})  will be exploited for MCMC estimation.
%Closed form expressions, where $ \xi^2_j$ is integrated out, are  available   for  
Marginalizing over $ \xi^2_j$ yields closed form  expressions for 
$p(\sqrt{\theta}_j|a^\xi,\kappa^2 )$ and %, consequently, for   
 %= p( \sqrt{\theta}_j|a^\xi,\kappa^2) /  \sqrt{\theta}_j$:
 $p(\theta_j|a^\xi,\kappa^2 )$:\footnote{Note that $F_{\theta_j} (c)= \Prob{\theta_j\leq c}= \Prob{-\sqrt{c} \leq \sqrt{\theta}_j\leq \sqrt{c}} = 2 F_{\sqrt{\theta}_j} (\sqrt{c})$,  where $F_{\theta_j} (\cdot)$ is  the cdf of  the random variable  $\sqrt{\theta}_j$. Therefore,  $p(\theta_j|a^\xi,\kappa^2 ) = p( \sqrt{\theta}_j|a^\xi,\kappa^2) /  \sqrt{\theta}_j$.}
\begin{eqnarray} \label{teetamarg}
p(\sqrt{\theta}_j|a^\xi, \kappa^2) &=&\frac{(\sqrt{a^\xi \kappa^2})^{a^\xi+1/2}}{\sqrt{\pi}2^{a^\xi-1/2} \Gamma(a^\xi)} |\sqrt{\theta}_j|^{a^\xi-1/2}
K_{a^\xi-1/2}( \sqrt{a^\xi \kappa^2}  |\sqrt{\theta}_j|), \\
 p( \theta_j|a^\xi,\kappa^2) &=& \frac{(\sqrt{a^\xi \kappa^2})^{a^\xi+1/2}}{\sqrt{\pi}2^{a^\xi-1/2}\Gamma(a^\xi)} ( \theta_j ) ^{a^\xi/2-3/4}
 K_{a^\xi-1/2}(  \sqrt{a^\xi \kappa^2 \theta_j}),  \nonumber
\end{eqnarray}
%\begin{equation} \label{betamarg}
% p(\sqrt{\theta}_j|\kappa^2) = \frac{1}{\sqrt{\pi}2^{a^\xi-1/2} \sqrt{a^\xi \kappa^2}^{a^\tau+1/2}\Gamma(a^\xi)} |\sqrt{\theta}_j|^{a^\xi-1/2} K_{a^\xi-1/2}(\sqrt{\theta}_j |\sqrt{a^\xi \kappa^2})
%\end{equation}
where $K_p(\cdot)$ is the modified Bessel function of the second kind with index $p$.
The display of   $ \log  p(\sqrt{\theta}_j|a^\xi,\kappa^2)$  for different values of $\kappa^2$  in  Figure~\ref{log_density_normal_gamma_prior}  shows  that the  double gamma prior with $a^\xi \leq  1$ is an  example of a  global-local shrinkage prior \citep{pol-sco:shr}.  A  pronounced spike at zero is present and the  mass placed close to zero strongly depends on  the global parameter $\kappa^2$.
 From  representation   (\ref{normaltwo})  we obtain that, marginally,  $\Ew{\theta_j} %= \E{ \E{\theta_j|\xi^{2}_j} =  \Ew{\xi^{2}_j}
% = \frac{2}{\kappa^2}$,
= 2/\kappa^2$, whereas
 \begin{eqnarray*} \label{eq:vartheta}
 \V(\theta_j) = \Ew{\theta_j^2}- \Ew{\theta_j}^2= % \Ew{2(\xi^{2}_j)^2/2+(\xi^{2}_j)^2} -   \frac{4}{\kappa^4}
 3 \Ew{ (\xi^{2}_j)^2} -   \frac{4}{\kappa^4}
 =\frac{12} {a^{\xi} \kappa^4} +   \frac{8}{\kappa^4} = \Ew{\theta_j}^2 (2+ 3/a^{\xi}).
 \end{eqnarray*}

\noindent Hence, independently of  $a^{\xi}$,  the hyperparameter $\kappa^2$ controls the global  level of shrinkage, which is the stronger,
the larger  $\kappa^2$.   At the same time, also $ \V(\theta_j)$ decreases, as $\kappa$ increases.
Therefore, the larger  $\kappa^2$, the  more mass is placed close to zero.
 On the other hand, the term $3/a^{\xi}$ -- which is  equal to the excess kurtosis of   $\sqrt{\theta}_j$  --
controls local adaption to the  global  level of shrinkage, with more local adaption, the smaller  $a^{\xi}$.
As $a^\xi$ decreases, the excess kurtosis of $\sqrt{\theta}_j$ increases and   the tails of $p(\sqrt{\theta}_j|a^\xi,\kappa^2)$  become thicker.
%, in particular for values  of $\kappa$ that are not too large. %,  see again Figure~\ref{log_density_normal_gamma_prior}.

It is also illuminating to investigate the joint marginal prior distribution of  $(\theta_1, \ldots,\theta_d)$ or (equivalently)  of $(\sqrt{\theta}_1, \ldots,\sqrt{\theta}_d)$ given $a^{\xi}$ and $\kappa^2$.  Since the random prior variances  $\xi^2_j$ in (\ref{normaltwo}) are drawn independently,
 also marginally  the double gamma prior is characterized  by    prior conditional independence of $(\theta_1, \ldots,\theta_d)$  given  fixed values of
  $a^\xi$ and $\kappa^2 $:
 % \begin{eqnarray*}
   $p( \theta_1, \ldots,\theta_d|a^\xi,\kappa^2)= \prod_{j=1}^d p(\theta_j|a^\xi,\kappa^2 )$.
 %\end{eqnarray*}

\begin{figure}[t!]%[ht]
\centering \vspace{-1.5cm}
\begin{tabular}{ccc}
\scalebox{0.2}{\includegraphics{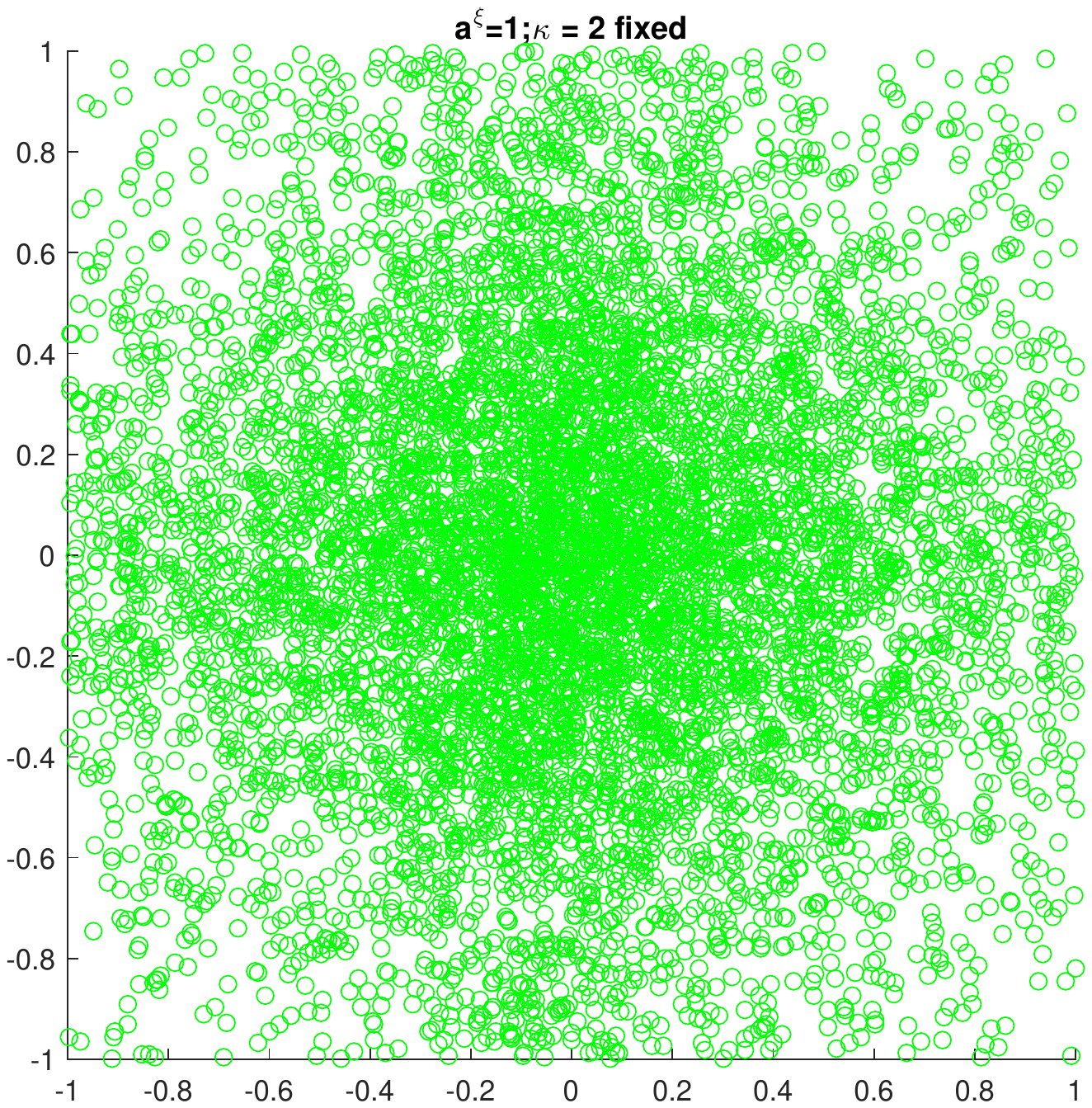}}
& \scalebox{0.2}{\includegraphics{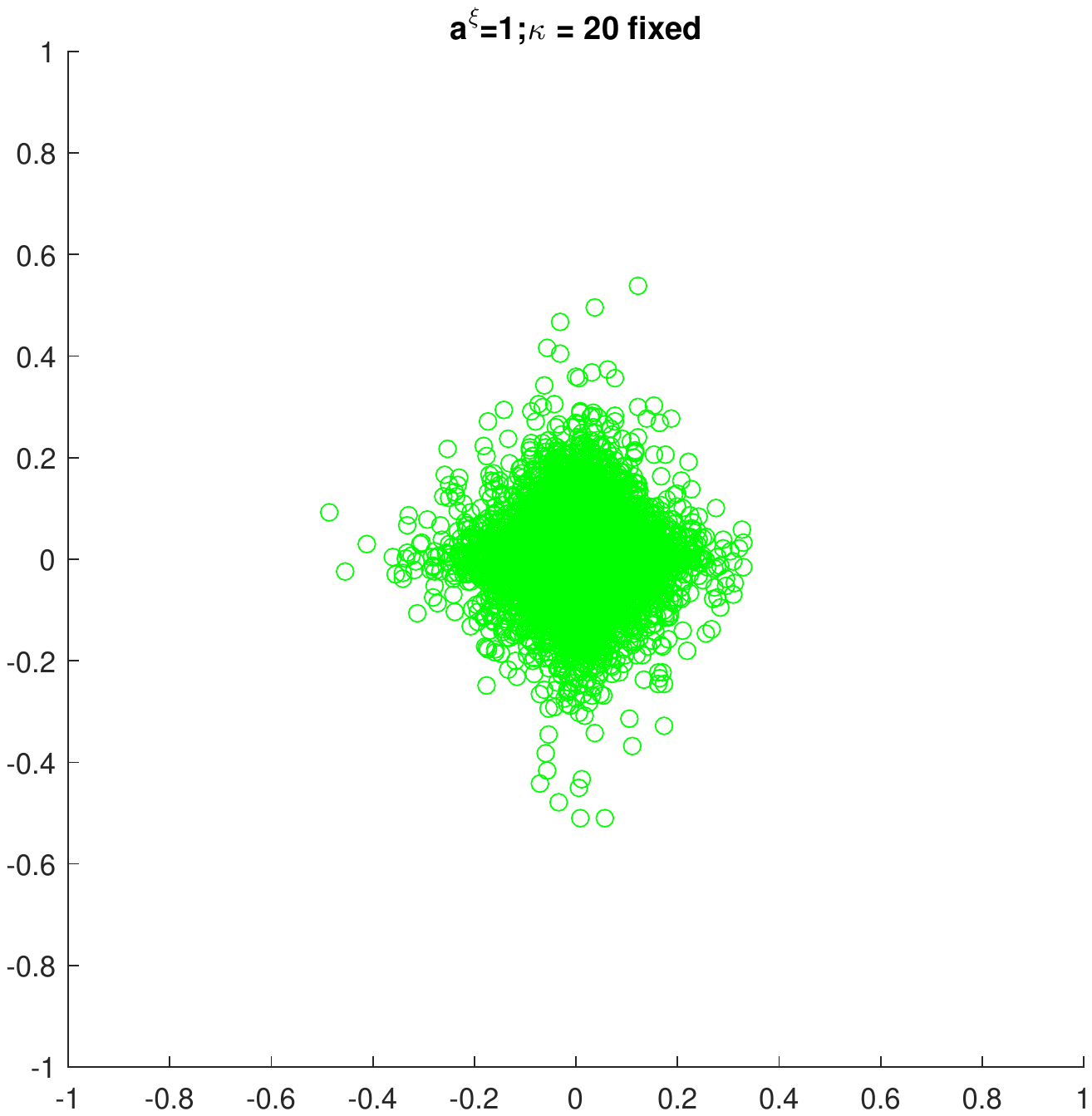}}
& \scalebox{0.2}{\includegraphics{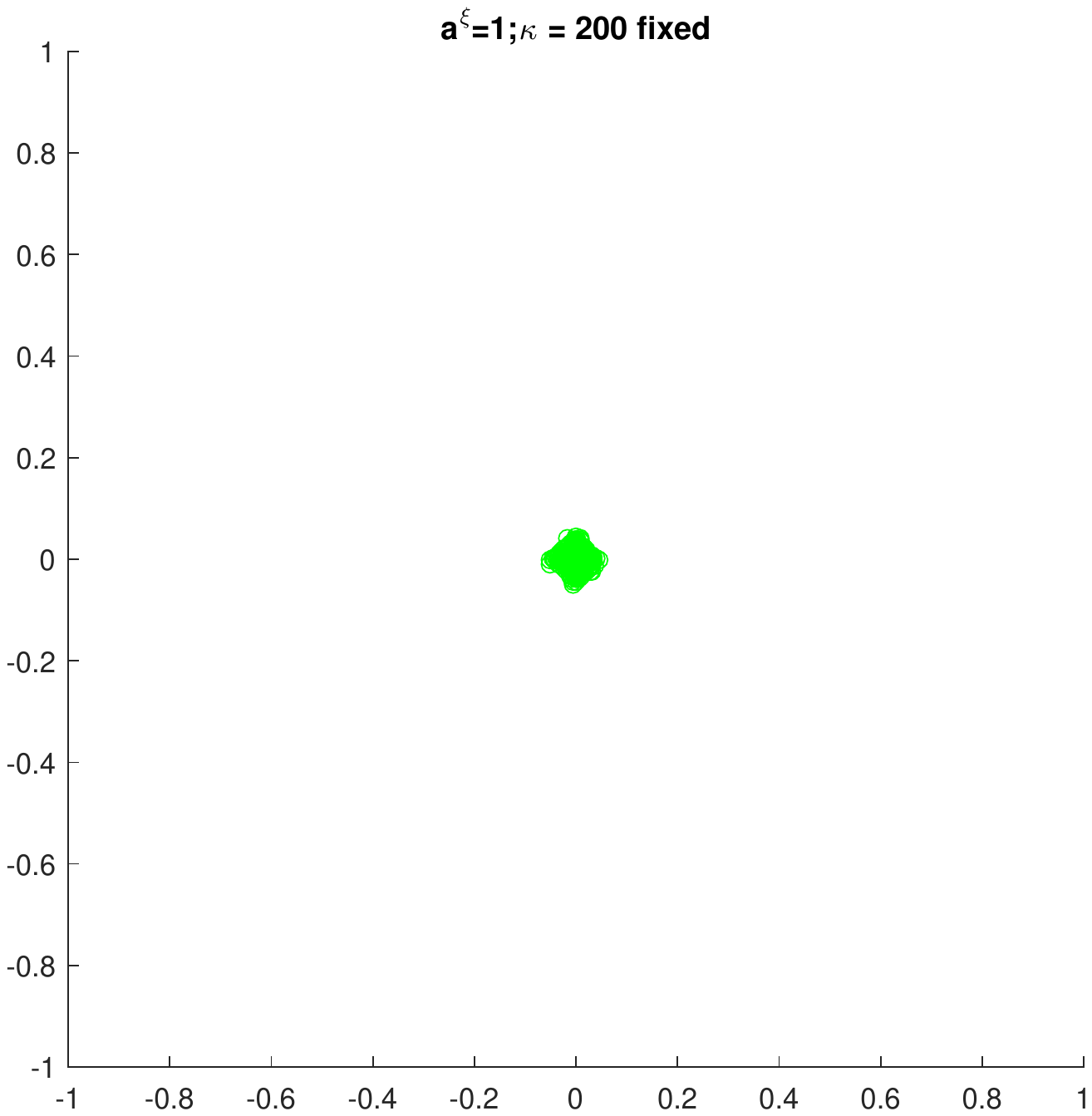}}\\[-3.2cm]
\scalebox{0.2}{\includegraphics{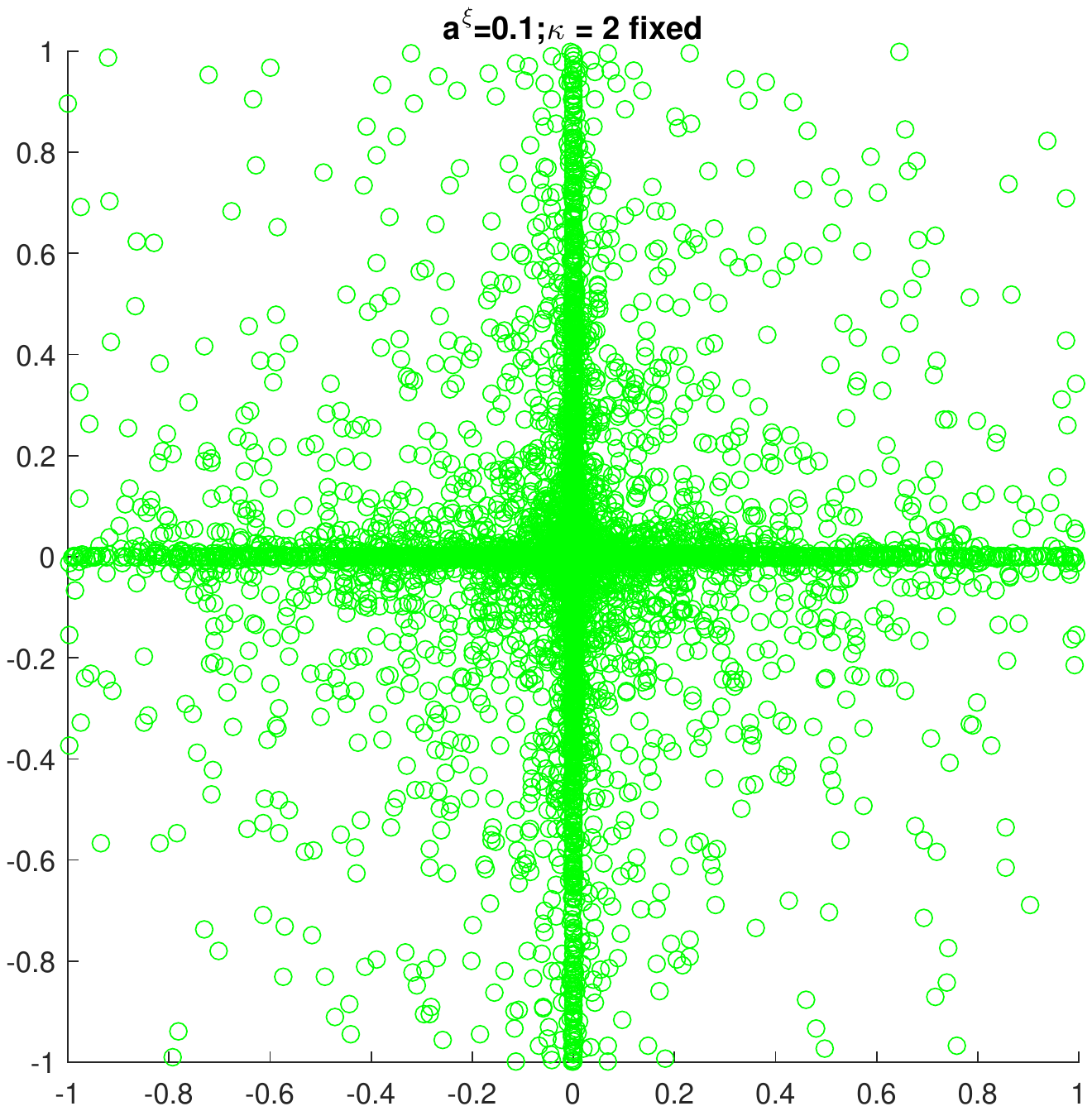}}
& \scalebox{0.2}{\includegraphics{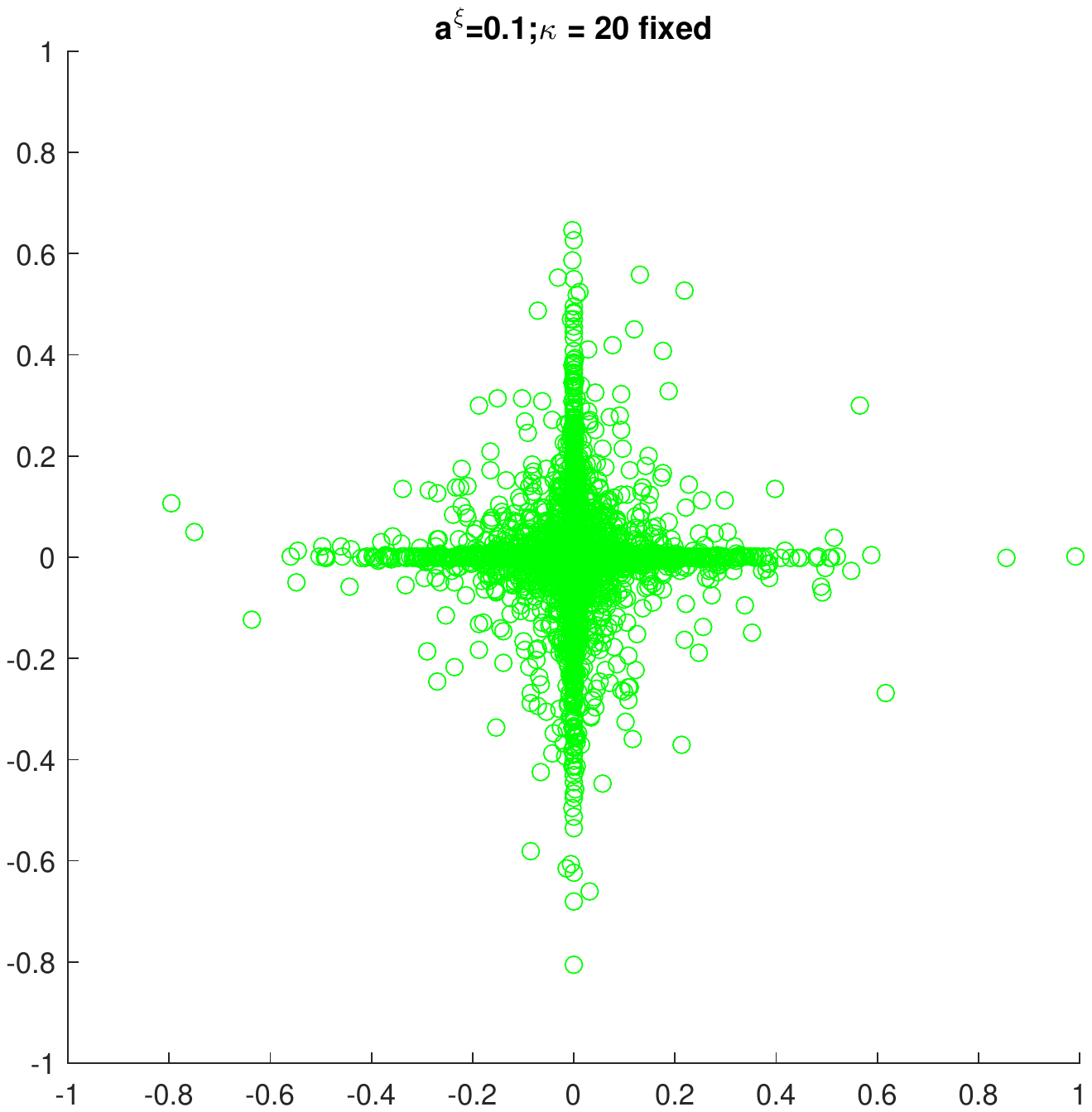}}
&\scalebox{0.2}{\includegraphics{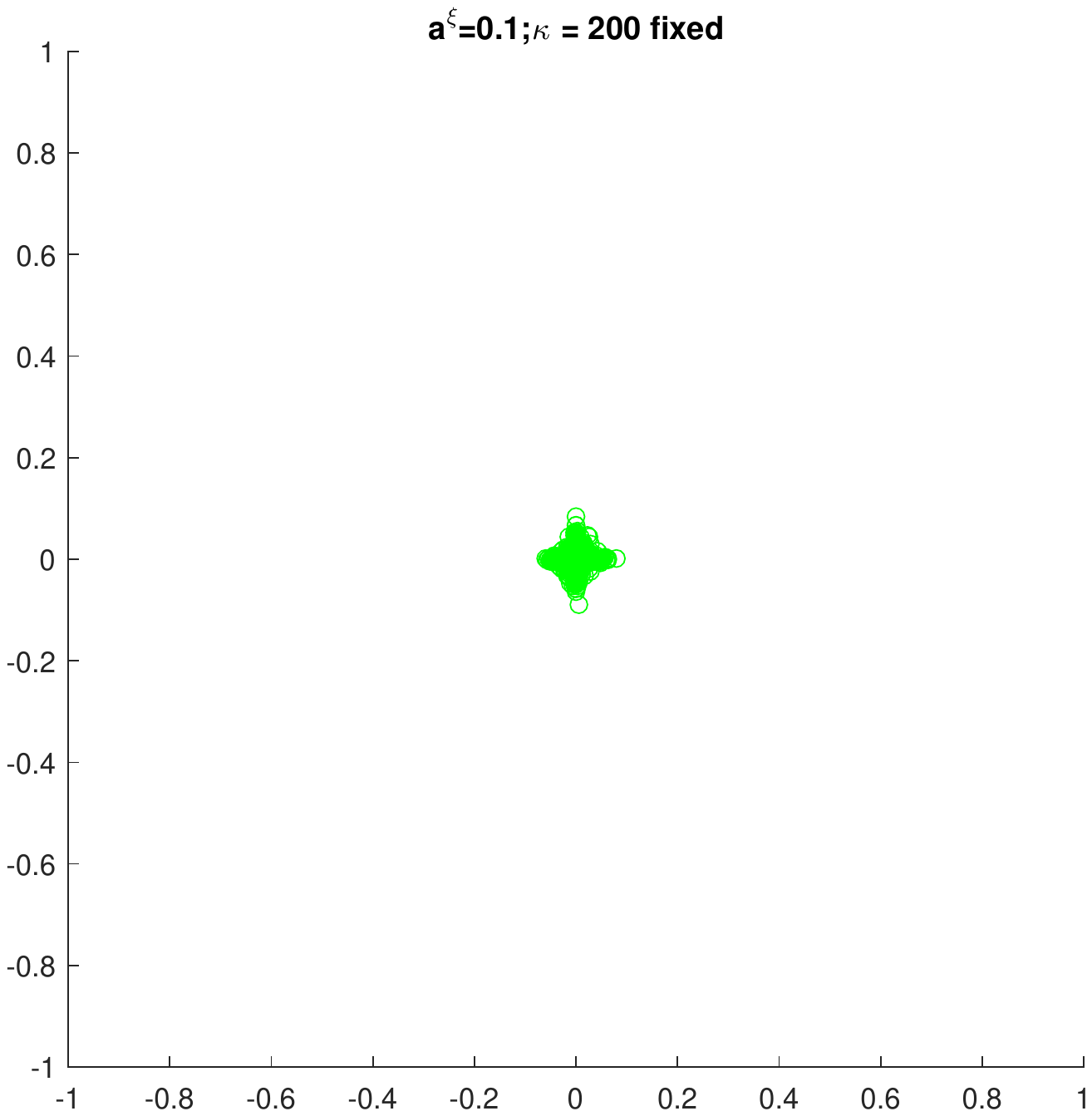}}\\[-1.5cm]
\end{tabular}
\caption{Simulations from the double gamma prior  $p( \sqrt{\theta}_1, \sqrt{\theta}_2|a^\xi,\kappa^2)$  for $a^\xi=1$  (top) and  $a^\xi=0.1$   (bottom) for different values of  $\kappa^2$ (left-hand side: $\kappa^2 =2$, middle: $\kappa^2 =20$, right-hand side: $\kappa^2 =200$). The plots at the top correspond to the Bayesian Lasso prior.}  \label{log_density_normal_gamma_biv}
\end{figure}

For illustration, Figure~\ref{log_density_normal_gamma_biv} shows  simulations from the joint prior $p( \sqrt{\theta}_1, \sqrt{\theta}_2|a^\xi,\kappa^2)$ for $d=2$ for various values of  $a^{\xi}$ and $\kappa^2$.   Not surprisingly from the previous discussions, for the same value of $\kappa^2$,   the double gamma with
$a^{\xi} =0.1$ has a pronounced spike at 0 with  fat tails  in both  directions  of
$\sqrt{\theta}_1$ and $ \sqrt{\theta}_2$  and  provides more flexible shrinkage compared to the Bayesian Lasso prior ($a^{\xi} =1$).  For the Bayesian Lasso prior,  large values of  $\kappa^2$ (e.g. $\kappa^2=200$) are needed to introduce strong shrinkage toward 0.
 %whereas   the normal-gamma prior  looses the  fat tails for large values for $\kappa^2$,  even if $a^\xi$ is small,  see also Figure~\ref{log_density_normal_gamma_prior}.  Thus, for different values of $a^\xi$ different values of $\kappa^2$ seem to be optimal.

To infer   appropriate values of  $a^{\xi}$ and $\kappa^2$   from the data,  hierarchical priors are employed.
We assume that   $\kappa^2$ follows  a gamma distribution with fixed hyperparameters $d_1$ and $d_2$:
  \begin{eqnarray} \label{conjkl}
\kappa^2 \sim \Gammad{d_1,d_2}. 
\end{eqnarray}
  For $a^{\xi}=1$ this corresponds to the  hierarchical Bayesian Lasso prior considered by \citet{bel-etal:hie-tv}.
In addition, we assume that  the shrinkage parameter $a^\xi$ follows  an exponential distribution  as in \citet{gri-bro:inf},
  \begin{eqnarray} \label{prioraxi}
a^\xi \sim \Exp{(\axipr)},
\end{eqnarray}
 with  a fixed hyperparameter $\axipr \geq 1$.
Combining  (\ref{normaltwo}) with (\ref{conjkl}) and (\ref{prioraxi})  defines the   hierarchical double gamma  prior.
%Apart from adapting  the hyperparameter $\kappa^2$ to the time series data at hand,  prior  (\ref{conjkl}) also
Given  the hyperparameters $d_1$, $d_2$, and $\axipr$, this hierarchical prior introduces prior  dependence among  $(\theta_1, \ldots,\theta_d )$  which is advantageous
in a shrinkage framework, as recently shown by \citet{gri-bro:hie}.
 Prior dependence  is  desirable  in situations, where  only a few  variances are  expected to be different from 0. In this case, whether a certain process variance is  shrunken toward 0  depends on how close the other process variances  are to 0.

%For illustration, Figure~\ref{log_density_normal_gamma_hier} shows  simulations from the joint prior $p( \sqrt{\theta}_1, \sqrt{\theta}_2|d_1,d_2)$ for $d=2$. For the same set of hyperparameters   $d_1=d_2$, the prior with $a^{\xi}=0.1$ shows a much more flexible shrinkage behaviour than the prior with $a^{\xi}=1$ which corresponds to the  hierarchical Bayesian lasso prior considered by \citet{bel-etal:hie-tv}.

 %\begin{figure}[ht]
%\centering
%\begin{tabular}{ccc}
%\scalebox{0.25}{\includegraphics{./plot_JE/sim_biv_e001_lasso}}
%& \scalebox{0.25}{\includegraphics{./plot_JE/sim_biv_e01_lasso}}
% &\scalebox{0.25}{\includegraphics{./plot_JE/sim_biv_e10_lasso}}\\[-3.2cm]
%\scalebox{0.25}{\includegraphics{./plot_JE/sim_biv_e001_ng}}
% & \scalebox{0.25}{\includegraphics{./plot_JE/sim_biv_e01_ng}}
% &  \scalebox{0.25}{\includegraphics{./plot_JE/sim_biv_e10_ng}}\\[-1.5cm]
%\end{tabular}
%\caption{Simulations from the triple gamma prior  $p( \sqrt{\theta}_1, \sqrt{\theta}_2|d_1,d_2)$  for  $a^\xi=1$  (top) and   $a^\xi=0.1$  (bottom) and different values of  $d_1$ and $d_2$ (left hand side: $d_1=d_2 =0.001$, middle: $d_1=d_2  =0.01$, right hand side: % $d_1=d_2  =0.1$ and $d_1=d_2  =1$).}  \label{log_density_normal_gamma_hier}
%\end{figure}

%\subsection{Shrinkage priors for the fixed regression coefficients} \label{sec:priorbeta}

Prior dependence also exists between  $\beta_{j0}$ and $\theta_j$, as  the size (but not the sign) of $\beta_{j0}-\beta_j$ depends on $\theta_j$ through $\V(\beta_{j0}-\beta_j|\theta_j)= \theta_j P_{0,jj}$.  If  $\theta_j$ is shrunken toward 0, then   $\beta_{j0}$ and all subsequent values  $\beta_{jt}$ are  pulled toward   $\beta_{j}$ for covariate $x_{tj}$. In high dimensions, where  many coefficients are expected  to be static,
it is of interest to allow a  practically constant  coefficient $\beta_{jt}$ to be  insignificant throughout the entire observation period. As these coefficients are characterized by a parameter setting where both  $\theta_{j}$ and $\beta_{j}$ are close to 0,  a   second normal-gamma prior is  employed as a shrinkage prior for   $\beta_{j}$ to allow shrinkage of $\beta_{j}$ toward 0:\footnote{A  closed form expression, comparable to
(\ref{teetamarg}),  is available for  $p(\beta_j|a^\tau,\lambda^2) $, with    expectation  $ \Ew{|\beta_j| }= \sqrt{\frac{4}{\pi a^\tau \lambda^2}} \frac{\Gamma(a^\tau+1/2)}{\Gamma(a^\tau)}$,  $\V(\beta_j)= \frac{2}{\lambda^2}$, while  the excess kurtosis is given by $\frac{3}{a^\tau}$.}
 \begin{eqnarray}  \label{equNG01}
  \beta_{j}|\tau^2_j \sim  \Normal{0,\tau^2_j}, \qquad \tau_j^2|a^\tau ,\lambda^2 \sim  \Gammad{a^\tau,a^\tau \lambda^2/2}.
 \end{eqnarray}
%, namely
%\begin{equation*} \label{betamarg}
 %p(\beta_j|\lambda^2) = \frac{\sqrt{a^\tau \lambda^2}^{a^\tau+1/2}}{\sqrt{\pi}2^{a^\tau-1/2} \Gamma(a^\tau)} |\beta_j|^{a^\tau-1/2} K_{a^\tau-1/2}(|\beta_j|\sqrt{a^\tau \lambda^2}).
%\end{equation*}
%where $K_p()$ is the modified Bessel function of the second kind with index $p$.
In this case,   any  (practically constant) coefficient $\beta_{jt}$   is  insignificant,
 whenever the corresponding fixed regression effect  $\beta_{j}$   %  ($\beta_{j}\neq 0 $)
is  zero.\footnote{It should be noted that  the data are not informative about  $\beta_{j}$, if   $\theta_{j} > 0$, but they  are always informative about the initial regression coefficient $\beta_{j0}$.  For $\theta_{j}= 0$,  $\beta_{j0}$ and $\beta_{j}$ coincide.}
%
%Again,  for $a^\tau =1$,  prior  (\ref{equNG01}) reduces to the prior discussed in \citet{bel-etal:hie-tv}.
%
%Since $\beta_1, \ldots,\beta_d|\lambda^2 $   are independent  apriori, if  $\lambda^2 $  is  fixed,
 % \begin{eqnarray}
 %p( \beta_1, \ldots,\beta_d|\lambda^2 )= \prod_{j=1}^d p(\beta_j|\lambda^2).
%\end{eqnarray}
 Similarly as for $\theta_j$,  another layer of hierarchy is added, by assuming that   $\lambda^2 \sim \Gammad{e_1,e_2} $
%  \begin{eqnarray} \label{priorlam}
 %\lambda^2 \sim \Gammad{e_1,e_2},
%\end{eqnarray}
% This introduces prior dependence,  and whether a fixed regression coefficient $\beta_j$ should be shrunken toward 0 or not also depends on how many other coefficients are shrunken toward 0.
and  $a^\tau \sim \Exp{(\ataupr)}$ with fixed hyperparameters  $e_1,e_2$ and  $\ataupr \geq 1$.
%hold as in the previous subsection.

\section{MCMC Estimation} \label{sec:mcmc}

To carry out Bayesian inference for  a sparse TVP model under the  shrinkage priors  introduced in Section~2, we develop an efficient scheme for  Markov chain Monte Carlo (MCMC) sampling,   given all hyperparameters,  % i.e.~$\ataupr, \axipr, d_1, d_2, e_1,e_2$
 i.e.~$e_1,e_2, \ataupr, d_1, d_2, \axipr$
in the priors for $\betav$ and $\Qm$,
$c_P, \nu_P$  in the prior  of  $P_{0,11}, \ldots, P_{0,dd}$,
% $P_{0,11}, \ldots, P_{0,dd} $ in the prior  of $\betav_0$,
as well as $c_0,g_0, G_0$ for homoscedastic variances $\sigma^2$
and $b_\mu,B_{\mu}, a_0,b_0,B_\sigma$ for parameters of   %$\sigma^2_t$,
  the SV model  (\ref{svht}).
 Bayesian inference operates in the latent variable formulation of the TVP  model
  and relies on data augmentation of the latent processes  $ \betav=( \betav_0,  \betav_1, \ldots,  \betav_T)$
 for the centered  and $\tilde \betav=(\tilde \betav_0, \tilde \betav_1, \ldots, \tilde \betav_T)$  for the  non-centered parameterization.
For the  SV model,   the log volatilities
$\bold h=(h_0,\dots,h_T)$  are introduced as additional  latent variables.

 For the centered parameterization under the common inverted gamma prior  (\ref{gaminv}) for the process variances $\theta_j$,
    Gibbs sampling is totally standard, see e.g. \citet{pet-etal:dyn}.
   However,  %like many  MCMC schemes which alternate between sampling from the full conditionals of the latent states and the model parameters,
    if some of the  process variances are small, then this MCMC scheme   suffers from  slow convergence and poor mixing of the sampler. %, see e.g.~\citet{fru:eff}.
    As shown by \citet{fru-wag:sto}, MCMC estimation based on the non-centered parameterization   proves to be useful,  in particular if process variances are close to 0.

%However, MCMC estimation based on different parameterizations will often be efficient in separate regions of the parameter space.
 \citet{fru:eff}  discusses the relationship between the various
 parametrizations for  a simple TVP model  %with $d=1$
 and the computational efficiency of the resulting MCMC samplers, see also  \citet{pap-etal:gen}.
 For TVP models with $d>1$,  MCMC estimation in the centered parameterization
is  preferable for all coefficients that are actually time-varying,
whereas the non-centered parametrization  is preferable for (nearly) constant coefficients.
   For practical time series analysis, both types of coefficients are likely to be present and choosing a computationally efficient parametrization in advance is not possible.  %Subsequently,

   We show how  these two data augmentation schemes can be combined   through the {\em ancillarity-sufficiency interweaving strategy} (ASIS)   introduced by  \citet{yu-men:cen} to obtain an efficient sampler  combining  the \lq\lq best of  both worlds\rq\rq .
ASIS provides a  principled way of  interweaving different data augmentation  schemes by  re-sampling certain parameters conditional on the latent variables in the alternative parameterization of the model. This strategy has been successfully employed to  univariate SV models \citep{kas-fru:anc},
 multivariate factor SV models \citep{kas-etal:eff}  and
dynamic linear state space models \citep{sim-etal:int}.
In the present paper,  ASIS  is applied  to interweave  the centered and the non-centered parameterization of a TVP model. More specifically,
  we use the non-centered parametrization as baseline, and interweave into the centered parameterization.
  This leads to the  MCMC sampling  scheme outlined  in Algorithm~\ref{facsvalg}
 which  increases posterior sampling efficiency considerably compared to conventional Gibbs sampling for  either of the two parameterizations.

  \begin{alg} \label{facsvalg}
 Choose %  appropriate
 starting values for
$\betav,\Qm $,    $\tauv=(\tau_1,\ldots, \tau_d)$, $\xiv=(\xi_1,\ldots, \xi_d)$,  $a^{\tau},\lambda^2, a^{\xi}$, $\kappa^2, \Pm_{0}$,
and  (for homoscedastic variances) $\sigma^2$ and $ C_0$  and repeat the following steps:
 \begin{enumerate}
  \item[(a)] Sample the states $  \tilde \betav =( \tilde \betav_0, \ldots, \tilde \betav_T)$ in the non-centered parametrization   from
  the multivariate Gaussian posterior $ \tilde \betav | \betav,\Qm , \Pm_{0},\sigma^2  \sim \Normult{(T+1)d}{\Omegav^{-1}\cm,\Omegav^{-1} }$ given in (\ref{postbetav}).
\item[(b)] Joint sampling of $\alphav  = (\beta_1, \dots,\beta_d, \sqrt{\theta_1},\dots,\sqrt{\theta_d})^{\prime}$ from the multivariate Gaussian posterior
$p(\alphav| \tilde \betav, \tauv, \xiv, \sigma^2, \ym) $ given in (\ref{app:allpar}).

\item[(c)] For each $j=1,\ldots,d$, redraw the constant coefficient $\beta_j$ and the square root of the process variance $\sqrt{\theta_j}$  through interweaving into the state equation of the centered parameterization:
    \begin{enumerate}
 \item[(c-1)]  Use the transformation (\ref{betatrans}) to match  the draws of the latent process
 $\tilde \beta_{j0}, \ldots, \tilde \beta_{jT}$ in the non-centered  %parameterization  (\ref{eq:solve}) and (\ref{eq:obs})
 to the  latent process $\beta_{j0},\ldots, \beta_{jT}$ in the  centered parameterization
 % (\ref{eq:TVP2}) and (\ref{eq:TVP}):
and  store the sign of $\sqrt{\theta_j}$.

\item[(c-2)] Update  $\beta_j$ and $\theta_j$ %, conditional on  the state process $\beta_{j0},\ldots, \beta_{jT}$
 in the centered parameterization by sampling $\theta_j\new$ % of each state equation
 from  the generalized inverse Gaussian  posterior $\theta_j|\beta_{j0},\ldots, \beta_{jT}, \beta_j, \xi^2_j,P_{0,jj}$, given in (\ref{gigtheta}), and 
 %sampling   
 $\beta_j \new$ from the Gaussian posterior  $  \beta_j | \beta_{j0}, \theta_j\new, \tau^2_j , P_{0,jj}$, given in  (\ref{inicoen}).

   \item[(c-3)]  Determine  $\sqrt{\theta \new_j}$  using  the same  sign as the old value $\sqrt{\theta_j}$.
Based on  $\sqrt{\theta \new_j}$  and $\beta_j \new$, the state process $\tilde \beta_{jt}$  in the non-centered parameterization  is updated  in a deterministic manner through the inverse of the transformation (\ref{betatrans}):
\begin{eqnarray*}
\tilde \beta _{jt} \new =  % \frac{(\beta_{jt} - \beta \new _j)}{\sqrt{\theta \new_j}},
(\beta_{jt} - \beta \new _j)/\sqrt{\theta \new_j},  \qquad  t=0,\ldots,T.
\end{eqnarray*}
\end{enumerate}

\item[(d)] Sample from $a^\tau| \beta_1,\ldots,\beta_d,\lambda^2$   and $ a^\xi| \sqrt{\theta_1}, \ldots, \sqrt{\theta_d}, \kappa^2 $   using a random walk Metropolis-Hastings (MH)  step based on proposing  $\log a^{\tau,\mathrm{new}} \sim \Normal{\log a^{\tau},c_\tau^2}$  and  $\log a^{\xi,\mathrm{new}} \sim \Normal{\log a^{\xi},c_\xi^2}$.

\item[(e)]  Sample the prior variances  $ \tau_j |\beta_j, a^\tau, \lambda^2 $   and  $ \xi_j|  \theta_j, a^\xi, \kappa^2$, for $j=1,\ldots,d$,  from conditionally independent  generalized inverse Gaussian distributions  given  in  (\ref{app:tau})   and   (\ref{app:xi}), respectively,
      and  update  the hyperparameters $\lambda^2|a^\tau, \tauv$ and  $\kappa^2|a^\xi,   \xiv$  from the gamma distributions given  in (\ref{lambdapost})  and
       (\ref{kappapost}).

\item[(f)]   Sample %  the error variance  
 $\sigma^2 |\tilde{\betav}, \boldsymbol{\alpha},C_0, \ym$ from the following  inverted gamma distribution
     \begin{eqnarray*}% \label{signing}
 \sigma^2 |  \tilde{\betav}, \boldsymbol{\alpha},C_0 , \ym  \sim  \Gammainv{c_0 + \frac{T}{2}, C_0 + \frac{1}{2} \sum _{t=1}^T (y_t -\mathbf{z}_t \boldsymbol{\alpha})^2}, \quad
 \end{eqnarray*}
 where  $ \mathbf{z}_t$ is defined in (\ref{regzt}), and sample  $C_0$ from % following gamma distribution:
   $C_0 | \sigma^2  \sim  \Gammad{g_0 + c_0, G_0 + \frac{1}{\sigma^2}}$.
\item[(g)]  Sample  the scale parameters of the initial distribution  for each $j=1,\ldots,d$,  from $P_{0,jj}|\tilde \beta_{j0}  \sim \Gammainv{\nu_P+\frac{1}{2}, (\nu_P-1)c_P + \frac{1}{2} \tilde \beta^2_{j0} }$.
\end{enumerate}
\end{alg}

  After discarding a certain amount of initial draws (the \emph{burn-in}), the  full conditional sampler  iterating Steps~(a) to  (g) of Algorithm~\ref{facsvalg} yields draws from the joint posterior distribution $p( \tilde \betav, \beta_1, \dots,\beta_d, \sqrt{\theta_1},\dots,\sqrt{\theta_d},  \tauv, \xiv,  a^\tau, \lambda^2, a^\xi, \kappa^2,  \Pm_{0},\sigma^2, C_0, |\ym)$ under the hierarchical shrinkage priors outlined in Section~\ref{sec:shrinkage}. % and \ref{sec:priorbeta}.

In Step~(a), we sample the latent states  $\tilde \betav=(\tilde \betav_0, \ldots, \tilde \betav_T)$  in the non-centered parametrization
conditional on known parameters $\betav,\Qm ,\Pm_{0}$ and known error variances $\sigma^2$. % $\sigma^2_1, \ldots, \sigma^2_T$.
As an alternative to the  commonly used  {\em Forward Filtering Backward Sampling} \citep{fru:dat,car-koh:ong},
 we  implemented  a multi-move sampling algorithm  in the spirit of \citet{mcc-etal:sim}
 which allows to sample the entire state process $\tilde \betav$
 \textit{all without a loop} (AWOL; \citet{kas-fru:anc}). Full details are provided in Appendix~\ref{section:FFBS}.

 In Step~(b), conditional on the latent states $\tilde \betav$, a regression type
 model results from the observation equation (\ref{eq:obs}) of the non-centered state space model.
Based on the Gaussian priors appearing  in   the hierarchical representations of the shrinkage  priors  (\ref{equNGtheta}) and (\ref{equNG01}),
we sample the parameters $\beta_1,\dots,\beta_d$ and $\sqrt{\theta_1},\dots,\sqrt{\theta_d}$ jointly from
the conditionally Gaussian posterior  given in (\ref{app:allpar});
see Appendix~\ref{stepB} for details.
One major advantage of working with the square root of the process variance $\sqrt{\theta_j}$, instead of  $\theta_j$, is that we avoid
boundary space problems for small variances, resulting in better mixing behaviour of the sampler.

 The interweaving Step~(c) turns out to be  instrumental for an efficient implementation of the hierarchical shrinkage priors introduced  in Section~\ref{sec:shrinkage}. % and \ref{sec:priorbeta}
In this step, we temporarily move from the non-centered to the centered parameterization to  resample $\beta_j$ and $\theta_j$.  To ensure that  the posterior distributions  obtained with and   without interweaving are identical,
   the priors between the non-centered and the centered parametrization are matched. Whereas the Gaussian prior $\beta_j|\tau^2_j $  for the initial value $\beta_j$ is  the same  for both parameterizations, we  transform the Gaussian  prior for $ \sqrt\theta_{j} |\xi_j^2$
 to the corresponding gamma prior for  $\theta_{j} |\xi_j^2$ in the centered parameterization, see (\ref{normal}).
 In Step~(c-2),  the  posteriors of $\theta_j$ and $\beta_j$  in the centered parameterization,
conditional on the state process $\beta_{j0},\ldots, \beta_{jT}$, are easily obtained.
First, the  conditional  posterior
\begin{eqnarray*}
p(\theta_j|\beta_{j0},\ldots, \beta_{jT} ,\beta_j,  \xi^2_j,P_{0,jj}) \propto  p(\theta_j|\xi^2_j) p(\beta_{j0}|\beta_j,\theta_j,P_{0,jj}) \prod_{t=1}^T  p(\beta_{jt}|\beta_{j,t-1}, \theta_j) ,
 \end{eqnarray*}
where % the prior equals
$\beta_{j0}|\beta_j,\theta_j ,P_{0,jj}  \sim \Normal{\beta_j,\theta_j P_{0,jj}  }$
and  % the likelihood results from
 $\beta_{jt}|\beta_{j,t-1},\theta_j \sim \Normal{\beta_{j,t-1},\theta_j}$, % for $t=1,\dots,T$,
% Hence,  the   conditional posterior  $p(\theta_j|\beta_j, \beta_{j0},\ldots, \beta_{jT} ,\xi^2_j,P_{0,jj})$
 is the density of a generalized inverse Gaussian distribution (GIG) with  following parameters:
\begin{eqnarray} \label{gigtheta}
\theta_j|\beta_{j0},\ldots, \beta_{jT} , \beta_j, \xi^2_j,P_{0,jj}  \sim  \GIG{-\frac{T}{2}}{\frac{1}{\xi^2_j}}{\sum^T_{t=1} (\beta_{jt} -\beta_{j,t-1})^2 +\frac{(\beta_{j0} -\beta_{j})^2}{{P_{0,jj}}}}.
 \end{eqnarray}
Note that sampling the process variance $\theta_j$ from this GIG posterior\footnote{To sample from the GIG distribution, we use  a  method proposed by \citet{hoe-ley:gen} which is  implemented in the R-package {\tt GIGrvg} \citep{hoe-ley:gig}. %, see Appendix~\ref{stepdd}.
  This method is especially reliable  for TVP models where the  scale parameters of the GIG distribution   can be extremely small due to shrinkage and other samplers tend to fail.}  deviates from the usual MCMC inference for the centered state space model, since the conditionally conjugate inverted gamma prior  (\ref{gaminv})  is substituted by a prior from the gamma distribution.
 Second,  the  posterior $p( \beta_j | \beta_{j0}, \theta_j, \tau^2_j, P_{0,jj})$ is a Gaussian distribution, obtained
 by combining   the prior $\beta_j  | \tau^2_j\sim \Normal{0,\tau^2_j}$
with the conditional likelihood
 $ \beta_{j0}|\beta_j,\theta_j  ,P_{0,jj} \sim  \Normal{\beta_j,\theta_j {P}_{0,jj} }$:
\begin{eqnarray}
\beta_j | \beta_{j0}, \theta_j , \tau^2_j ,P_{0,jj}
%&\propto& \frac{1}{\sqrt{\theta_j {P}_{0,jj}  }} \exp \left(-\frac{(\beta_{j0}-\beta_j)^2}{2\theta_j {P}_{0,jj}  }   \right)    \frac{1}{\sqrt{\tau^2_j}} \exp \left(  -\frac{\beta_j^2}{2\tau^2_j}  \right),    \\
      \propto \Normal{ \frac{\beta_{j0}  \tau^2_j }{\tau^2_j + \theta_j P_{0,jj}}   , \frac{\tau^2_j \theta_j P_{0,jj} }{\tau^2_j + \theta_j P_{0,jj}}}.    \label{inicoen}
\end{eqnarray}
Sampling the parameters  $a^\tau$ and  $a^\xi$ in Step~(d) is performed without conditioning on   $ \tau_1, \ldots, \tau_d$ and $ \xi_1, \ldots, \xi_d$.
The acceptance probability for   $a^{\xi,\mathrm{new}}$ reads:
\begin{equation*}
\min \bigg\{ 1, \frac{p(a^{\xi,\mathrm{new}})  a^{\xi,\mathrm{new}}}{p(a^{\xi})   a^{\xi}}\prod_{j=1}^d \frac{p(\sqrt{\theta}_j|a^{\xi,\mathrm{new}},\kappa^2) }{p(\sqrt{\theta}_j|a^{\xi},\kappa^2)} \bigg\},
% \min \bigg\{ 1,\frac{p(\sqrt{\theta}_1,\dots, \sqrt{\theta}_d|a^{\xi,\mathrm{new}},\kappa^2) p(a^{\xi,\mathrm{new}})  a^{\xi,\mathrm{new}}}{p(\sqrt{\theta}_1,\dots, \sqrt{\theta}_d|a^{\xi},\kappa^2) p(a^{\xi})   a^{\xi}} \bigg\},
\end{equation*}
based on the marginal prior (\ref{teetamarg}). A similar acceptance probability holds  for   $a^{\tau,\mathrm{new}}$.

  Sampling  the latent  prior variances    $\tau_j^2$  and  $\xi^2_j$
  of  the hierarchical shrinkage  priors  (\ref{equNG01})  and  (\ref{equNGtheta})   for  $\beta_j$   and $\sqrt{\theta_j}$ in  Step~(e)
  is less standard and we briefly discuss sampling $\xi^2_j$ (full details  are given in Appendix~\ref{stepdd}).
 The conditionally normal prior  $\sqrt \theta_j|\xi_j^2$ in  (\ref{equNGtheta})
leads to a  likelihood  for $\xi^2_j$ which is the  kernel of an inverted gamma
density. In combination with the gamma prior for $\xi^2_j|a^\xi, \kappa^2$,
this  leads to a posterior  distribution arising from a
 generalized inverse Gaussian (GIG) distribution: $ \xi^2_j| \theta_j, a^\xi, \kappa^2  \sim  \GIG{a^\xi-1/2}{a^\xi \kappa^2}{\theta_j}$.

 Finally,  Step~(f)    has to be modified  %if the error variance   follows
for the SV model defined in  (\ref{svht}).   To  sample $(h_{0}, \ldots, h_T)$ as well as  $\mu$, $\phi$, and $\sigma_\eta^2$,   we rely  on \citet{kas-fru:anc}
who developed  an interweaving strategy for boosting  MCMC estimation of SV models.\footnote{This step is easily incorporated into Algorithm~\ref{facsvalg} using the R-package {\tt stochvol} \citep{kas:dea}. }

\section{Comparing shrinkage priors through log predictive density scores}   \label{sec:forecasting}

Log predictive density scores (\LPS ) are an often used scoring rule %proposed by \citet{goo:rat}
 to compare models; see, e.g., \citet{gne-raf:str}. %see, e.g., \citet{gne-ran:com,die-etal:eva,dik-etal:lik,gne:mak}.
 \citet{gew-kea:smo} % and \citet{vil-etal:reg}
  introduced \LPS\ for model comparison of econometric models,
 see also \citet{gew-ami:com} for an excellent review of Bayesian predictive analysis. In the present paper, we use  log predictive density scores as a means of evaluating and comparing different shrinkage priors.

As common in this framework, the first $t_0$ time series observations  $\ytr=(y_{1},\ldots, y_{t_0})$ are used as a \lq\lq training sample\rq\rq , while
evaluation is performed for the remaining time series observations $y_{t_0+1}, \ldots, y_T$, based on  the log predictive density:
\begin{eqnarray} \label{marlikr}
\LPS = \log  p(y_{t_0+1}, \ldots, y_T| \ytr ) =
 \sum_{t=t_0+1}^T  \log   p(y_{t}|  \ym^{t-1} ) =   \sum_{t=t_0+1}^T  \LPSo{t}.
\end{eqnarray}
In (\ref{marlikr}),   $  p(y_{t}|  \ym^{t-1}) $ is the one-step ahead   predictive density for time $t$  given  $ \ym^{t-1}=(y_1,\ldots, y_{t-1})$
which is evaluated at the observed value $y_t$.
%where  is the one-step ahead log predictive density score.
The (individual) log predictive density scores  $\LPSo{t}=  \log  p(y_{t}| \ym^{t-1}) $   provide a tool to analyze  performance separately  for each observation
$y_t$, whereas $\LPS $ is an aggregated measure of   performance  for the entire time series.

As shown by  \citet{fru:bay} in the context of  selecting time-varying and fixed components for a basic structural
  state space model,  $\LPS $ %defined in  (\ref{marlikr}) 
  can be %additional 
  interpreted as
  a log marginal likelihood based on the training sample prior $p(\thmod|  \ytr)$,  since
  \begin{eqnarray*}
  p(y_{t_0+1}, \ldots, y_T| \ytr) = \int p(y_{t_0+1}, \ldots, y_T | \ytr,\thmod) p(\thmod|  \ytr) d \, \thmod,
\end{eqnarray*}
 where $\thmod$ summarises the unknown model parameters, e.g.  $\thmod = (\beta_1, \ldots,  \beta_d$,  $\sqrt{\theta_1} ,\ldots,\sqrt{\theta_d}, \sigma^2)$
 for the homoscedastic state space model.  This provides a sound and coherent foundation  for  using
 the log predictive density score  for model -- or, in our context, rather prior -- comparison.

To approximate the one-step ahead predictive density $p(y_{t}|\ym^{t-1})$,  we use Gaussian sum approximations, which are derived from the MCMC draws
$(\thmod^{(m)}, m=1,\dots,M)$ from the posterior distribution $p(\thmod |\ym^{t-1})$  given  information up to $\ym^{t-1}$, i.e:
\begin{eqnarray} \label{LPDSexa}
 \LPS^*_{t} =  \log \, p(y_{t}|\ym^{t-1})= \log \, \int p(y_{t}|\ym^{t-1},\thmod)  p(\thmod |\ym^{t-1}) d \, \thmod
	  \approx \log \, \left( \frac{1}{M} \sum^M_{m=1}  p(y_{t}|\ym^{t-1},\thmod^{(m)}) \right),
\end{eqnarray}
where the one-step ahead predictive density $p(y_{t}|\ym^{t-1},\thmod)$ is Gaussian conditional on knowing  $\thmod$.

We derive an  approximation,  called the \emph{conditionally optimal Kalman mixture approximation}, which exploits the fact that the TVP model is a conditionally Gaussian state space model  given $\thmod = (\beta_1, \ldots,  \beta_d,  \sqrt{\theta_1} ,\ldots,\sqrt{\theta_d}, \sigma^2_{t})$.\footnote{Alternative approximations are discussed in Appendix~\ref{sec:onestep}.}
 For each   draw $\thmod^{(m)}=(\beta_1^{(m)}, \ldots,  \beta_d^{(m)}$,  $\sqrt{\theta_1}^{(m)} ,\ldots,\sqrt{\theta_d}^{(m)}$, $ \sigma^{2(m)}_{t})$ from the
posterior  $p(\thmod|\ym^t)$,  we  determine
the \emph{exact} predictive density $p(y_{t}|\ym^{t-1},\thmod^{(m)} )$  given by the normal distribution
%\begin{equation*}
 $ y_{t}|\ym^{t-1},\thmod^{(m)} 	\sim \Normult{d}{ \hat y_{t}^{(m)} ,{S}^{(m)}_{t}}$,
 % \end{equation*}
where $ \hat{y}_{t}^{(m)}$ and $  S^{(m)}_{t}  $  are obtained  from the prediction step of the Kalman filter (see Appendix~\ref{kalffbs}),
 based on the filtering density $ \tilde{\betav} _{t-1}|\ym^{t-1},\thmod^{(m)}  \sim \Normult{d}{\mv_{{t-1}}^{(m)},\Cm_{{t-1}}^{(m)}}$:
\begin{eqnarray*}
  \hat y_{t}^{(m)} &=& \xm_{t}\betav^{(m)}+ \Fm^{(m)}_{t} \mv^{(m)}_{t-1}, \\
{S}^{(m)}_{t} &=& \Fm^{(m)}_{t} (\Cm^{(m)}_{t-1} + \mathbf{I}_d) \Fm^{\prime(m)}_{t}  + \sigma^{2(m)}_{t}, \label{Sexcat}
 \end{eqnarray*}
where $ \Fm^{(m)}_{t}=\xm_{t} \Diag{\sqrt{\theta_1}^{(m)},\dots,\sqrt{\theta_d}^{(m)}}$ and $\mathbf{I}_d$ is the $d\times d$ identity matrix.
This yields the following Gaussian mixture approximation for $p(y_{t}|\ym^{t-1})$:
\begin{equation}
 p(y_{t}|\ym^{t-1})  % = \int p(y_{t}|\ym^t,\thmod) p(\thmod|\ym^t) d \thmod
 \approx \frac{1}{M} \sum^M_{m=1} \mathnormal{f}_N \left(y_{t};   \hat{y}_{t}^{(m)},  {S}^{(m)}_{t}  \right).
\end{equation}
Draws from %the posterior  distribution
$p(\thmod|\ym^{t-1})$ are obtained  by running the Gibbs sampler outlined in Algorithm~\ref{facsvalg} for the reduced sample $\ym^{t-1} = (y_{1},y_{2},\dots,y_{t-1})$.
For a homoscedastic error specification, $\sigma^{2(m)}_{t} \equiv \sigma^{2(m)}$, whereas
 $\sigma^{2(m)}_{t}$ is forecasted in the following way for the SV model  (\ref{svht}). Given the posterior draw $h^{(m)}_{t-1}$,
we  simulate $h^{(m)}_{t}$  from a  conditional  normal
 distribution with mean $\mu^{(m)}+\phi^{(m)}(h^{(m)}_{t-1} -\mu^{(m)} )$ and
 variance $\sigma^{2(m)}_{\eta}$ and define $\sigma^{2(m)}_{t} = \e^{h^{(m)}_{t}}$.

\section{Extension to multivariate time series}  \label{sec:mult_TVP}

\subsection{Sparse TVP models for multivariate time series}  \label{sec:mult_TVP_model}

The methods introduced in the previous sections are easily extended to TVP models for multivariate time series,  such as  time-varying parameter {VAR}s, see
 e.g.  \citet{eis-etal:sto} who analyze  the response of macro variables to fiscal shocks,
 and  time-varying  structural {VAR}s, see  e.g.  \citet{pri:tim} for a monetary policy application.
Consider, as illustration, the following  TVP model for an $\dimy$-dimensional  time series  $\ym_{t}$,
\begin{eqnarray}
%\label{eq:TVP2}
%\Bm_{t} &=& \Bm_{t-1}+\boldsymbol{\omegav}_{t}, \quad \boldsymbol{\omegav}_{t} \sim \Normult{d}{\mathbf{0},\mathbf{Q}}, \\
\label{eq:TVPmult}
\ym_{t} &=& \Bm_t \xm_{t} +\errorv_{t},\qquad \errorv_{t} \sim  \Normult{\dimy}{\bfz,\Sigmam_{t}} ,   %{0,\Dm_{t}},
\end{eqnarray}
where $\xm_{t}$ is a  \emph{column}  vector of $d$ regressors,  and $\Bm_t $ is a  time-varying  $\dimmat{\dimy}{d}$  matrix with coefficient $\beta_{ij,t}$  in row $i$ and column $j$, potentially containing structural zeros or constant values such that  $\beta_{ij,t} \equiv c$ apriori.  The  (apriori) unconstrained time-varying coefficients  $\beta_{ij,t}$ are assumed to follow  independent random walks as in the univariate case:
\begin{eqnarray} \label{eq:betaij}
\beta_{ij,t} = \beta_{ij,t-1} + \omega_{ij,t}, \quad \omega_{ij,t} \sim \Normal{0,\theta_{ij}},
\end{eqnarray}
 with initial value %\begin{eqnarray}
$\beta_{ij,0} \sim \Normal{\beta_{ij},\theta_{ij} {P_{0,ijj}}}$, where $P_{0,ijj} \sim \Gammainv{\nu_P,(\nu_P-1)c_P}$ as before.
Both the fixed regression coefficients  $\beta_{ij}$ as well as  the  process variances  $\theta_{ij}$ are assumed to be unknown.

  Each of the  apriori  unconstrained   coefficients $\beta_{ij,t}$ is potentially constant, with the corresponding  process variance  $\theta_{ij}$ being 0.  A constant coefficient  $\beta_{ij,t} \equiv \beta_{ij}$  is potentially insignificant, in which case   $\beta_{ij}=0$. Hence, shrinkage priors as introduced in Section~\ref{sec:shrinkage} %  and  \ref{sec:priorbeta}
for the univariate case,  are imposed  on  the $\theta_{ij}$s
and $\beta_{ij}$s  to  define a sparse TVP model for
identifying  which of these scenarios holds for each coefficient  $\beta_{ij,t}$.

 For $i=1, \ldots,\dimy $, the hierarchical double gamma prior for the process variances $\theta_{ij}$ of the coefficients in the $i$th row of a multivariate TVP model  reads: % for each $j$:
\begin{eqnarray} \label{normultwo}
\theta_{ij}|\xi^{2}_{ij}  \sim \Gammad{\frac{1}{2},\frac{1}{2\xi_{ij}^{2}}},  \quad \xi_{ij}^2|a^\xi_i,\kappa_i^2  \sim  \Gammad{a_i^\xi,a_i^\xi \kappa_i^2/2},  \quad \kappa^2_i \sim \Gammad{d_1,d_2},  \quad a_i^\xi \sim \Exp (\axipr)   ,
 \end{eqnarray}
 with  prior expectation  $\xi_{ij}^2$ for each process variance $\theta_{ij}$. Similarly,  an individual prior variance  $\tau_{ij}^2$ is introduced for each fixed regression coefficient  $\beta_{ij}$ as in   (\ref{equNG01}):
 \begin{eqnarray}  \label{equNGtwo}
  \beta_{ij}|\tau^2_{ij} \sim  \Normal{0,\tau^2_{ij}}, \quad \tau_{ij}^2|a_i^\tau ,\lambda_i^2 \sim  \Gammad{a_i^\tau,a_i^\tau \lambda_i^2/2}, \quad
   \lambda_i^2 \sim \Gammad{e_1,e_2},  \quad a_i^\tau \sim \Exp (\ataupr).
 \end{eqnarray}
   By choosing  $a_i^\tau =a^\xi_i=1$ in  (\ref{normultwo})  and  (\ref{equNGtwo}),  a hierarchical Bayesian Lasso prior for multivariate TVP models results.

We assume row specific hyperparameters  $a_i^\tau, \lambda^2_i $ and $a^\xi_i,  \kappa ^2_i $,  drawn  from   common hyperpriors 
 with fixed hyperparameters  $ e_1,e_2,\ataupr$  and $  d_1,d_2, \axipr$.
This  leads  to prior independence across the  $\dimy$ rows of the observation equation (\ref{eq:TVPmult}) and
   is advantageous for computational reasons,
in particular, if the errors $\errorv_{t}$ are uncorrelated,  i.e.~$\Sigmam_{t}$ %=\Dm_{t}$
is a diagonal matrix. In this case,  the multivariate TVP model  has a representation as $  \dimy$ independent univariate TVP models as in Section~\ref{sec:model} and MCMC estimation using Algorithm~\ref{facsvalg}  can be performed independently for each of the  $\dimy$ rows of the system, e.g. in a parallel computing environment.

  If  $\Sigmam_{t}$ is a full covariance matrix, then the rows are not independent, because of the correlation among the various components in  $\errorv_{t}$.
However, as shown by \citet{lop-etal:par}, a Cholesky decomposition of $\Sigmam_{t}$ leads to such a representation, see also \citet{eis-etal:sto} and
\citet{zha-etal:dyn}. Further details are provided in the next subsection. %Section~\ref{sec:chol}.

\subsection{The sparse TVP  Cholesky SV model} \label{sec:chol}

 \citet{lop-etal:par} demonstrate how  a multivariate time series  $\ym_t\sim \Normult{\dimy}{\bfz,\Sigmam_t}$  with time-varying covariance matrix $\Sigmam_t$  %  following
  %\begin{equation*} \label{eq:multNormal}
 %\ym_t\sim \Normult{\dimy}{\bfz,\Sigmam_t},
 %\end{equation*}
 can be transformed into a system  of $\dimy$ independent equations  using the  time-varying Cholesky decomposition
 $\Sigmam_t = \Am_t \Dm_t \Am_t'$ , where $\Am_t \Dm_t^{1/2}$  is the lower triangular Cholesky decomposition of $\Sigmam_t$.
 $\Am_t$ is lower triangular with ones
on the main diagonal, while $ \Dm_t$ is a time-varying diagonal matrix.
It follows that
 %\begin{equation} \label{eq:mult2}
 $\Am_t^{-1}  \ym_t \sim  \Normult{\dimy}{\bfz,\Dm_t}$.
 %\end{equation}
 Denoting the elements of $\Am^{-1}_t$ as  $\Phi_{ij,t}$, for $j< i$, % equation~(\ref{eq:mult2})
 this can be expressed as
\begin{equation*}\label{eq:mult}
\left( \begin{matrix}
1 			& \dots 	& & &0  \\
{\Phi_{21,t}} 		& 1 		& & & 0 \\
 		&  		& \ddots & & 0 \\
\vdots 			& 		& & 1 & 0\\
\Phi_{r1,t} 		& \Phi_{r2,t} & \dots & \Phi_{r,r-1,t}  & 1   \\
 \end{matrix}\right)
\left( \begin{matrix}
y_{1t} 		\\ y_{2t} 		\\ \vdots \\ y_{rt} 		\\
 \end{matrix} \right)
 \sim \Normult{\dimy}{\bfz,\Dm_t} ,
 \end{equation*}
which can be written as  in (\ref{eq:TVPmult}):
\begin{equation} \label{eq:mult2}
 \ym_t \sim  \Normult{\dimy}{ \Bm_t \xm_{t},\Dm_t},
\end{equation}
where $\Bm_t$ is a   $\dimy \times (\dimy-1)$  matrix with elements $\beta_{ij,t}=-\Phi_{ij,t}$, $\Dm_t$ is a diagonal matrix and the $(\dimy-1)$-dimensional vector
$\xm_{t}=(y_{1t},\ldots,y_{\dimy -1,t})'$ is a regressor derived from $\ym_t$.
Thus the  distribution of $\ym_t$ can be represented by  a system of $\dimy$ independent TVP models  as in  Section~\ref{sec:mult_TVP_model}, where each  time-varying coefficient  $\beta_{ij,t}, j< i$, $i=1 , \ldots, \dimy$, follows a random walk as in (\ref{eq:betaij}).
Employing the prior      (\ref{equNGtwo}) for  $ \beta_{ij}$  and   (\ref{normultwo})  for  $\theta_{ij}$  yields the sparse
TVP Cholesky SV model.

To capture conditional heteroscedasticity, the  matrix  $\Dm_t=\Diag{\e^{h_{1t}},\ldots,\e^{h_{rt}}}$ is assumed to be time-varying, where for each row $i=1,\ldots, \dimy$, the log volatility $h_{it}$ is assumed to follow an individual SV model as in  (\ref{svht}),  with row specific parameters   $\mu_i$, $\phi_i$, and  $\sigma^2_{\eta,i}$:
\begin{eqnarray*} \label{svhtmult}
 h_{it} | h_{i,t-1}, \mu_i , \phi_i , \sigma^2_{\eta,i} \sim \Normal{\mu_i +\phi_i (h_{i, t-1}- \mu_i),\sigma^2_{\eta,i}}.
\end{eqnarray*}
For $\dimy = 3$, for instance, the TVP Cholesky SV model reads:
\begin{align*}
y_{1t} &= \varepsilon_{1t},  						&\varepsilon_{1t} 	\sim \Normal{0, \e^{h_{1t}}}, \\
y_{2t} &= \beta_{21,t}y_{1t} +  \varepsilon_{2t}, 			&\varepsilon_{2t}	\sim \Normal{0, \e^{h_{2t}}}, \\
y_{3t} &= \beta_{31,t}y_{1t} + \beta_{32,t}y_{2t} +  \varepsilon_{3t},  	&\varepsilon_{3t}	\sim \Normal{0, \e^{h_{3t}}}.
\end{align*}
 No intercept is present in  these TVP models. For the  TVP  model in the first row,  no regressors are present and only the time-varying volatilities $h_{1t}$ have to be estimated.
 In the $i$-th  equation,   $i-1$ regressors  are present and
   $d=i-1$ time-varying  regression coefficients  $\beta_{ij,t}$  as well as the time-varying volatilities $h_{it}$ need  to be estimated.
 Each of these equations is transformed  into a non-centered TVP model and the MCMC scheme %developed
 in Algorithm~\ref{facsvalg} is  applied  to perform Bayesian inference % under the hierarchical shrinkage priors on   $\beta_{ij}$ and $\theta_{ij}$.
independently for each row $i$.

\begin{figure}[t!]%[h]
\centering
\includegraphics[width=0.7\textwidth,height=6.5cm]{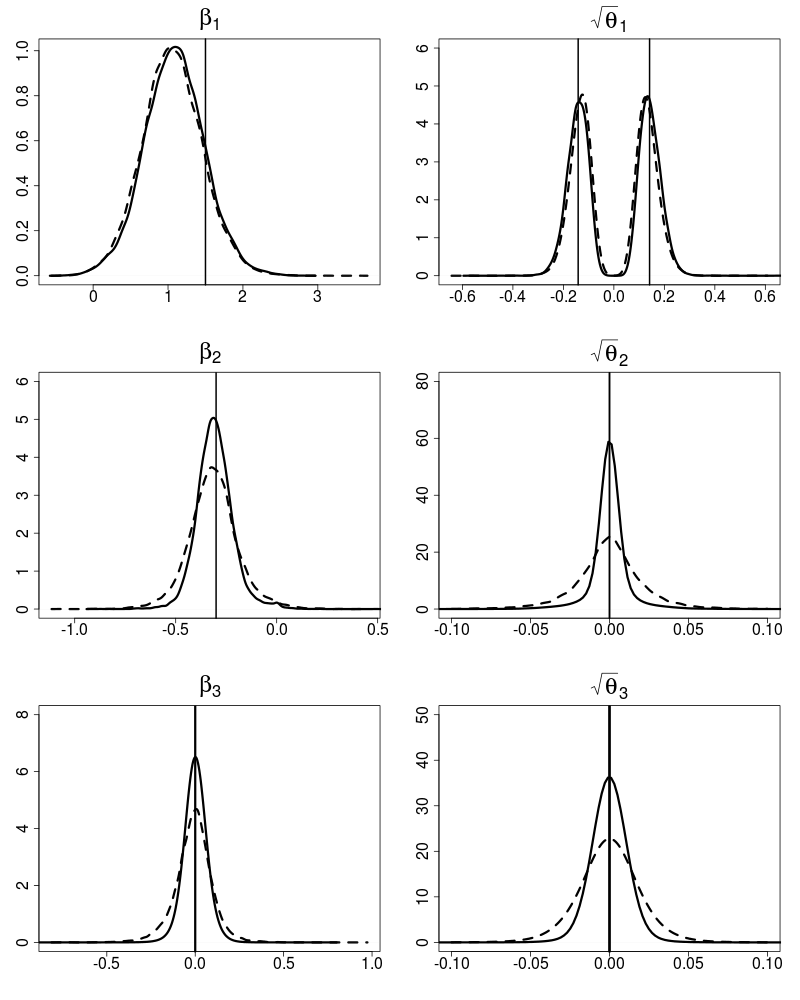}
  \caption{Simulated data.  Posterior densities of $\beta_j$ (left-hand side) and  $\sqrt{\theta_j}$ (right-hand side) together with the true values (indicated by the vertical lines),  based on the hierarchical double gamma prior
 with $a^\tau \sim \Exp (10)$ and $a^\xi \sim \Exp (10)$ (solid line)  and  the hierarchical Bayesian Lasso prior (dashed line).}  %VERSION 10. Februar 2018}
% \comment{TO DO ANGELA: ERROR in plottting? Lasso more shrinkage than normal gamma?}}
 \label{fig:SIM_DATA_plot_beta_sqrt_theta}
\end{figure}

\section{Illustrative  Application to Simulated Data}   \label{sec:sim_demo1}

To  illustrate our  methodology for  simulated data,
we generated 100 univariate time series of length $T=200$ from  a TVP model where  $d=3$,  $\{x_{1t}\} \equiv 1$,  $\{x_{jt}\} \sim \Normal{0,1}$ for $j=2,3$, $\sigma^2=1$,  $(\beta_1, \beta_2, \beta_3) = (1.5,-0.3,0)$ and  $(\theta_1, \theta_2, \theta_3) = (0.02,0,0)$.  For each time series, $\beta_{1t}$  is a strongly time-varying coefficient,  $\beta_{2t}$
  is a constant, but significant coefficient,    and $\beta_{3t}$ is an insignificant coefficient.  As shrinkage priors on   $\beta_j$ and $\sqrt{\theta_j}$, we  consider  the hierarchical double gamma prior
 with $a^\tau \sim \Exp (10)$ and $a^\xi \sim \Exp (10)$  and  the hierarchical Bayesian Lasso prior  (that is  $a^\tau=a^\xi=1$)
 under the hyperparameter setting  $d_1=d_2=e_1=e_2 = 0.001$.  For each of the 100 simulated time series,  MCMC estimation  is based on Algorithm~\ref{facsvalg} by drawing $M=30,000$ samples after a  burn-in of length 30,000.\footnote{The Bayesian Lasso prior is combined with the ASIS strategy by fixing   $a^\tau=a^\xi=1$ and skipping Step~(d).}

In Figure~\ref{fig:SIM_DATA_plot_beta_sqrt_theta} we compare  the posterior densities for $\beta_j$ and
$\sqrt\theta_j$ for one such time series under both shrinkage priors.
In general, we want to distinguish three types of  coefficients: %$\beta_{jt}$:
 time-varying,   static but significant, and  insignificant.
One way to achieve  a classification  is by visual inspection of the posterior distributions of  $\beta_j$ and $\sqrt{\theta}_j$.
The posterior density of the scale parameter $\sqrt{\theta_j}$ is symmetric around zero by definition. Thus, if the unknown
variance $\theta_j$ is   different from zero, then the posterior density of $\sqrt{\theta_j}$ is likely to be bimodal.
 If we find that the posterior density of $ \sqrt{\theta_j}$ is unimodal, then the unknown variance is likely to be zero.

\begin{figure}[t!]% [h]
\centering
  \includegraphics[width=0.7\textwidth,height=8cm]{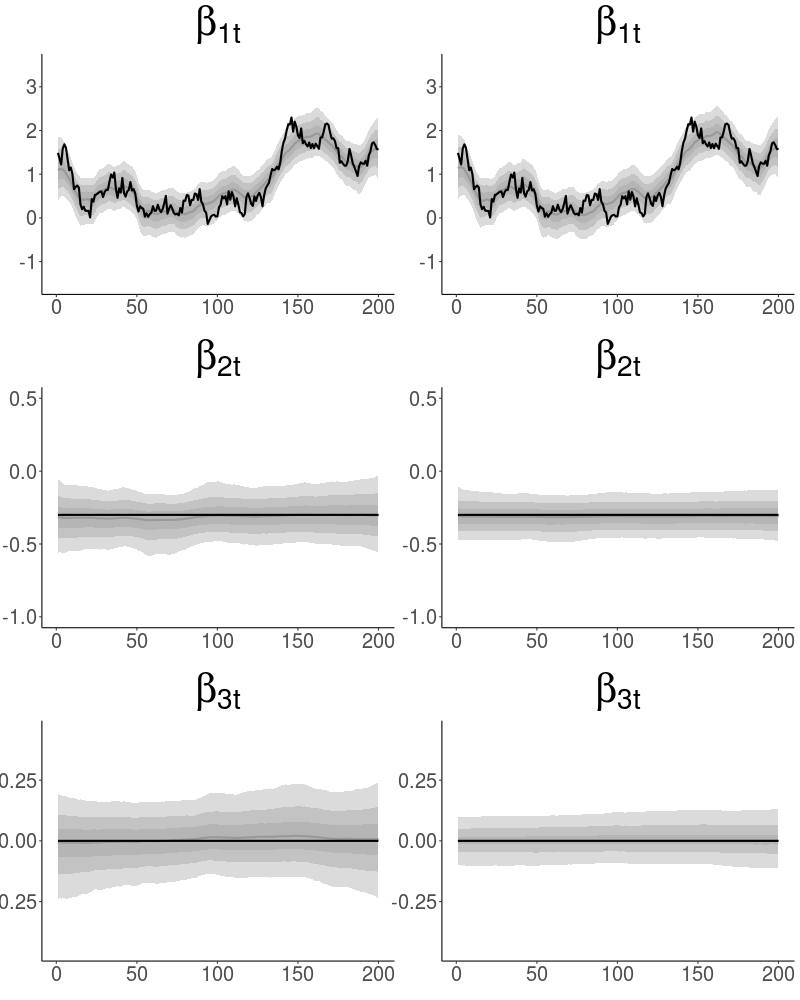}
  \caption{Simulated data. Pointwise $(0.025,0.25,0.5,0.75,0.975)$-quantiles of the posterior paths
  $\beta_{jt} = \beta_j + \sqrt{\theta_j} \tilde \beta_{jt}$ in the centered parametrization in comparison to the true paths (thick black line) for one of the simulated
  time series; left-hand side: hierarchical Bayesian Lasso prior, right-hand side: hierarchical double gamma prior with $a^\tau \sim \Exp (10)$ and $a^\xi \sim \Exp (10)$. } %VERSION 2.Februar 2018.
%  \comment{TO DO ANGELA: Remove values for $a^\tau , a^\tau$ from the caption and center  $\beta_{jt}$. Legends of both axis invisible. Match the fonts of the caption and the axis.}}
 \label{fig:paths_sim_data_centered_beta_test_11082016_de_1}
\end{figure}

While such a bimodal structure of  $p(\sqrt{\theta_j}| \ym)$ is well pronounced for the first coefficient
 where  $\sqrt\theta_1=0.141$,   $p(\sqrt{\theta_j} | \ym)$ is indeed shrunken toward  zero for the two coefficients with zero variances  $\theta_2=\theta_3=0$.
For the third coefficient, where in addition $\beta_3=0$, also the posterior    $p(\beta_3|\ym)$ is shrunken toward zero.
Further, we show the posterior paths of $\beta_{jt}$ in
Figure~\ref{fig:paths_sim_data_centered_beta_test_11082016_de_1}. Evidently, shrinkage priors are able to detect the
 time-varying  coefficient $\beta_{1t}$, the constant but significant coefficient $\beta_{2t}$ and the insignificant coefficient  $\beta_{3t}$.
 In both figures, the advantage of  the
double gamma prior compared to the Bayesian Lasso prior is reflected by increased efficiency in identifying coefficients that are not time-varying.

Table~\ref{table:average_MSE_sim_data}  summarises  the average mean
squared error ($avMSE$), the average squared bias ($avBIAS^2$) and the average variance ($avVAR$)   for the parameters $\beta_1$, $\beta_2$, $\beta_3$, $|\sqrt\theta_1|$, $|\sqrt\theta_2|$, and $|\sqrt\theta_3|$ over the  100 simulated time series.\footnote{Given $M$ draws $\thc\im{i1}, \ldots, \thc\im{iM}$, of a parameter $\thc$ for each time series $i$,
these measures are defined as  $avMSE =  avVAR +  avBIAS^2$, where  $avVAR= \frac{1}{100} \sum^{100}_{i=1} \V _i$  and $avBIAS^2= \frac{1}{100}\sum^{100}_{i=1} (\E_i - \thc ^{{\footnotesize \rm true}} )^2$ with $ \E_i = \frac{1}{M}\sum^{M}_{m=1} \thc\im{im}$ and $ \V _i =  \frac{1}{M}\sum^{M}_{m=1} (\thc\im{im}- \E_i )^2$.}
%\begin{eqnarray*}
%& avMSE =  avVAR +  avBIAS^2, \qquad avVAR= \frac{1}{100} \sum^{100}_{i=1} \V _i  , \qquad  avBIAS^2= \frac{1}{100}\sum^{100}_{i=1} \left(\E_i - \thc ^{{\footnotesize \rm true}} \right)^2, &\\ & \E_i = \frac{1}{M}\sum^{M}_{m=1} \thc\im{im}, \qquad  \V _i =  \frac{1}{M}\sum^{M}_{m=1} \left(\thc\im{im}- \E_i \right)^2. &
%\end{eqnarray*}}
 Heavier shrinkage introduced by the hierarchical double gamma prior leads to
reduced $avMSE$ compared to the hierarchical Bayesian Lasso prior, in particular for the two coefficients which are not  time-varying.

\begin{table}[t!]
\centering
\begin{tabular}{crrrrrr}
  \hline
%&  \multicolumn{3}{c}{Double Gamma prior} &  \multicolumn{3}{c}{Hierarchical Bayesian Lasso} \\
&  \multicolumn{3}{c}{$a^\tau \sim \Exp (10)$, $a^\xi \sim \Exp (10)$} &  \multicolumn{3}{c}{$a^\tau= a^\xi=1$}  \\
\hline
 & $avMSE$ & $avVAR$ & $avBIAS^2$  & $avMSE$ & $avVAR$ & $avBIAS^2$\\
  \hline
  $\beta_1$ & 3.30E-01 & 1.67E-01 & 1.63E-01 & 3.60E-01 & 1.57E-01 & 2.03E-01 \\
  $\beta_2$ & 8.18E-03 & 8.11E-03 & 6.47E-05 & 1.56E-02 & 1.55E-02 & 1.77E-04 \\
  $\beta_3$ & 2.10E-03 & 2.10E-03 & 1.36E-06 & 1.14E-02 & 1.13E-02 & 1.31E-04 \\
  $|\sqrt{\theta_1}|$ & 1.81E-03 & 1.79E-03 & 2.50E-05 & 1.61E-03 & 1.56E-03 & 5.32E-05 \\
  $|\sqrt{\theta_2}|$ & 1.14E-04 & 9.33E-05 & 2.11E-05 & 5.02E-04 & 2.47E-04 & 2.55E-04 \\
  $|\sqrt{\theta_3}|$ & 4.33E-05 & 3.53E-05 & 7.97E-06 & 3.10E-04 & 1.44E-04 & 1.66E-04 \\
  \hline
\end{tabular}
\caption{Simulated data. Average mean
squared error ($avMSE$),  average variance ($avVAR$),
and  average squared bias ($avBIAS^2$) over 100 simulated time series for the hierarchical double gamma prior with $a^\tau \sim \Exp (10)$ and $a^\xi \sim \Exp (10)$ and the hierarchical Bayesian Lasso prior with $a^\tau= a^\xi=1$.  % \comment{VERSION 16. J\"anner 2018.}
}%\footnote{
\label{table:average_MSE_sim_data}
\end{table}

\section{Applications in Economics and Finance}   \label{sec:appl}

\subsection{Modelling EU area inflation}  \label{inflatiodata}

As a first application, we reconsider EU-area inflation data analyzed in \citet{bel-etal:hie-tv} and
consider the generalized Phillips curve specification, where inflation $\pi_t$ depends on (typically $p=12$) lags of
inflation and other predictors $\zv_t$:
\begin{equation} \label{phillips}
 \pi_{t+h} = \sum_{j=0}^{p-1} \phi_{jt} \pi_{t-j} + \zv_t  \gammav_t      % + \gamma_t x_t
 + \varepsilon_{t+h},   \quad \varepsilon_{t+h} \sim  \Normal{0,\sigma^2}.
\end{equation}
This set-up has been discussed by \citet{stock2012generalized}, among others, for forecasting
the annual inflation rate, that is  $h=12$.
% \footnote{\citet{stock2012generalized} also consider the monthly inflation rate $(h=1)$ and our method works equally for these data.}
%Concerning, we consider a homoskedastic  error model.
 Data are monthly and range from February 1994 until November 2010, i.e. $T=190$.  % comment:  (202-12), as we considert the annual inflation rate
We list precise definitions of all variables in Appendix~\ref{sec:data_ecb}. As the time series  are not seasonally adjusted,  we include   monthly dummy variables as covariates  in  (\ref{phillips}) to account for seasonal patterns. Thus we are estimating in total $d=37$ possibly time-varying coefficients, consisting of the intercept, 13 regressors like the \textit{unemployment rate} and  the \textit{1-month interest rate}, 12 lagged values of inflation and 11 seasonal dummies.

\begin{figure}[t!]%[h]
\centering
 \includegraphics[width=0.7\textwidth,height=6cm]{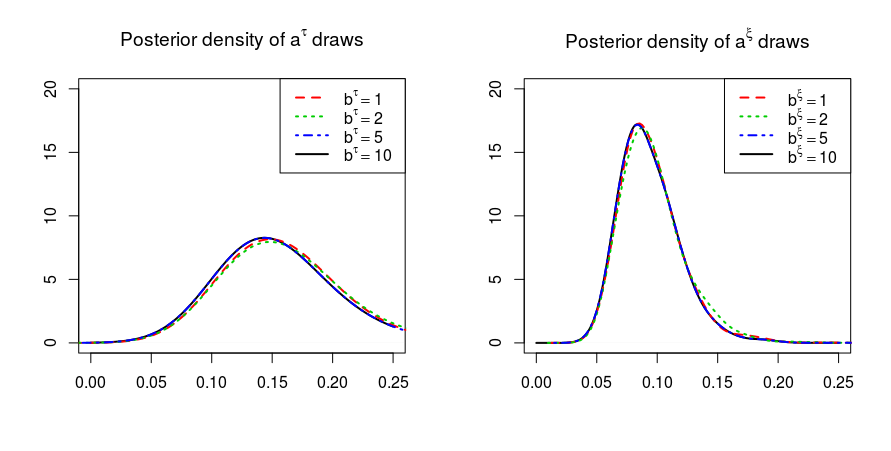}
 \caption{ECB data. Posterior density of  $a^\tau$   (left-hand side) and   $a^\xi$ (right-hand side). %VERSION Februar 2018.
% \comment{TO DO ANGELA: Legende an die neue Notation  $\ataupr$ and  $\axipr$ statt $\atauprold$ and  $\axiprold$  anpassen.}
 } \label{fig:SIM_DATA_posta}
\end{figure}

As shrinkage priors on   $\beta_j$  and   $\sqrt{\theta_j}$, we  consider  the hierarchical double gamma prior
 with $a^\tau \sim \Exp (\ataupr)$ and $a^\xi \sim \Exp ( \axipr)$
 under the hyperparameter setting  $d_1=d_2=e_1=e_2 = 0.001$  and compare it with  the hierarchical Bayesian Lasso prior (that is  $a^\tau=a^\xi=1$)
  applied by \citet{bel-etal:hie-tv}.   For each prior, MCMC inference  is based on Algorithm~\ref{facsvalg}  with $M=$100,000 draws after a burn-in of the same size. For the hierarchical double gamma prior, we considered  various hyperparameters  $\ataupr$ and  $\axipr$   and the corresponding  posterior densities
of  $a^\tau$ and $a^\xi$ are shown in Figure~\ref{fig:SIM_DATA_posta}, with posterior summaries being provided  in Table~\ref{ECB_postaxi}. The acceptance probability for the  random walk MH algorithm in Step~(d) of Algorithm~\ref{facsvalg} lies in the range of 0.24 to  0.26. For these data,  the posteriors of $a^\tau$ and  $a^\xi$ clearly point at the double gamma prior  rather than the Bayesian Lasso prior. The choice of the hyperparameters $\ataupr$ and  $\axipr$  does not play a significant role and  the following results are presented for  $\ataupr=\axipr=10$.

\begin{table}[t!]
\centering
\begin{tabular}{ccccccccc}
  \hline
  & \multicolumn{4}{c}{${p(a^\tau|\ym)}$} &   \multicolumn{4}{c}{${p(a^\xi|\ym)}$} \\
$\ataupr=\axipr$  & 1st Qu. & Median & Mean & 3rd Qu. & 1st Qu. & Median & Mean & 3rd Qu. \\
  \hline
1   & 0.128 & 0.153 & 0.158 & 0.181    & 0.078 & 0.091 & 0.094 & 0.107    \\
  2   & 0.127 & 0.151 & 0.158 & 0.181     & 0.078 & 0.092 & 0.096 & 0.109    \\
  5   & 0.124 & 0.147 & 0.153 & 0.174     & 0.076 & 0.090 & 0.094 & 0.107   \\
  10  & 0.122 & 0.145 & 0.150 & 0.172     & 0.074 & 0.088 & 0.090 & 0.103  \\
 \hline
%\begin{tabular}{crrrrrrrrrrrr}
%  \hline
%  & \multicolumn{6}{c}{${a^\tau}$} &   \multicolumn{6}{c}{${a^\xi}$ } \\
%$\ataupr=\axipr$ & Min. & 1st Qu. & Median & Mean & 3rd Qu. & Max.  & Min. & 1st Qu. & Median & Mean & 3rd Qu. & Max.\\
%  \hline
%1 & 0.063 & 0.128 & 0.153 & 0.158 & 0.181 & 0.453 & 0.036 & 0.078 & 0.091 & 0.094 & 0.107 & 0.253  \\
%  2 & 0.065 & 0.127 & 0.151 & 0.158 & 0.181 & 0.431 & 0.041 & 0.078 & 0.092 & 0.096 & 0.109 & 0.338  \\
%  5 & 0.063 & 0.124 & 0.147 & 0.153 & 0.174 & 0.419 & 0.039 & 0.076 & 0.090 & 0.094 & 0.107 & 0.258 \\
 % 10 & 0.059 & 0.122 & 0.145 & 0.150 & 0.172 & 0.396 & 0.032 & 0.074 & 0.088 & 0.090 & 0.103 & 0.195 \\
 %\hline
\end{tabular}
\caption{ECB data. Posterior summaries for  $p(a^\tau| \ym)$  and $p(a^\xi|\ym)$   for various values  $\ataupr=\axipr$.} \label{ECB_postaxi}
\end{table}

\begin{figure}[t!]%
  \centering
 \includegraphics[height=9cm,width=0.9\linewidth]{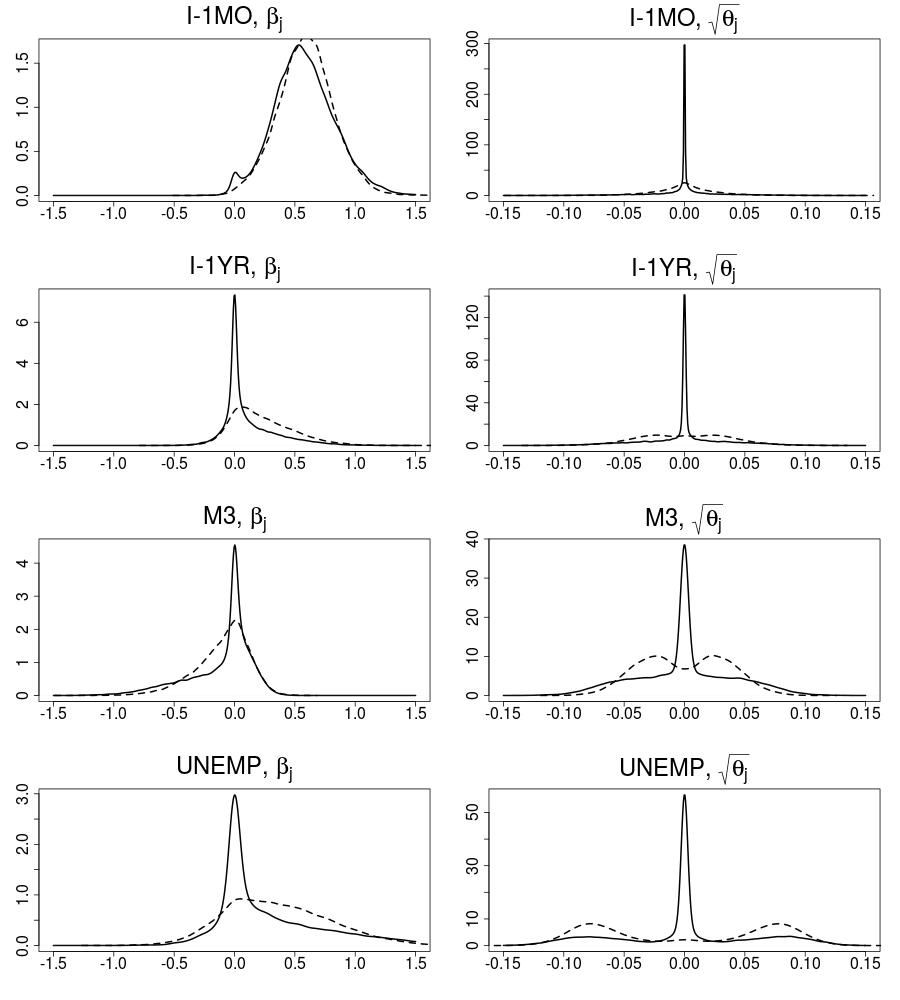}
 \caption{ECB data.~Posterior densities of $\beta_j$ (left-hand side) and  $\sqrt{\theta_j}$ (right-hand side),
  based on   the   hierarchical double gamma prior with $a^\tau \sim \Exp (10)$ and $a^\xi \sim \Exp (10)$
   (solid line)  and  the hierarchical Bayesian Lasso prior  (dashed line)  for  following predictors (from top to bottom):
  \emph{1-month interest rate}, % ($j=14$),
 \emph{1-year interest rate}, % ($j=15$),
 \emph{M3}, % ($j=22$),
 and \emph{unemployment rate}. % ($j=26$).
} % \comment{Version August 2017}
  \label{fig:post_dens_beta_sq_theta}
\end{figure}

\begin{figure}[t!]%
  \centering
 \includegraphics[height=7cm,width=0.9\linewidth]{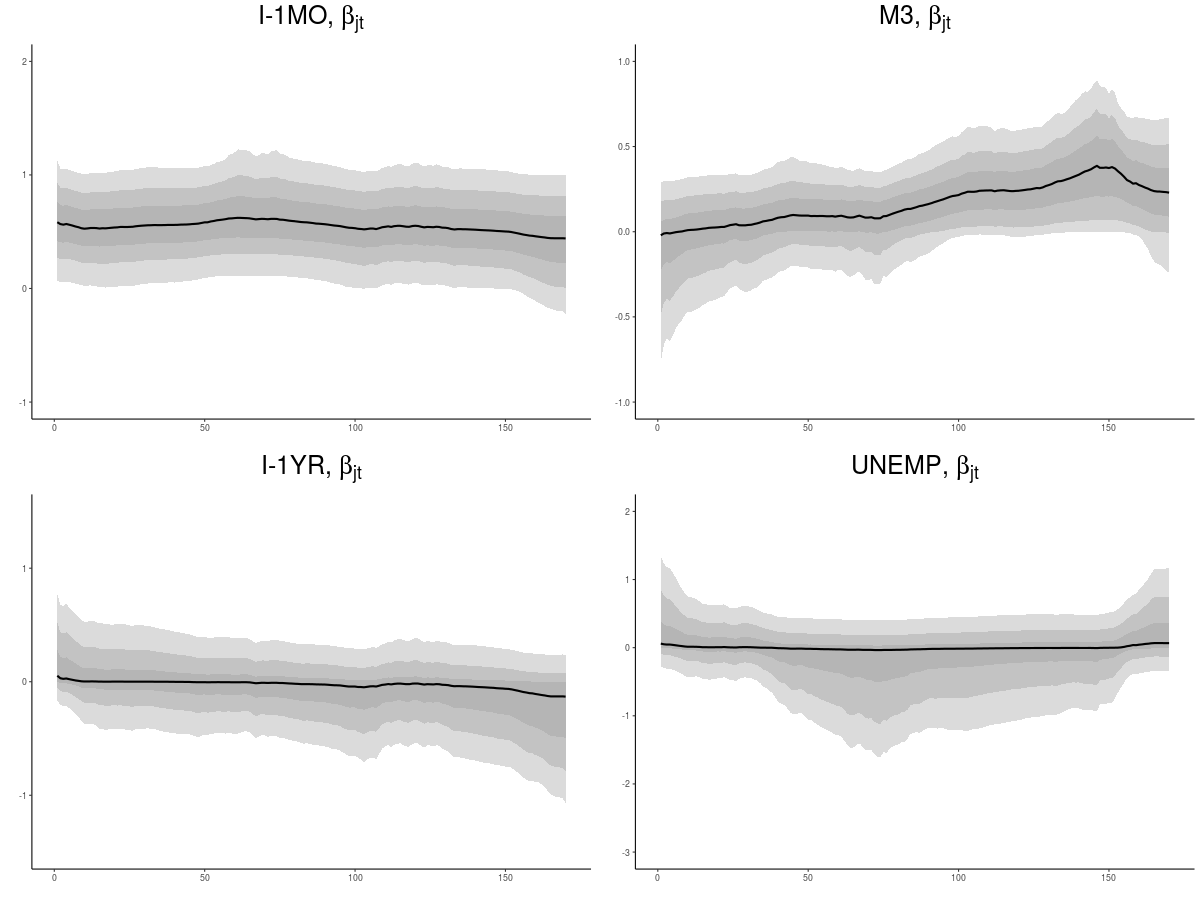}
  \caption{ECB data.~Pointwise $(0.025,0.25,0.5,0.75,0.975)$-quantiles of the posterior paths of
  $\beta_{jt} = \beta_j + \sqrt{\theta_j} \tilde \beta_{jt}$,  based on the  hierarchical double gamma prior with $a^\tau \sim \Exp (10)$ and $a^\xi \sim \Exp (10)$;   left-hand side: \emph{1-month interest rate} (top) and
\emph{1-year interest rate} (bottom); right-hand side:  \emph{M3} (top)  and \emph{unemployment rate} (bottom).  %\comment{Version August 2017}.
} \label{fig:paths_ECB_02}
\end{figure}

 Summary statistics of $p(\beta_j|\ym)$ and    $p(\sqrt{\theta_j}|\ym)$ for  the hierarchical double gamma prior
 with $a^\tau \sim \Exp (10)$ and $a^\xi \sim \Exp (10)$   are given in
Table~\ref{table:SV_FALSE_02_ASIS_TRUE} in Appendix~\ref{results:data_ecb}.
The easiest combination to spot is the case where both parameters $\beta_j$ and $\sqrt{\theta}_j$ are shrunken toward zero and the corresponding posterior
densities  exhibit peaks at zero. This is the case for most of the 37  covariates. The posterior median of  $\sqrt{|\theta_j|}$ in
Table~\ref{table:SV_FALSE_02_ASIS_TRUE} is smaller than $10^{-3}$ for 34  regression coefficients, among them the lagged values of inflation and the monthly dummy variables. In addition, for these coefficients  the 95\%-confidence regions for $\beta_j$ obtained from $p(\beta_j|\ym)$
are reported in Table~\ref{table:SV_FALSE_02_ASIS_TRUE} and show that none of these variables is \lq\lq significant\rq\rq .

The four variables in Table~\ref{table:SV_FALSE_02_ASIS_TRUE} with a posterior mean of  $\sqrt{|\theta_j|}$ larger than $10^{-2}$
 are  the  \emph{1-month interest rate} ($j=14$),
the \emph{1-year interest rate} ($j=15$), \emph{M3} ($j=22$),  and the \emph{unemployment rate} ($j=26$).
For illustration,  we present the corresponding posterior
densities  of $\beta_j$ and $\sqrt \theta_j$ in Figure~\ref{fig:post_dens_beta_sq_theta}  under the hierarchical double gamma prior
 with $a^\tau \sim \Exp (10)$ and $a^\xi \sim \Exp (10)$  and  the hierarchical Bayesian Lasso prior with $a^\tau=a^\xi=1$.
 The corresponding posterior paths $\beta_{jt} = \beta_j + \sqrt{\theta_j} \tilde \beta_{jt}$
are reported  in Figure~\ref{fig:paths_ECB_02}   for the hierarchical double gamma prior.
%s how that these parameters are, indeed, time-varying.
% As expected, we find a bimodal structure in some cases and
 A time-varying behaviour  is
visible for  \emph{M3}  and the \emph{unemployment rate}.
 The path of the \emph{1-month interest rate} is
significantly different from zero, but the
posterior density of $\sqrt{\theta_j}$ exhibits a peak at zero and indicates a constant coefficient. The \emph{1-year interest rate}  is basically shrunken toward zero and can be regarded as insignificant.

\begin{figure}[t!]%
  \centering
\includegraphics[height=6cm,width=0.7\linewidth]{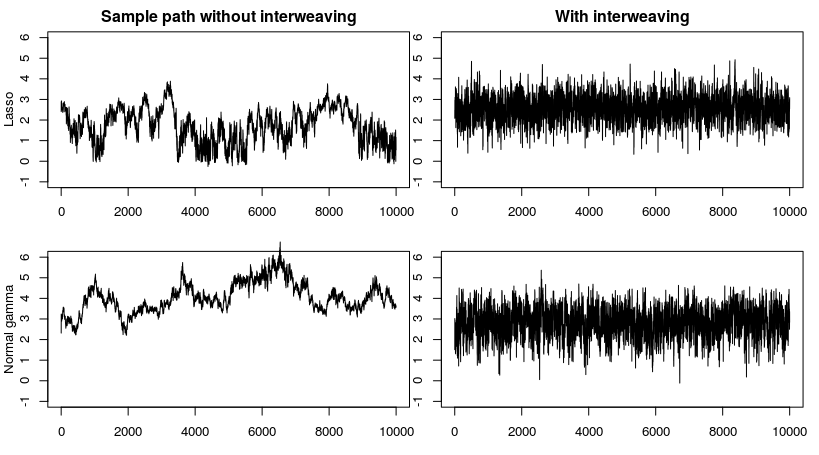}
  \caption{ECB data.~Sample paths of $\beta_1$  comparing the MCMC schemes  without interweaving   (left-hand side) and with interweaving  (right-hand side) for   $a^\tau= a^\xi=1$ (top row)  and    $ a^\tau \sim \Exp(10),  a^\xi \sim \Exp(10)$   (bottom row). $M=100,000$ draws, only every tenth draw is shown.
  %\comment{TO DO ANGELA: OLD Version October 2016; substitute $a^\xi=a^\tau=0.2$ by adaptive $a^\xi \sim \Exp(10), a^\tau \sim \Exp(10)$. }
  }  \label{fig:interweaving}
\end{figure}

\begin{table}[t!]
\centering
\begin{tabular}{rrrrrcrrrr}
  \hline
 &  \multicolumn{4}{c}{$a^\tau \sim \Exp (10), a^\xi \sim \Exp (10)$}  & &  \multicolumn{4}{c}{$a^\tau=a^\xi=1$}  \\
 &   \multicolumn{2}{c}{no ASIS}  &     \multicolumn{2}{c}{ASIS} &&  \multicolumn{2}{c}{no ASIS}  &  \multicolumn{2}{c}{ASIS}   \\ \hline
$j$   &     $ \beta_j$ & $ | \sqrt{\theta_j}| $ & $\beta_j$ & $ | \sqrt{\theta_j}|  $ & &  $\beta_j$ & $ | \sqrt{\theta_j}| $ & $\beta_j$ &$  | \sqrt{\theta_j}| $ \\
   \hline
   1  &
	  4368 & 271 & 86 & 105  		& &  1464& 185& 72& 71\\
  14 &
	  535 & 367 & 77 & 231 			& &  53& 57& 28& 57\\
  15 &
	  116 & 280 & 109 & 245 		& &  143& 197& 45& 76\\
  22  &
	  1192 & 328 & 110 & 136 		& &  64& 95& 45& 56\\
  26 &
	  450 & 672 & 231 & 441 		& &  266& 203& 79& 100\\
  \hline
\end{tabular}
 \caption{ECB data.~Inefficiency factors  of MCMC posterior draws of  selected parameters, obtained from Algorithm~\ref{facsvalg} with  and without interweaving under   the   hierarchical double gamma prior with $a^\tau \sim \Exp (10)$ and $a^\xi \sim \Exp (10)$ and the hierarchical Bayesian Lasso prior. % \comment{Version August 2017}.
 } \label{table:ECBIF}
\end{table}

For this data set,  full conditional MCMC sampling  turned out to be extremely inefficient  and  motivated us to include the interweaving step in the Gibbs sampler outlined  in Algorithm~\ref{facsvalg}.  For  illustration,    Figure~\ref{fig:interweaving} shows MCMC paths obtained for $\beta_1$
 with and without interweaving. As illustrated for selected parameters in Table~\ref{table:ECBIF},
adding the   interweaving step  leads to substantial  improvement of the mixing behaviour of MCMC sampling,  with  considerably reduced
inefficiency factors.\footnote{Inefficiency factors were computed  using the function {\tt effectiveSize} from the R package {\tt coda} \citep{plu-etal:cod}.}

\begin{figure}[t!]%
  \centering
 \begin{tabular}{c}
\includegraphics[height=5cm,width=0.7\linewidth]{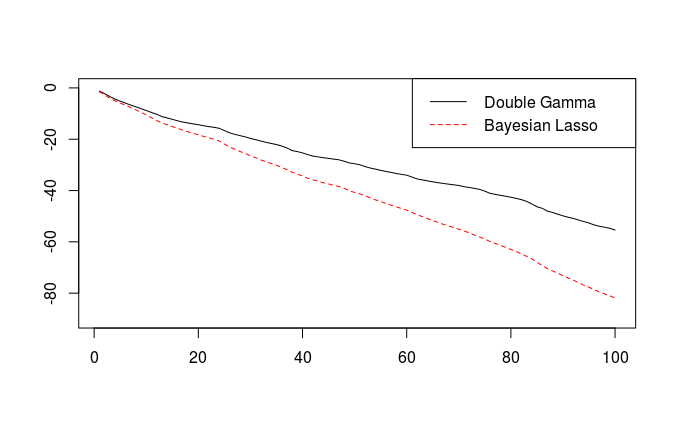}
\end{tabular}
  \caption{ECB data.~Cumulative log predictive scores for the last 100  time point (labelled with time index $t-t_0$, where $t_0=90$)  % 190-100
    under  the hierarchical double gamma prior  with $a^\tau \sim \Exp (10)$ and $a^\xi \sim \Exp (10)$ (solid line)  and  the hierarchical Bayesian Lasso  prior (dashed  line).}  % version Februar 2018
    %\comment{TO DO ANGELA for Bayesian Lasso: Version October 2016 has missing values at  $t-t_0= 40,    86,    90,    91,    92,    93,    94,    95,    96,    97,    98,    99,   100$; please provide a complete set of LPDS for the Bayesian Lasso}.}
  \label{fig:forecasting}
\end{figure}

Finally,  as discussed in Section~\ref{sec:forecasting}, we use log predictive density scores (LPDS)
to evaluate the various shrinkage priors.  Figure~\ref{fig:forecasting} shows  cumulative LPDS over  the last 100 time points,  using the conditionally optimal Kalman mixture approximation derived in Section~\ref{sec:forecasting}.\footnote{For numerical reasons, it is essential to use  the conditionally optimal Kalman mixture approximation rather than the naive approximation   to approximate the predictive density, see  Appendix~\ref{sec:predecb} for details.}
 Evidently, for this time series,   the hierarchical double gamma  prior is clearly preferable   to  the hierarchical Bayesian Lasso prior  applied by \citet{bel-etal:hie-tv}.

\subsection{Sparse TVP Cholesky SV modelling of DAX returns} \label{DAXdata}

As a second real world data application,  we  fit the sparse TVP Cholesky SV model introduced in  Section~\ref{sec:chol} to
 29 indices from the German Stock Index DAX, see Appendix~\ref{sec:data_dax} for more details on the data. The ordering of the indices is alphabetical and our data set spans roughly 2500 daily stock returns from September 4th, 2001 until August 31st, 2011.\footnote{As any model based on a Cholesky decomposition, inference is not invariant with respect to reordering the indices.  While inference for the elements of $\Sigmam_t$ was fairly robust, we observed sensitivity to the ordering of the data for  functionals of  $\Sigmam_t ^{-1}$,~ e.g. the time-varying global minimum variance portfolio  weights derived from  $\Sigmam_t ^{-1}$.}
 %\footnote{The DAX (Deutscher Aktienindex (German stock index)) is a blue chip stock market index consisting of the 30 major German companies trading on the Frankfurt Stock Exchange. }

 \begin{figure}[t!]%[h]
 \centering
 \begin{tabular}{cc}
 \includegraphics[width=0.4\textwidth,height=5cm]{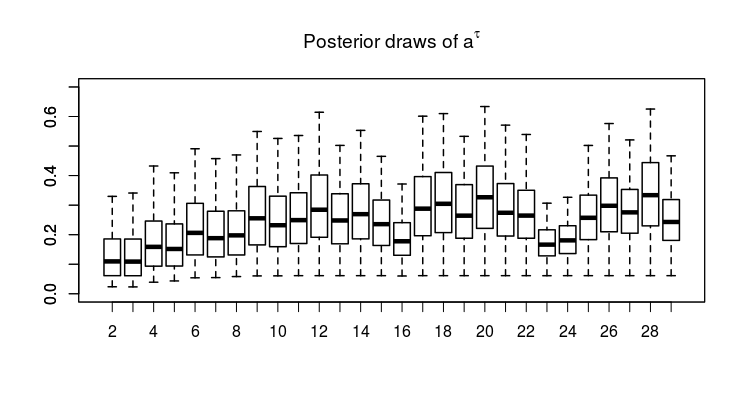}
 &  \includegraphics[width=0.4\textwidth,height=5cm]{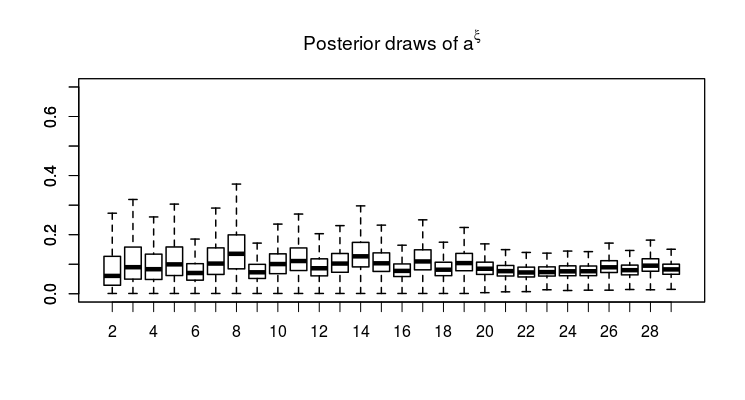}\\
  \end{tabular}
  \caption{DAX data.  Posterior densities of  $a_i^\tau$ (left-hand side)  and   $a_i^\xi$  (right-hand side) under a  hierarchical double gamma prior  with $a_i^\tau \sim \Exp (10)$ and $a_i^\xi \sim \Exp (10)$, represented by  box plots for
  %\comment{TO DO ANGELA: $i=1, \ldots, 28$ not correct- Change
   $i=2, \ldots, 29$. Whiskers correspond to the  0.05 and the 0.95 quantile. % \comment{Updated Februar 2018.}
  }  \label{fig:DAX_DATA_plot_axi}
\end{figure}

Due to the nature of the TVP Cholesky SV model, a representation of the multivariate model in terms of 29 independent equations exists.
Exploiting representation (\ref{eq:mult2}),  we estimate  a pure stochastic volatility model for the first index and
28 TVP models with SV error specification for the remaining indices, with the dimension $d$   increasing from 1 to 28. To estimate the resulting 406
potentially time-varying coefficients $\beta_{ij,t}$ in an efficient manner,  % and 29 log volatilities $h_{it}$
we apply  the hierarchical double gamma priors introduced in     (\ref{normultwo})  and  (\ref{equNGtwo})
 with $a^\tau_i \sim \Exp (10)$ and $a^\xi_i \sim \Exp (10)$  as well as  the hierarchical Bayesian Lasso  prior (that is $a_i^\tau=a^\xi_i=1$)
 under the hyperparameter setting  $d_1=d_2=e_1=e_2 = 0.001$. In addition to these shrinkage priors, we  apply the usual conditionally conjugate prior, i.e. $\theta_{ij} \sim \IGammad{s_0,S_0}$   for all
process variances $\theta_{ij}$ and $\beta_{ij} \sim \Normal{0,A_0} $  for all fixed regression coefficients $\beta_{ij}$, with  prior
setting as in \citet{pet-etal:dyn}, namely $s_0 = 0.1$, $S_0= 0.001$ and $A_0=10$.

 For all TVP models and all priors, MCMC inference is performed using Algorithm~\ref{facsvalg} with $M=50,000$ draws after a burn-in of  $50,000$.\footnote{MCMC estimation under the inverted gamma prior requires a minor modification of Algorithm~\ref{facsvalg}. We sample $\theta_{ij}$ only in the centered parameterization from the conditional posterior $\theta_{ij}|\betav  \sim \IGammad{s_0+ \frac{T+1}{2},S_0 + \frac{1}{2} \sum _{t=1}^T (\beta_{ij,t}-  \beta_{ij,t-1})^2 + \frac{( \beta_{ij,0}-\beta_{ij})^2}{2 P_{0,ijj}}}$.}  The acceptance probability for the  MH algorithm in Step~(d)  lies in the range of 0.24 to 0.26.
 Posterior densities of  $a_i^\tau$   and   $a_i^\xi$  under the  hierarchical double gamma prior  are provided in Figure~\ref{fig:DAX_DATA_plot_axi}. The posterior medians of $a_i^\tau$ are in the range of 0.11 to 0.37,  whereas the posterior  medians of  $a_i^\xi$ lie between 0.07 and 0.15.

% OLD  prior setting, where paths wehre rather wiggly: $z_0 = 0.5$ and $Z_0= 0.2275$.  Furthermore, we assume $\beta_{ij} \sim \Normal{0.9}$.}
%%\citet{bel-etal:hie-tv} use the following specification for non-shrinkage versions: For models without the shrinkage prior on the time-varying coefficients $\sqrt{\theta_j}$, they use a non-hierarchical prior $\sqrt \theta _j \sim \Normal{0,I}$ and thus neglect $\tau^2_j$ and $\kappa^2$. For models without the shrinkage prior on the constant coefficients, they use a non-hierarchical prior $\beta_j \sim \Normal{0,9 \times I}$ and thus neglect $\xi^2_j$ and $\lambda^2$.} %
 % For the estimation of the log volatilities, we have chosen $b_\mu=0$, $b_\mu=100$, $a_0=20$, $b_0=1.5$ and $B_\sigma=1$. Alternative hyperparameter settings have basically resulted in similar findings. ALREADY DEFINED IN SUBSECTION 2.2.

\begin{figure}[t!]%[h]
\centering
\begin{tabular}{c}
 \includegraphics[width=0.8\textwidth,height=15cm]{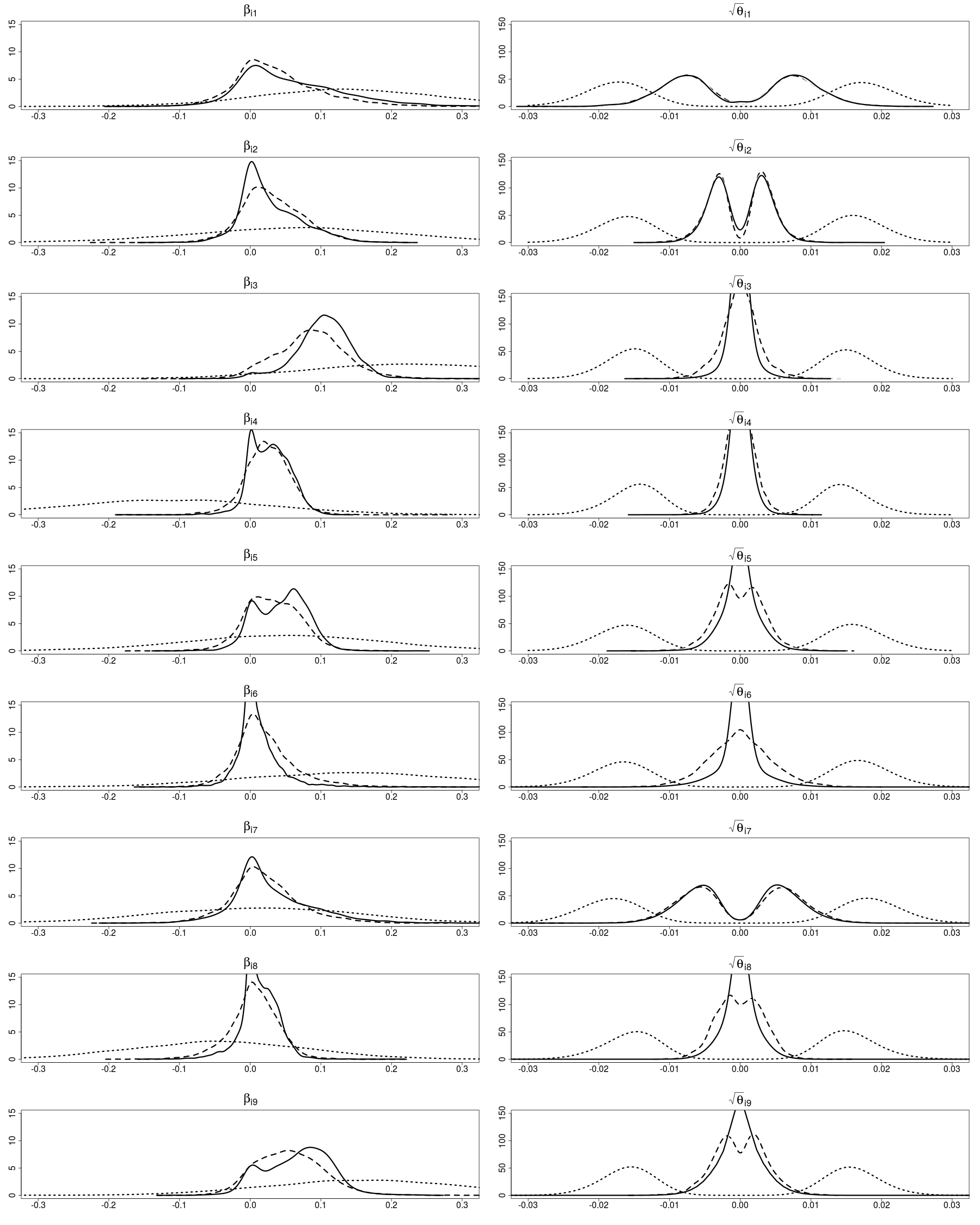}\\
  \end{tabular}
  \caption{DAX data.  Posterior densities of  $\beta_{ij}$ (left-hand side) and $\sqrt{\theta_{ij}}$  (right-hand side)
  for  $i=10$ and  $j=1,\dots,9$ (from top to bottom), derived  under the conditionally conjugate prior  $\beta_{ij} \sim \Normal{0,10} $ and
   $\theta_{ij} \sim \IGammad{0.1,0.001}$      (dotted line),   the hierarchical Bayesian Lasso prior with $a_i^\tau=a_i^\xi=1$ (dashed line) and
   a  hierarchical double gamma prior  with $a_i^\tau \sim \Exp (10)$ and $a_i^\xi \sim \Exp (10)$ (solid line). % Updated februar 2018.
   % \comment{TO DO ANGELA: as requested by a reviewer, remove values for $a^\tau , a^\tau$ from the caption and center  $\beta_{ij}$ and $\theta_{ij}$. Match the fonts of the caption and the axis.}
    }    \label{fig:DAX_DATA_plot_beta_final_22082016_1_5}
\end{figure}

Exemplarily,  detailed results are presented for the tenth time-varying regression
%In Figure~\ref{fig:DAX_DATA_plot_beta_final_22082016_1_5}  and Figure~\ref{fig:DAX_DATA_plot_sqrt_theta_final_22082016_6_9} we present, respectively,
 in Figure~\ref{fig:DAX_DATA_plot_beta_final_22082016_1_5},  where we compare the posterior densities
of $\beta_{ij}$ and $\sqrt \theta_{ij}$, for $i=10$ and $j=1,\dots,9$, obtained  under the different priors. As expected, % \citep{fru-wag:sto},
under  the inverted gamma prior all posteriors  distributions  of $\sqrt \theta_{ij}$  are bounded away from 0,
  with the position of the symmetric posterior modes  (roughly $\pm$0.015)
 being more or less the same for all coefficients.\footnote{The location of the posterior distribution is mainly driven by the prior --
   for the alternative hyperparameters  $s_0 = 0.5$ and $S_0= 0.2275$ (not shown in the figure)  the posterior modes  shift  to around $\pm$0.1.}
 As opposed to this, both shrinkage priors  allow the posterior distribution of $\sqrt \theta_{ij}$ to concentrate at 0, if appropriate, and in this way
allows to distinguish between coefficients that  are time-varying  ($j=1,2,7$)  and the remaining coefficients which turn out to be static.
When comparing both shrinkage priors,  %in  Figure~\ref{fig:DAX_DATA_plot_beta_final_22082016_1_5},
the influence of the increased shrinkage introduced by the double gamma prior %compared to  $a^\tau=a^\xi=1$
is evident for static coefficients, with the posterior of   $\sqrt \theta_{ij}$  showing  a much more pronounced spike at 0 than   the Bayesian Lasso prior. % $a^\tau=a^\xi=1$, which leads to increased efficiency in identifying static components.

\begin{figure}[t!]%[h]
  \centering
\begin{tabular}{c}
\includegraphics[width=0.9\textwidth,height=10cm]{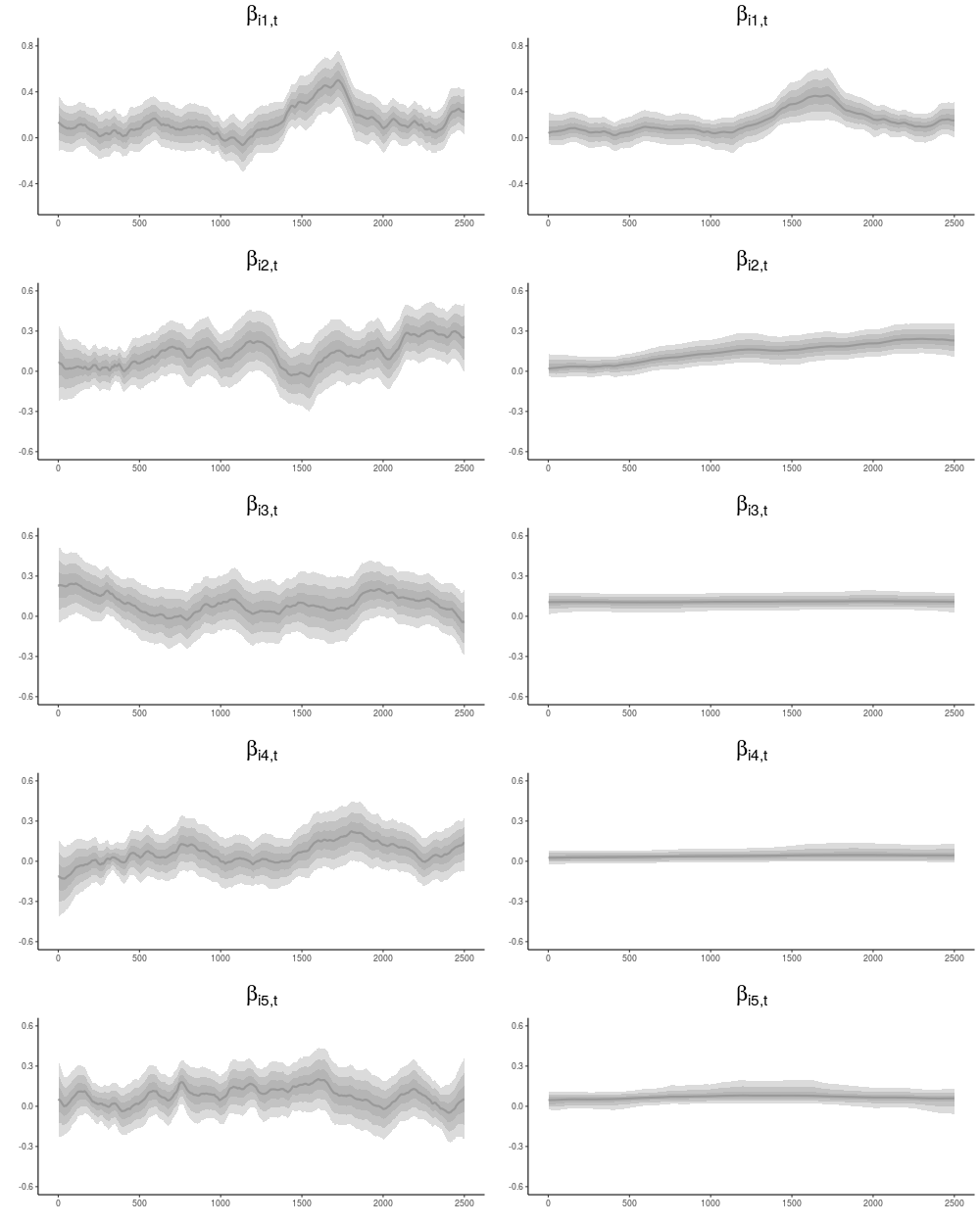}\\
\includegraphics[width=0.9\textwidth,height=8cm]{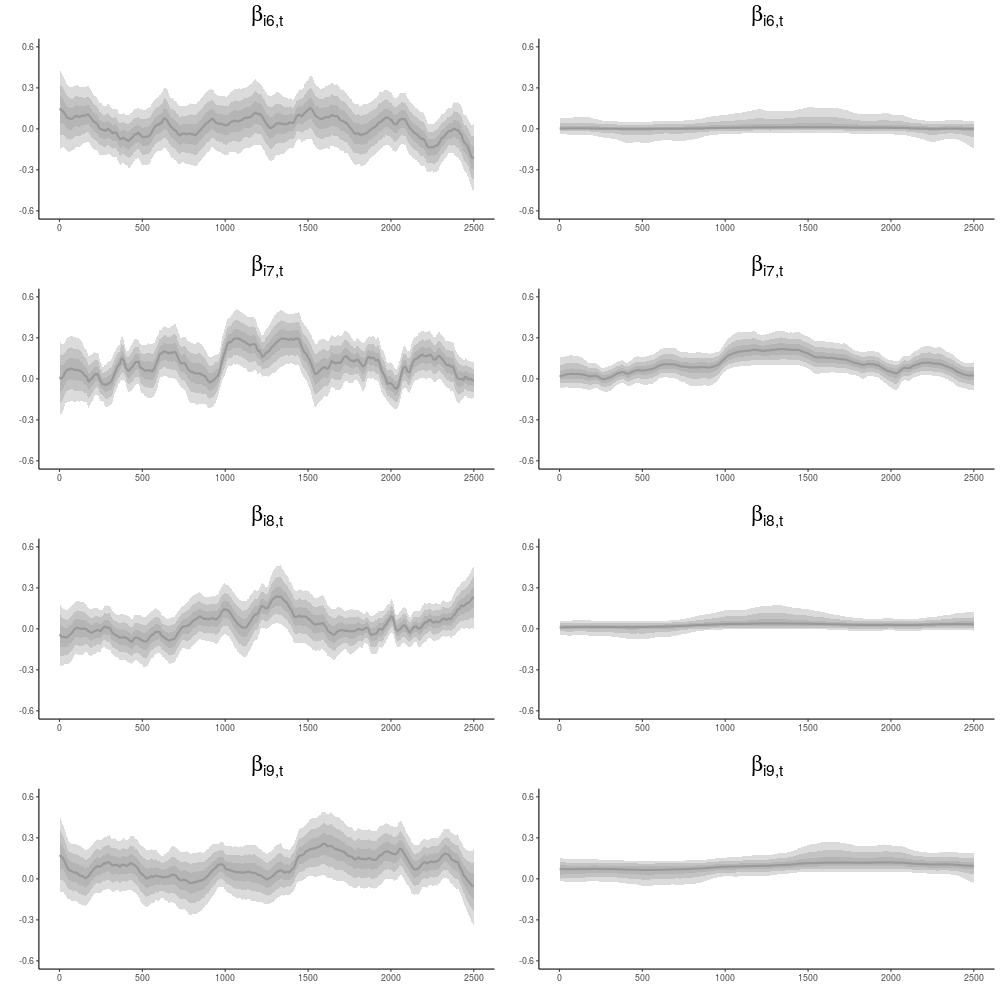}\\
  \end{tabular}
  \caption{DAX data.~Pointwise $(0.025,0.25,0.5,0.75,0.975)$-quantiles of the posterior paths
  $\beta_{ij,t} = \beta_{ij} + \sqrt{\theta_{ij}} \tilde \beta_{ij,t}$  for  $i=10$
  and  $j=1,\dots,9$ (from top to bottom);
 derived  under the  conditionally conjugate prior  $\beta_{ij} \sim \Normal{0,10} $ and
   $\theta_{ij} \sim \IGammad{0.1,0.001}$     (left-hand side)  and a   hierarchical double gamma prior with $a^\tau \sim \Exp (10)$ and $a^\xi \sim \Exp (10)$ (right-hand side).  % Version Februar 2017
  % \comment{TO DO ANGELA: remove Non-shrunken/shrunken from the caption and center  $\beta_{ij,t}$ and $\theta_{ij}$. Match the fonts of the caption and the axis.}
   }     \label{fig:paths_DAX_1}
\end{figure}

For the static coefficients, the posterior distributions  of  $\beta_{ij}$ indicate that
 some coefficients are   significant, in particular when $j=3$ and $j=9$, whereas others are clearly insignificant, e.g. when $j=6$.
These findings are confirmed by the corresponding posterior paths of
$\beta_{ij,t} = \beta_{ij} + \sqrt{\theta_{ij}} \tilde \beta_{ij,t}$   displayed  in
Figure~\ref{fig:paths_DAX_1} under the hierarchical double gamma prior   and the inverted gamma prior.
For the double gamma prior, the coefficients  $\beta_{i1,t}$, $\beta_{i2,t}$, and $\beta_{i7,t}$ are the only ones that are  time-varying, whereas
   $\beta_{i3,t}$ and  $\beta_{i9,t}$ are constant, but   shifted away from 0.    Figure~\ref{fig:paths_DAX_1} also demonstrates a dramatic gain in statistical efficiency, in terms of  dispersion of the  posterior distribution of $\beta_{ij,t}$ for each point in time,  compared to the  inverted gamma prior. This holds   in particular for coefficients which are static, but significant such as  $\beta_{i3,t}$  and $\beta_{i9,t}$. In addition,  the estimated paths are much smoother under the double gamma prior, which facilitates  the interpretation of   the time-varying components  $\beta_{i1,t}$, $\beta_{i2,t}$, and $\beta_{i7,t}$.  The coefficient $\beta_{i2,t}$, for instance, shows a trending behaviour, which is not apparent  under the  inverted gamma prior.

\begin{figure}[t!]%[ht]
\vspace*{1cm}\begin{center}
\begin{minipage}[b]{0.45\linewidth}
    \includegraphics[scale=0.18,page=1]{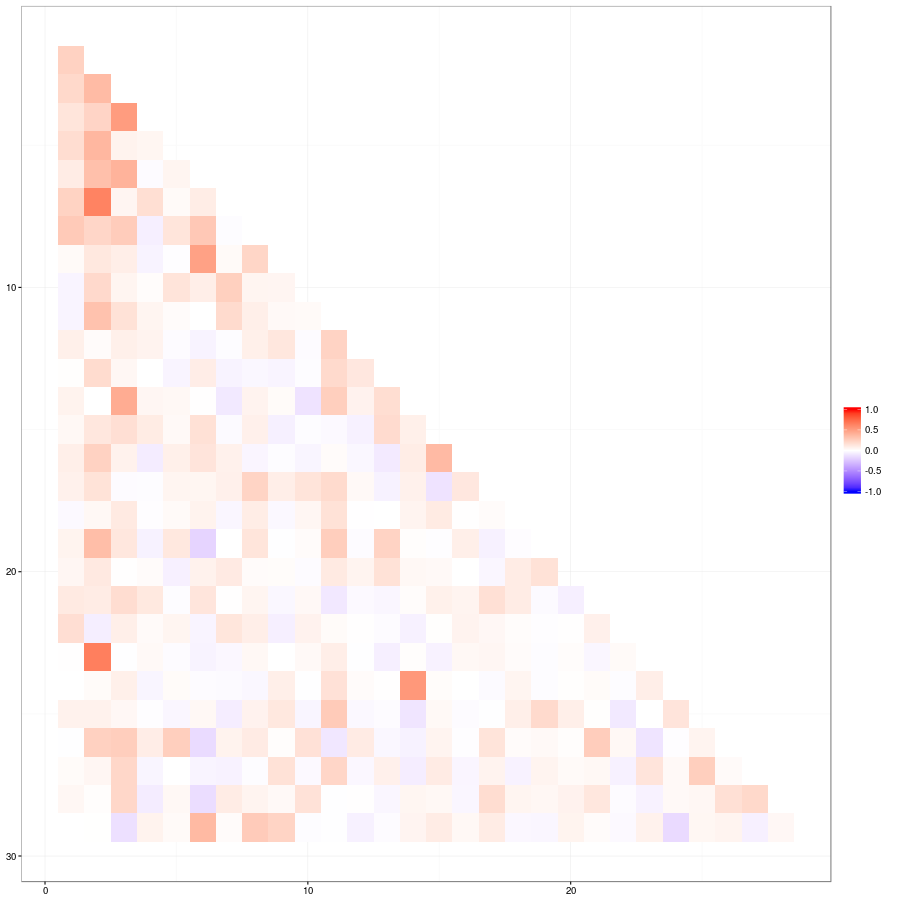}
%\%subcaption{ \textit{Non-shrunken version.}}
\label{graph:vergleich_IG}
\end{minipage}
\quad
%\begin{minipage}[b]{0.45\linewidth}
 %   \includegraphics[scale=0.18,page=1]{./plots_DAX/Xohne_text_plot_atau_1_1_08092016}
%%\subcaption{ \textit{$a^\tau=1$}}
%\label{graph:vergleich_1}
%\end{minipage} \\
%\begin{minipage}[b]{0.45\linewidth}
%    \includegraphics[scale=0.18,page=1]{./plots_DAX/Xohne_text_plot_atau_01_1_08092016}
%%\subcaption{ \textit{$a^\tau=0.1$}}
%\label{graph:vergleich_01}
%\end{minipage}
\begin{minipage}[b]{0.45\linewidth}
    \includegraphics[scale=0.18,page=1]{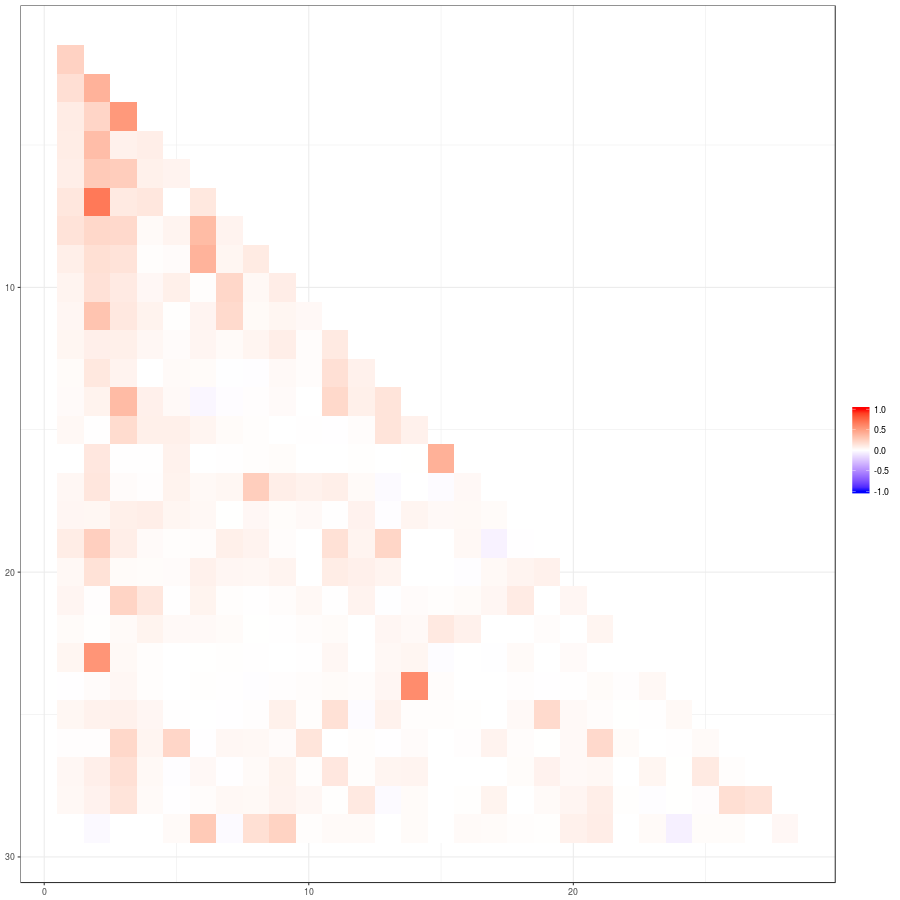}
%%\subcaption{ \textit{$a^\tau=0.1$}}
\label{graph:vergleich_01}
\end{minipage}
%\quad
%\begin{minipage}[b]{0.45\linewidth}
  %  \includegraphics[scale=0.18,page=1]{./plots_DAX/Xohne_text_plot_atau_005_1_08092016}
%%\subcaption{ \textit{$a^\tau=0.05$}}
%\label{graph:vergleich_005}
%\end{minipage} \\
\end{center}
\caption{DAX data.~Heat plot of the posterior median  of the 29$\times$28 Cholesky factor matrix  $\Bm_t$  at $t= 1150$,
derived  under the  conditionally conjugate prior  $\beta_{ij} \sim \Normal{0,10} $ and
   $\theta_{ij} \sim \IGammad{0.1,0.001}$     (left-hand side)  and a   hierarchical double gamma prior with $a^\tau \sim \Exp (10)$ and $a^\xi \sim \Exp (10)$ (right-hand side).  Values shrunken to zero are white.   }
% derived  under the conditionally conjugate prior  $\beta_{ij} \sim \Normal{0,10} $ and
  % $\theta_{ij} \sim \IGammad{0.1,0.001}$     (top left) and shrinkage priors with  $d_1 =d_2 =e_1=e_2 = 0.001$
  % and  $a^\tau=a^\xi=1$ (top right), $a^\tau=a^\xi=0.1$ (bottom left) and $a^\tau=a^\xi=0.05$ (bottom right). Red and blue indicates, respectively,  positive %and negative values.
  %\comment{TO DO ANGELA: substitute OLD Version October 2016 by adaptive prior on $a^\tau,a^\xi$.}
  \label{fig:vergleich_shrinkage}
\end{figure}

 Similar impact of our shrinkage method can be observed  for the remaining 27 equations in the TVP   model. %s  for $i=2,\ldots, 9, 11, \ldots  , 29$.
Overall, we investigated  all 406 posterior paths $\beta_{ij,t}$, together with the corresponding posterior distributions of $\beta_{ij}$ and
 $\sqrt \theta_{ij}$,  and found that a large fraction  of these coefficients is not significant.
For illustration, we display  in Figure~\ref{fig:vergleich_shrinkage} one (out of  2500) heat maps of the posterior median of  the $29 \times 28$ Cholesky factor matrix $\Bm_t$  at $t=1150$. Whereas the majority of the estimated  coefficients $\hat{\beta}_{ij,t}$  is different from zero for the
 inverted gamma prior, only a small part is significantly different from zero for the double gamma prior.

%%%CHECK  \clearpage

 \begin{figure}[t!]%[ht]
\begin{center}
 \begin{tabular}{cc}
 \includegraphics[scale=0.4]{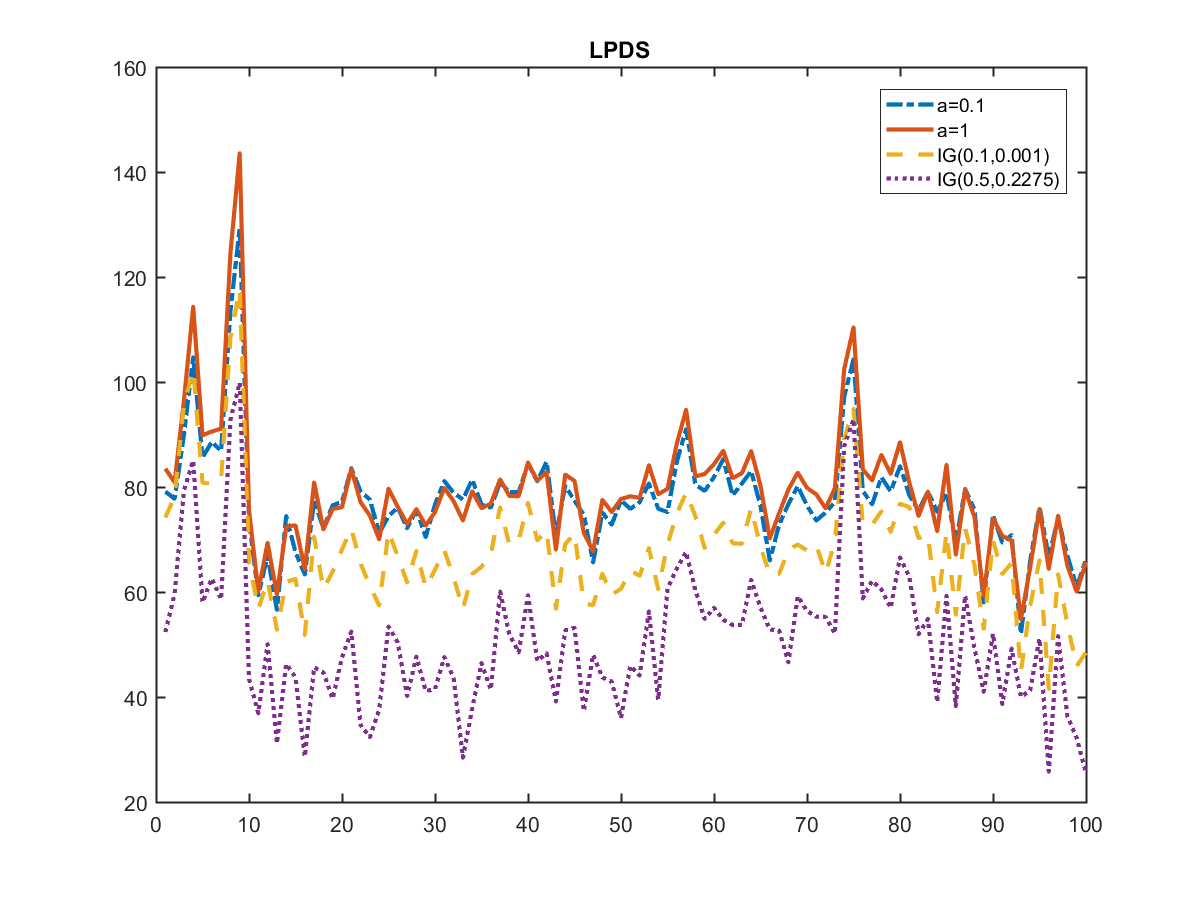}&
 \includegraphics[scale=0.4]{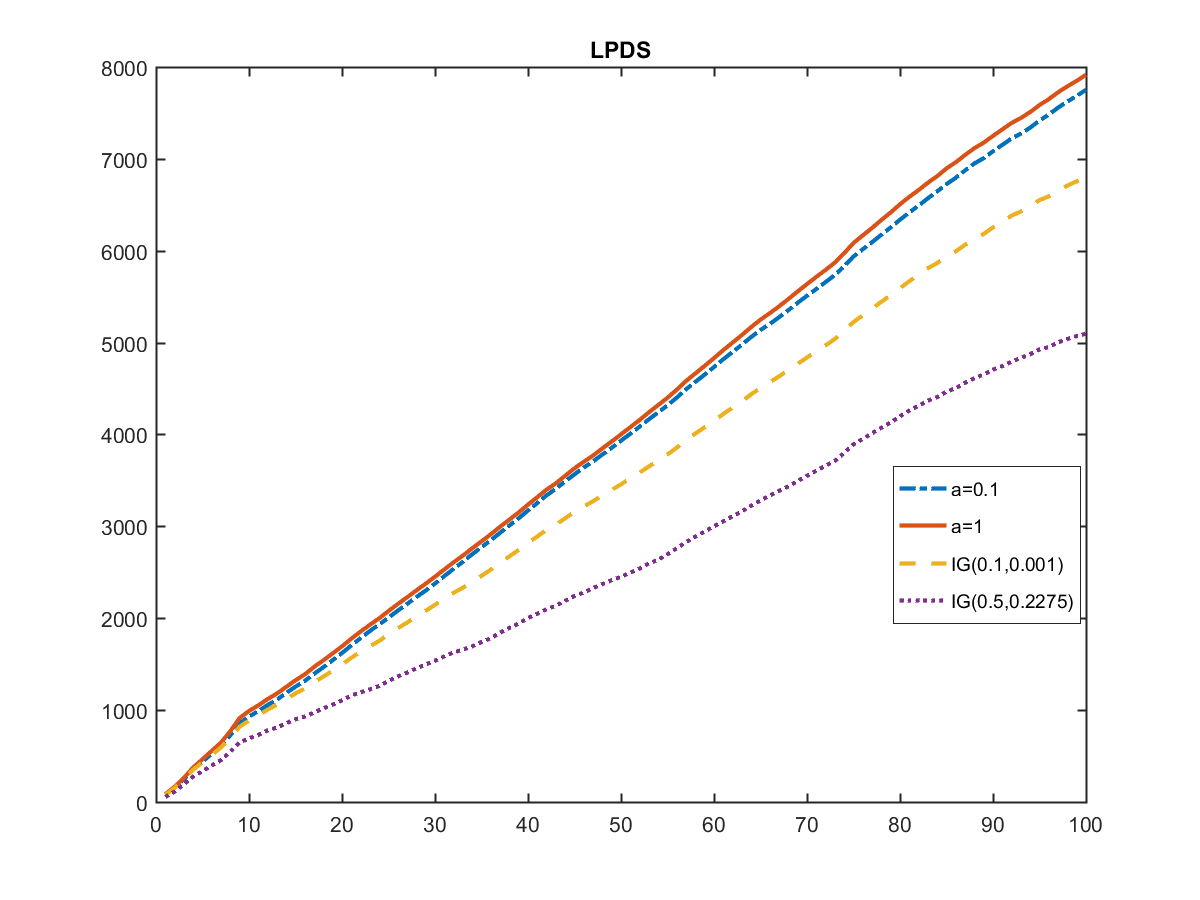}
 % based on Sylvia's MATLAB programm
 %DAX_p15_cum_log_pred_scores_12_08_2016}
 \end{tabular}
 \end{center}
 \caption{DAX data.~Individual (left-hand side) and cumulative (right-hand side) log predictive density scores for the last 100 time points using the last 400 observations as training sample. Shrinkage prior with $a^\tau=a^\xi=1$ (full line) and $a^\tau=a^\xi=0.1$ (dash-dotted line)
 in comparison to   the inverted gamma priors    $\theta_{ij} \sim \IGammad{0.1,0.001}$  (dashed line)  and $\theta_{ij} \sim \IGammad{0.5,0.2275}$  (dotted line). %  \Sylvia{Correct caption. LPDS!}
 }  \label{fig:cum_log_pred_scores_12_08_2016}
 % and $\beta_{ij} \sim \Normal{0,10} $.}
%  under   the   hierarchical double gamma prior with $a^\tau \sim \Exp (10)$ and $a^\xi \sim \Exp (10)$ and the hierarchical Bayesian Lasso
 \end{figure}

Finally, we  compare the various priors using LPDS  for the last 500 returns, with the first 400 observations serving as training sample.
Very conveniently, the triangular structure of the model allows  to  decompose  the  29-dimensional predictive density as $p(\ym_{t}|\ym^{t-1})= \prod_{i=1}^r p(y_{i,t}|\ym^{t-1}) $. Hence, the overall  log predictive density score $ \LPS^*_{t}$  at time $t$ results as the sum of the  individual log predictive density scores $\LPS^*_{i,t}= \log \, p(y_{i,t}|\ym^{t-1})$,  derived  independently for each of the  $r=29$ TVP models:
\begin{eqnarray}  \label{cumLPD}
 \LPS^*_{t} =  \log \, p(\ym_{t}|\ym^{t-1})= \sum_{i=1}^r \log \, p(y_{i,t}|\ym^{t-1}) =  \sum_{i=1}^r  \LPS^*_{i,t} .
\end{eqnarray}
The  individual log predictive density scores $ \LPS^*_{i,t}$ are approximated using  the conditionally optimal Kalman mixture approximation introduced in Section~\ref{sec:forecasting} and the cumulative log predictive scores  are shown in Figure~\ref{fig:cum_log_pred_scores_12_08_2016} for the various priors.
 We find overwhelming evidence in favour of  using shrinkage priors instead of the popular inverted gamma prior. For the later, the choice of the hyperparameters  (see $\theta_{ij} \sim \IGammad{0.1,0.001}$  versus $\theta_{ij} \sim \IGammad{0.5,0.2275})$  exercises tremendous influence on the log predictive  density scores.
Figure~\ref{fig:cum_log_pred_scores_12_08_2016} also compares the hierarchical Bayesian Lasso prior with  the hierarchical double gamma prior with
fixed values $a^\tau=a^\xi=0.1$.  Although the posteriors of  $a^\tau$ and $ a^\xi$ in Figure~\ref{fig:DAX_DATA_plot_axi} clearly are bounded   away from the values $a^\tau=a^\xi=1$ corresponding to the Bayesian Lasso prior,  the log predictive  density  scores  are very similar for both shrinkage priors.\footnote{The posteriors in Figure~\ref{fig:DAX_DATA_plot_axi} are based on the entire time series, but similar figures result for the last 500 observations.}
 Evidently, the major predictive gain comes from substituting the popular inverted gamma prior for the process variances by a sensible shrinkage prior  that allows posterior concentration of the process variances  at zero (see again Figure~\ref{fig:DAX_DATA_plot_beta_final_22082016_1_5}). As long as these priors behave sensibly  at zero,  the data contain little information to discriminate between them due to   the  small signal-to-noise ratio inherent in financial time series.

\section{Conclusion} \label{sec:con}

In the present paper,  shrinkage for time-varying parameter (TVP) models was investigated within a Bayesian framework both for
univariate and  multivariate time series, with the aim to automatically reduce time-varying  parameters to static ones, if the model is overfitting. This goal was  achieved by formulating  shrinkage priors for  the process variances based on the  normal-gamma prior \citep{gri-bro:inf}, extending
 previous work  using spike-and-slab priors  \citep{fru-wag:sto} and  the  Bayesian Lasso  prior  \citep{bel-etal:hie-tv}.
 As a  major computational contribution, an efficient MCMC estimation scheme was   developed, exploiting the  ancillarity-sufficiency interweaving   strategy of \citet{yu-men:cen}.

 Our applications  %in economics and finance
 included  EU area inflation modelling based on a TVP generalized Phillips curve  and  estimating a time-varying covariance matrix  based on a sparse TVP Cholesky SV model for a multivariate time series of returns of  the DAX-30 index.   We investigated  different prior settings, including the popular  inverted gamma prior for the process variances,  using  log predictive density scores.  Overall, our findings suggest that the family of  double gamma priors introduced in this paper for  sparse TVP models is successful in avoiding overfitting, if  coefficients are, indeed, static or even insignificant.
 The   framework  developed in this paper is  very general  and  holds the promise to be useful for introducing  sparsity in  other  TVP   and  state space models  in many different settings. In particular,
 sparse time-varying parameter VAR models  result by  straightforward extensions of the methods discussed  in this paper.

 The underlying strategy of  using the non-centered parametrization of  a state space model to extend shrinkage priors introduced for variable selection in regression models  to variance selection in a state space model is very generic and many alternative shrinkage priors for variance selection seem worth to be investigated.
 As pointed out by  a reviewer,  extending the normal-gamma-gamma prior,  introduced recently  for highly structured regression models \citep{gri-bro:hie}, to variance selection  is a particularly promising venue for future research. This strategy  leads  to the following \lq\lq triple gamma prior\rq\rq\  in the context of variance selection for state space models for univariate time series:
 \begin{eqnarray*} \label{normalhorse}
\theta_j|\xi^{2}_j  \sim \Gammad{\frac{1}{2},\frac{1}{2\xi^{2}_j}},  \quad \xi_j^2|a^\xi,\kappa_j^2  \sim  \Gammad{a^\xi, \kappa_j^2/2},
\quad    \kappa_j^2 \sim   \Gammad{c^\xi,d^\xi},
 \end{eqnarray*}
 with three hyperparameters  $a^\xi$, $c^\xi$,  and $d^\xi$.  The special case  where  $a^\xi = c^\xi =1/2$ is of particular  interest, as it extends
 the horseshoe prior \citep{car-etal:hor} to variance selection for  state space  models which is very popular in regression analysis for its outstanding properties, see e.g. \cite{bha-etal:las}.

\subsection*{Acknowledgements}

We are extremely grateful  to the Guest Editors Sylvia Kaufmann and Herman van Dijk and  two anonymous referees for numerous  valuable comments and suggestions that improved the paper considerably.
We acknowledge  inspiring comments
 from participants of  ESOBE Meetings (Vienna 2012, Oslo 2013, Gerzensee  2015) and NBER-NSF Time Series Conferences (St.~Louis 2014, Vienna 2015, New York  2016). We owe special thanks to Mauro Bernardi, discussant of the paper at the 2016 Bolzano Workshop on Forecasting in Finance and Macroeconomics. We are grateful for  comments on preliminary versions  by participants  of  CFE 2011, the 1st EFaB Workshop (2013), the 7th Rimini Bayesian Econometrics Workshop (2013), %the 6th Trondheim Symposium in Statistics (2013), the 1st Bayesian Young Statisticians Meeting (2013),
 the 1st Vienna Workshop on High Dimensional Time Series in Macroeconomics and Finance (2013), %the 4th MCMSki Meeting (2014),
 and the 12th ISBA World Meeting (2014).

%----------------------------------------------------------------------------------------
%	REFERENCE LIST
%----------------------------------------------------------------------------------------

\bibliographystyle{apalike}
\bibliography{mybib}

%\include{mainTVP}
%\input{mainTVP}

\include{appendixTVP}

\end{document}

%% file: newcommands.tex
\newcommand{\atauprold}{\lambda^{[a^{\tau}]}}  % ANGELA
\newcommand{\axiprold}{\lambda^{[a^{\xi}]}}    % ANGELA
\newcommand{\ataupr}{b^{\tau}}  % Sylvia
\newcommand{\axipr}{b^{\xi}}    % Sylvia
\newcommand{\gammav}{\boldsymbol{\gamma}}

\newcommand{\e}{\mbox{\rm e}}
\newcommand{\dimy}{r}                     % dimension of y_t
\newcommand{\tr}{{\tiny \mbox{\rm tr}}}
\newcommand{\ytr}{\ym^{\tr}}
\newcommand{\LPSo}[1]{\LPS^{\star}_{#1}}

\newcommand{\weg}[1]{}
\newcommand{\dx}{\,\mathrm{d}x}
\newcommand{\E}{\mathrm{E}}
\newcommand{\V}{\mathrm{V}}
\newcommand{\B}{\mathrm{B}}
\newcommand{\Exp}{\mathcal{E}}
\newcommand{\Cm}{{\mathbf C}}
\newcommand{\Bm}{{\mathbf B}}
\newcommand{\Am}{{\mathbf A}}
\newcommand{\Dm}{{\mathbf D}}
\newcommand{\cm}{{\mathbf c}}
\newcommand{\Fm}{{\mathbf F}}
\newcommand{\ystar}{y^\star}
\renewcommand{\arraystretch}{1.2}
\renewcommand{\vfill}{\vspace*{\fill}}
\newcommand{\nb}{\small}
\newcommand{\spc}{\phantom{$-$}}
% general parameters
\newcommand{\rvYm}{\mathbf{Y}}
\newcommand{\EVfs}{\mathrm{EV}}
\newcommand{\Gamfun}[1]{\Gamma (#1)}
\newcommand{\kfRc}{  R}  % symbol for the cov of the obs noise
\newcommand{\kfR}{{\mathbf \kfRc}}  % symbol for the cov of the obs noise
\newcommand{\pitrue}{\pi ^{\rm true}}
\newcommand{\thc}{\vartheta} %notation used to summarize  univariate unknown model parameters
\newcommand{\thmod}{{\mathbf{\boldsymbol{\thc}}}} %notation used to summarize  multivaraite   unknown vector
\newcommand{\thetav}{{\mathbf{\boldsymbol{\theta}}}} %notation used to summarize  multivaraite   unknown vector
%%%% Specific new commands for this paper
\newcommand{\iid}{\mbox{\rm i.i.d.}}
\newcommand{\bfz}{{\mathbf{0}}}
\newcommand{\ypro}{u}
\newcommand{\ypros}{y^{\star}}
\newcommand{\yprov}{\ym^{u}}
\newcommand{\yprodiff}{z}
\newcommand{\yprodiffv}{\mathbf{\yprodiff}}
\newcommand{\tm}{\mathbf{t}}
\newcommand{\errordiff}{\varepsilon}
\newcommand{\errordiffv}{\boldsymbol{\varepsilon}}
\newcommand{\psiv}{\boldsymbol{\psi}}
\newcommand{\tauv}{\boldsymbol{\tau}}
\newcommand{\xiv}{\boldsymbol{\xi}}
\newcommand{\Omegav}{\boldsymbol \Omega}
\newcommand{\sigmaerr}{R}
\newcommand{\alphad}{r} %sdimension of regression coefficient
\newcommand{\scale}{\omega} % scaling factor for variance heterogeneity
\newcommand{\scalev}{\boldsymbol{\scale}}
\newcommand{\im}[1]{^{(#1)}}
\newcommand{\labset}{L}
\newcommand{\Rv}{\mathbf{R}} % all mixture indicators
\newcommand{\Pm}{\mathbf{P}}
\newcommand{\Qm}{\mathbf{Q}}
\newcommand{\Sm}{\mathbf{S}}
\newcommand{\mv}{\mathbf{m}}
\newcommand{\sa}{Q}
\newcommand{\ev}{\mathbf{e}}
\newcommand{\z}{\phantom{0}}
\newcommand{\zz}{\phantom{00}}
\newcommand{\zzz}{\phantom{000}}
\newcommand{\Probsym}{\mbox{\rm Pr}}
\newcommand{\Prob}[1]{\Probsym (#1)}
\newcommand{\betac}{\beta} %symbol for regression parameter
\newcommand{\betad}{d} %sdimension of regression coefficient
\newcommand{\betar}{\boldsymbol{\betac}} % vector of regression coefficients
\newcommand{\betav}{\betar} % vector of regression coefficients - notation used in some chapter
\newcommand{\betai}[1]{\indiv{\betav}_{#1}} %notation for a subject specific parameter
\newcommand{\betavi}{\indiv{\betav}} %notation for a subject specific parameter
\newcommand{\betavitilde}{\indiv{\tilde{\betav}}}
\newcommand{\betainoncen}{b}
\newcommand{\zetai}[1]{\indiv{\betainoncen}_{#1}} %notation for a subject specific parameter
\newcommand{\zetaitilde}[1]{\indiv{\tilde{\betainoncen}}_{#1}} %notation for a subject specific parameter
\newcommand{\zetavi}[1]{\indiv{\mathbf{\betainoncen}}_{#1}} %notation for a subject specific parameter
\newcommand{\zetavitilde}[1]{\indiv{\tilde{\mathbf{\betainoncen}}}_{#1}} %notation for a subject specific parameter
\newcommand{\Psim}{\boldsymbol{\Psi}}
\newcommand{\Xbeta}{{  \mathbf \Xz}}
\newcommand{\Xz}{x}% symbol used for the row in a regressor matrix.
\newcommand{\Normult}[2]{ \mathcal{N} _{#1}\left(#2\right)}
\newcommand{\Mulnom}[1]{\mbox{\rm MulNom}\left(#1\right)}
\newcommand{\bs}{b} %mean of prior and posterior of mu
\newcommand{\Bs}{B} %variance- prior and posterior of variance
\newcommand{\br}{{\mathbf{\bs}}} %mean of prior and posterior of beta
\newcommand{\Br}{{\mathbf{\Bs}}} %variance-covariance of prior and posterior of beta (independence prior)
\newcommand{\yc}{y}
\newcommand{\ydens}{\yc}  % argument of a univariate rv in a density
\newcommand{\ym}{{\mathbf \yc}} % y multivariate observation
\newcommand{\yv}{\ym}
\newcommand{\constant}{\mbox{\rm constant}}
\newcommand{\pdf}[3]{f_{ {\footnotesize #1}}(#2;#3)}
\newcommand{\Normalpdfa}[2]{\pdf{\mathcal{N}}{#1}{#2}}
\newcommand{\Normal}[1]{ \mathcal{N}\left(#1\right)}
\newcommand{\NoGo}[1]{ \mathcal{NG}\left(#1\right)}
\newcommand{\Normalpdf}{\varphi}
\newcommand{\Gammad}[1]{ \mathcal{G}\left(#1\right)}
\newcommand{\IGammad}[1]{ \mathcal{IG}\left(#1\right)}
\newcommand{\GIG}[3]{ \mathcal{GIG}\left(#1,#2,#3\right)}
\newcommand{\Lap}[3]{ Lap\left(#1,#2\right)}
\newcommand{\InvGau}[2]{ \mathcal{InvGau}\left(#1,#2\right)}
\newcommand{\xm}{\mathbf{x}} % vector of regression coefficients

\newcommand{\LPS}{{\mbox{\rm LPDS}}}

\newcommand{\Fd}[1]{\mbox{\rm F}\left(#1\right)}
\newcommand{\Td}[2]{\mbox{\rm T}_{#1}\left(#2\right)}
\newcommand{\indic}[1]{I\{#1\}}
\newcommand{\trans}[1]{#1 ^{'}} % transposed sign in linear algebra
\newcommand{\EV}{\mathcal{EV}}
\newcommand{\KS}{\mathcal{KS}}
\newcommand{\Kv}{{\mathbf  K}}
\newcommand{\Logistic}{\mathcal{LO}}
\newcommand{\GenLogistic}{\mathcal{LG}}
\newcommand{\Real}{\Re}
\newcommand{\Betadis}[1]{\mathcal{B}\left(#1\right)}
\newcommand{\Betafun}[1]{B (#1)}
\newcommand{\new}{^\mathrm{new}}
\newcommand{\old}{^\mathrm{old}}
\newcommand{\minusindex}[1]{ - #1 }
\newcommand{\Chi}{\chi^2}
\newcommand{\Chisqu}[1]{\chi^2 _{#1}}
\newcommand{\Xbetar}{\Xbeta^r} % design matrix for the random effects - vector
\newcommand{\Xbetaf}{\Xbeta^f} % design matrix for the fixed effects - vector
\newcommand{\alphaf}{\alpha} % notation for fixed effects
\newcommand{\alphav}{\boldsymbol{\alphaf}} % vector of regression coefficients
\newcommand{\Sigmam}{\boldsymbol{\Sigma}} % vector of regression coefficients
\newcommand{\omegav}{\boldsymbol{\omega}} % vector of regression coefficients
\newcommand{\Edis}[1]{\mbox{\rm E}\left(#1\right)} % Expectation of a rv in displaymode
\newcommand{\Bincoef}[2]{\left( \begin{array}{c}   #1 \\#2 \end{array}\right)}
\newcommand{\indivsymb}{s}
\newcommand{\indiv}[1]{ #1 ^{\indivsymb}} %notation for a subject specific parameter
\newcommand{\kfwc}{ w}
\newcommand{\kfw}{{\mathbf{\kfwc}}}  % symbol for the state noise
\newcommand{\wt}[1]{\kfw_{#1}}
\newcommand{\Qrcm}{{\mathbf{\kfQ}}} % covariance in random effects model
\newcommand{\mum}{\boldsymbol{\mu}} % multivariate mean of a normal distribtuion-
\newcommand{\Cholrcm}{{\mathbf{C}}} % group mean  in mixture model
\newcommand{\kfQc}{ Q }  % symbol for the cov of the state noise
\newcommand{\kfQ}{{\mathbf{\kfQc}}}  % symbol for the cov of the state noise
\newcommand{\Xbetamat}{{  \mathbf X}}
\newcommand{\Xbetarmat}{\Xbetamat^r} % design matrix for the random effects - matrix
\newcommand{\Xbetafmat}{\Xbetamat^f} % design matrix for the fixed effects - matrix
\newcommand{\error}{\varepsilon} % symbol for error
\newcommand{\errorv}{\boldsymbol{\error}}
\newcommand{\Ferror}{F_{\error}}
\newcommand{\ferror}{f_{\error}}
\newcommand{\Vrcm}{{\mathbf{V}}} % error variance in the marginal model
\newcommand{\Diag}[1]{\mbox{\rm Diag}\left(#1\right)}
\newcommand{\dimmat}[2]{(#1\times #2)} % Dimension of a matrix
\newcommand{\Bino}[1]{\mbox{\rm BiNom}\left(#1\right)}
\newcommand{\zs}{z}% symbol used for explanatory variable
\newcommand{\zv}{\mathbf{\zs}}% symbol used for explanatory variable
\newcommand{\betaci}[1]{\indiv{\betac}_{#1}} %notation for a subject specific parameter
\newcommand{\Gammainv}[1]{\mathcal{G}^{-1} \left(#1\right)}

\newcommand{\Uniform}[1]{\mathcal{U}\left[#1\right]}
\newcommand{\Ew}[1]{\mbox{\rm E}(#1)}   % Expectation of a rv
\newcommand{\pl}{\pi}  %notation used for the probabilities in a logit model
\newcommand{\plv}{{\mathbf{\boldsymbol{\pl}}}}  %notation used for the probabilities in a logit model
\newcommand{\plt}[1]{\pl_{#1}} %probability with index for a binary model
\newcommand{\plsw}[2]{\pl_{#1,#2}} %swithcing probability, single component; first index:class, second
\newcommand{\identm}{{\mathbf I}}
\newcommand{\identy}[1]{{\identm}_{#1}}
\newcommand{\kfW}{{\mathbf{W}}}  % symbol for the weight matrix
\newcommand{\dm}{{\mathbf{d}}}  % symbol for the weight matrix
\newcommand{\minusmcmc}[1]{_{\minusindex{#1}}} % exclude MCMC
\newcommand{\Studmult}[2]{t _{#1}\left(#2\right)}
\newcommand{\rv}{{\mathbf{r}}}

\newcommand{\Lihood}[2]{L \left(#1 |#2 \right) }
\newcommand{\expND}[0]{-\frac{1}{2} \sum_{i=1}^{n}\left(\xm_{i} - \mu \right)' \rho \left(\xm_{i} - \mu   \right)}
\newcommand{\expNDtwo}[0]{         \sum_{i=1}^{n}\left(\xm_{i} - \mu \right)' \rho \left(\xm_{i} - \mu    \right)}
\newcommand{\expNDtwominusrho}[0]{ \sum_{i=1}^{n}\left(\xm_{i} - \mu \right)' \left(\xm_{i} - \mu         \right)}
\newcommand{\expNDthree}[0]{\sum_{i=1}^{n}\left(\xm_{i} - \bar \xm \right)' \rho \left(\xm_{i} - \bar \xm \right)}
\newcommand{\expNDfour}[0]{\sum_{i=1}^{n}\left(\xm_{i} - \bar \xm \right)'  \left(\xm_{i} - \bar \xm      \right)} 

%% file: appendixTVP.tex
%%%%%%%%%%%%%%%
% Appendix

\appendix

\begin{center}
{\LARGE Achieving Shrinkage in a Time-Varying Parameter Model Framework} \\[3mm]
{\LARGE  Webappendix} \\[5mm]

{\large   Angela Bitto and Sylvia Fr\"uhwirth-Schnatter}\\[5mm]

{\large Institute for Statistics and Mathematics}\\
 {\large  Department of Finance, Accounting and Statistics}\\
{\large WU Vienna University of Economics and Business, Vienna, Austria}\\
{\large Email: {\tt  angela.bitto@wu.ac.at, sfruehwi@wu.ac.at}}\\

\end{center}

% Change equation numbering
\setcounter{equation}{0}
\setcounter{figure}{0}
\setcounter{table}{0}
\setcounter{page}{1}

\renewcommand{\thetable}{A.\arabic{table}}
\renewcommand{\thefigure}{A.\arabic{figure}}
\renewcommand{\thesection}{A.\arabic{section}}
\renewcommand{\theequation}{A.\arabic{equation}}
%\numberwithin{table}{section}
%\numberwithin{figure}{section}
%\numberwithin{equation}{section}

\section{Computational Details} \label{detmcmc}

\subsection{Details on the MCMC scheme in Algorithm~\ref{facsvalg}} \label{detmcmcsub}

 \subsubsection{Step (a): Sampling the latent states}
  \label{section:FFBS}

Step~(a) of Algorithm~\ref{facsvalg}  samples the latent states  $ \tilde \betav =( \tilde \betav_0, \tilde \betav_1, \dots, \tilde \betav_T)$  conditional on known parameters using either
{\it Forward Filtering Backward Sampling} (FFBS), as discussed in \citet{fru:dat} and  \citet{car-koh:ong},  or
the faster alternative known as \textit{all without a loop} (AWOL),  discussed in \citet{mcc-etal:sim} and \citet{kas-fru:anc}.
Subsequently, we provide details how the AWOL algorithm is implemented for TVP models.

The algorithm is  implemented for a slight  modification of the  non-centered TVP model  (\ref{eq:solve}) and  (\ref{eq:obs}), given by:
\begin{eqnarray*}
\tilde\beta_{jt}&=&\tilde\beta_{j,t-1}+\tilde\omega_{jt},\quad \tilde\omega_{jt}  \sim \Normal{0,1}, \\
y^\star_t &=& \Fm_t \tilde \betav_{t}+\varepsilon_{t},  \quad \varepsilon_{t} \sim  \Normal{0,\sigma^2_{t}},
\end{eqnarray*}
with %initial value $\tilde \betav_0 = (\tilde\beta_{10},\dots,\tilde\beta_{d0})' \sim \Normult{d}{0,\mathbf{P}_{0}}$ and, for each $t=1,\dots,T$,
outcome  $y^\star_t = y_t - \mathbf{x}_t \betav$  and  $\Fm_t=\xm_{t}\Diag{\sqrt{\theta_{1}},\dots,\sqrt{\theta_{d}}}$ for $t=1,\dots,T$.
 The homoscedastic model results with $\sigma^2_1=\ldots = \sigma^2_T=\sigma^2$.

  % \subsubsection{AWOL}  \label{awol}
Conditional on all other variables,  the joint density for the state process
   $ \tilde \betav =( \tilde \betav_0, \tilde \betav_1, \dots, \tilde \betav_T)$ is multivariate normal.
This distribution can be written in
   terms of the tri-diagonal precision matrix $\Omegav$  and the covector $\cm$, see also \citet{rue-hel:gau}:
   \begin{eqnarray} \label{postbetav}
    \tilde \betav  | \betav,\Qm , \Pm_{0} , \sigma^2 % \equiv (\tilde \betav_0, \tilde \betav_1, \dots, \tilde \betav_T)
   \sim \Normult{(T+1)  d }{\Omegav^{-1}\cm,\Omegav^{-1} },
   \end{eqnarray}
   where:
   \begin{equation*}
\Omegav  =   \begin{bmatrix}
\Omegav_{00}&\Omegav_{01}  & 0 & & \\
\Omegav^{\prime}_{01} &  \Omegav_{11}&\Omegav_{12} &0&0&\\
0 &  \Omegav^{\prime}_{12}&\Omegav_{22}&\Omegav_{23}&\ddots&\vdots\\
&  0&\Omegav^{\prime}_{23}&\ddots&\ddots&0 \\
&  \vdots & \ddots & \ddots &\Omegav_{T-1,T-1}&\Omegav_{T-1,T} \\
&  0&\ldots &0 &\Omegav^{\prime}_{T-1,T}&\Omegav_{TT} \\
  \end{bmatrix},  \quad
 \cm  =  \begin{bmatrix}
  \cm_0\\
  \cm_1\\
 \cm_2 \\
  \vdots \\
  \cm_T	
  \end{bmatrix} .
\end{equation*}
In this representation, each submatrix $\Omegav_{ts}$ is a matrix of dimension $d \times d$  defined as
\begin{eqnarray*}
 \Omegav_{00} &=&  \Diag{1/P_{0,11}\cdots 1/P_{0,dd}} % \bold{P}^{-1}_0
  + \bold{I}_d,\\
%  \Omegav_{11} &\equiv& \mathbf{v}^{\prime}_1 A_{1}  \mathbf{v}_1 + I_d + P^{-1}_0,\\
  \Omegav_{tt} &=& \Fm^{\prime}_t   \Fm _t /\sigma^2_t +   2 \bold{I}_d,\quad t= 1,\dots, T-1,\\
%  \Omegav_{tt} &\equiv& \mathbf{v}^{\prime}_t  A_{11,t} \mathbf{v}_t +  A_{22,t} + A_{22,t-1},\quad t= 1,\dots, T-1,\\
   % \Omegav_{TT} &\equiv& \mathbf{v}^{\prime}_T  A_{11,T}   \mathbf{v}_T  + A_{22,T},\\
      \Omegav_{TT} &=& \Fm^{\prime}_T   \Fm_T /\sigma^2_T  + \bold{I}_d,\\
   \Omegav_{t,t+1} &=& -  \bold{I}_d, \quad t=0, \dots, T-1,
\end{eqnarray*}
where $\bold{I}_d$ is the $d\times d$ identity matrix and
 $\cm_{t}$ is a  column vector of  dimension $d \times 1$, defined as
\begin{eqnarray*}
\cm_{0}  =  \bfz, \qquad
   \cm_{t}  =  (\Fm^{\prime}_t /\sigma^2_t) y_t^{\star},	\quad t= 1,\dots, T .
\end{eqnarray*}
 The specific structure of $\Omegav$
   allows sampling the state process $ \tilde \betav = (\tilde \betav_0, \dots, \tilde \betav_T)$,
    \textit{all without a loop}  from
   %$p(\tilde \betav_0, \dots, \tilde \betav_T|\betav, \sqrt{\theta_1},\dots,\sqrt{\theta_d},\sigma^2_1,\dots,\sigma^2_T)$,
    the $(T+1)  d$-variate normal posterior distribution $ \tilde \betav | \betav,\Qm , \Pm_{0} , \sigma^2  \sim \Normult{(T+1)  d }{ \Omegav^{-1}\cm, \Omegav^{-1} }$.
 %    \begin{equation}
Due to the band structure of $\Omegav$, calculating the Cholesky
decomposition $  \Omegav = \bold{LL'} $ is computationally inexpensive.
Based on a draw $\boldsymbol \epsilon \sim \Normult{(T+1)  d }{\bfz,\bold{I}}$,
we solve $\bold{ La}=\cm$ for $\bold a$ and $\bold L' \tilde  \betav = \bold a+ \boldsymbol \epsilon$ for $\tilde \betav$ by using back-band substitution
instead of actually calculating $\bold L^{-1}$. %Finally the initial value can be sampled from $\tilde \betav_0|$.
Further details on this method can be found in \citet{mcc-etal:sim}.

 \subsubsection{Step (b): Sampling the constant coefficients $\beta_j$ and the square root of the process variances $\sqrt{\theta_j}$}  \label{stepB}

Conditional on the state process $\tilde \betav=(\tilde \betav_0,  \tilde \betav_1,  \ldots, \tilde \betav_T)$, the observation equation (\ref{eq:obs}) of the non-centered state space model defines an
expanded regression model:
 \begin{equation*}  \label{eq:step2}
  y_t = \mathbf{z}_t \boldsymbol{\alpha} + \varepsilon_t, \quad \varepsilon_t
  \sim \Normal{0,\sigma^2_t},
 \end{equation*}
with regression coefficient $\boldsymbol{\alpha} = (\beta_1, \ldots,\beta_d, \sqrt{\theta_1},\ldots,\sqrt{\theta_d})^{\prime}$ and
covariate vector  $\mathbf{z}_t$ defined as:
\begin{equation}  \label{regzt}
\mathbf{z}_t = \left(\begin{array}{cccccc}
  x_{t1}   & \cdots & x_{td} &  x_{t1}  \tilde \beta_{1t}  & \cdots & x_{td}  \tilde \beta_{dt} \\
  \end{array}\right).
 \end{equation}
%Under the conjugate prior $\boldsymbol{\alpha} \sim \Normult{2d}{\mathbf{a}_0,\mathbf{A}_0}$, where $\mathbf{a}_0 =\bfz$ and
Under the conjugate prior $\boldsymbol{\alpha}| \tauv, \xiv \sim \Normult{2d}{\bfz,\mathbf{A}_0}$, where
$\mathbf{A}_0 = \text{Diag}(\tau^2_1,\dots,\tau^2_d$, $\xi^2_1,\dots,\xi^2_d)$,
it follows that the conditional posterior distribution  $p ( \alphav|  \tilde \betav,\tauv, \xiv, \sigma_1^2,\ldots, \sigma^2_T,  \ym )$  is a multivariate normal distribution,
\begin{eqnarray}   \label{app:allpar}
  \alphav|  \tilde \betav, \tauv, \xiv, \sigma_1^2,\ldots, \sigma^2_T, \ym
 \sim \Normult{2d}{\mathbf{a}_T,\mathbf{A}_T},
\end{eqnarray}
 with
 \begin{equation*}
 %\mathbf{a}_T= \mathbf{A}_T \left(\mathbf{\tilde W} \ym  + \mathbf{A}^{-1}_0 \mathbf{a}_0 \right), \qquad
 \mathbf{a}_T= \mathbf{A}_T \mathbf{\tilde W} \ym  , \qquad
 \mathbf{A}_T = \left( \mathbf{\tilde W  W } + \mathbf{A}^{-1}_0 \right)^{-1},
 \end{equation*}
 where $\ym=(y_1, \dots, y_T)'$ and $\mathbf{W}$ is a
 $(T \times 2d)$ regressor matrix with the $t$-th row being equal to $\mathbf{z}_t$ and
 $ \mathbf{\tilde W} = \mathbf{ W'} \text{Diag} (1/\sigma^2_1,\dots,1/\sigma^2_T)$.

In a shrinkage framework, some of the  variances  $\tau^2_j$   and $\xi^2_j$ may be close to zero, leading to conditional priors
$p(\beta_j|\tau_j^2)$ and   $p(\sqrt\theta_j|\xi_j^2)$  with huge prior information equal to   $1/\tau^2_j$   and $1/\xi^2_j$.
 In order to overcome numerical difficulties with such very informative priors, we make use of the following alternative computation of the posterior covariance matrix $\mathbf{A}_T$:
 \begin{eqnarray*}
& \mathbf{A}_T = \mathbf{A}^{1/2}_0 \mathbf{A}^*_T \mathbf{A}^{1/2}_0, & \label{psotAA} \\
& %\nonumber  \\
  \mathbf{A}^{1/2}_0   =  \text{Diag}(\tau_1,\dots,\tau_d,\xi_1,\dots,\xi_d), \quad   \mathbf{A}^*_T  =  \left(\mathbf{A}^{1/2}_0 \mathbf{\tilde W} \mathbf{ W} \mathbf{A}^{1/2}_0 + \mathbf{I}_{2d}    \right)^{-1}.   \nonumber
 \end{eqnarray*}
%% PROOF:
% (\ref{psotAA}) follows immediately from:
 %\begin{eqnarray*}
% \mathbf{A}_T &=& \left( \mathbf{\tilde W  W } + \mathbf{A}^{-1}_0 \right)^{-1}  \\
 %     &=&	\left(\mathbf{A}^{-1/2}_0 ( \mathbf{A}^{1/2}_0  \mathbf{\tilde W  W } \mathbf{A}^{1/2}_0 + \mathbf{I}_{2d}) \mathbf{A}^{-1/2}_0  \right)^{-1},  \\
%	      &=&	\mathbf{A}^{1/2}_0 ( \mathbf{A}^{1/2}_0  \mathbf{\tilde W  W } \mathbf{A}^{1/2}_0 + \mathbf{I}_{2d})^{-1} \mathbf{A}^{1/2}_0.
%\end{eqnarray*}

\subsubsection{Step (c): Sampling the prior variances}  \label{stepdd}

For the normal-gamma hierarchical priors given in  (\ref{equNGtheta}) and (\ref{equNG01}),
 it follows that the conditionally normal prior $\beta_j|\tau^2_j$ ($\sqrt \theta_j|\xi^2_j$) leads to a
posterior for the variance $\tau^2_j|\beta_j,a^\tau, \lambda^2$ ($\xi^2_j| \theta_j ,a^\xi, \kappa^2$), where the likelihood is the  kernel of an inverted gamma
density in $\tau^2_j$ ($\xi^2_j$). In combination with the gamma prior $\tau^2_j|a^\tau, \lambda^2  \sim \Gammad{a^\tau,a^\tau \lambda^2/2}$
($\xi^2_j  | a^\xi, \kappa^2  \sim \Gammad{a^\xi,a^\xi \kappa^2/2}$), this leads to
 a generalized inverse Gaussian distribution for $\tau^2_j|\beta_j,a^\tau, \lambda^2 $ ($\xi^2_j| \theta_j,a^\xi, \kappa^2 $):
 \begin{eqnarray}
    \tau^2_j|\beta_j, a^\tau, \lambda^2  &\sim&  \GIG{a^\tau-1/2}{a^\tau \lambda^2}{\beta^2_j}, \label{app:tau} \\
     \xi^2_j|\theta_j, a^\xi, \kappa^2 &\sim&  \GIG{a^\xi-1/2}{a^\xi \kappa^2}{\theta_j}.   \label{app:xi} 
 \end{eqnarray}
  The generalized inverse Gaussian distribution, $Y \sim \GIG{p}{a}{b}, a > 0, b > 0$  is a three parameter family with support on $y \in \mathbb{R}^+$.
 The density is given by
\begin{eqnarray*}
 f(y) = \frac{(a/b)^{p/2}}{2 K_p(\sqrt{ab})} y^{p-1} e^{-(a/2)y} e^{-b/(2y)},
 \end{eqnarray*}
where  $K_p(\cdot)$ is the modified Bessel function of the second kind. The $k$th moment $\mu_k = \E(Y^k)$ is given as
\begin{eqnarray*}
 \mu_k = \frac{K_{p+k}(\sqrt{ab})}{K_{p}(\sqrt{ab})} \left(\sqrt{\frac{b}{a}}\right)^k. %, \quad k \in R.
\end{eqnarray*}
 \citet{hoe-ley:gen} propose a new generation method for the cases where $p < 1, \sqrt{ab} < 0.5$, which is
 especially useful in the time-varying parameter case.
A very stable generator is implemented in the  R-package {\tt GIGrvg} \citep{hoe-ley:gig}.
%--

 As we have equipped   $\lambda^2$  and $\kappa^2$  with hyperpriors  $\lambda^2 \sim \Gammad{e_1,e_2}$  and
$ \kappa^2 \sim \Gammad{d_1,d_2}$, we need to sample these parameters  from the corresponding conditional posteriors:
 \begin{eqnarray}
\lambda^2|a^\tau, \tau^2_1, \ldots,  \tau^2_d \sim  \Gammad{e_1 + a^\tau  d ,  e_2 + \frac{\overline{\tau^2}}{2} a^\tau d  },  \label{lambdapost}\\
 \kappa^2|a^\xi,    \xi^2_1, \ldots,  \xi^2_d \sim \Gammad{d_1 + a^\xi  d , d_2 + \frac{\overline{\xi^2}}{2} a^\xi d }, \label{kappapost}
  \end{eqnarray}
 where $\overline{\tau^2}$   and $\overline{\xi ^2}$  are the averages of the  variances in the shrinkage  priors:
 \begin{eqnarray*}
\overline{\tau^2}=  \frac{1}{d}\sum^{d}_{j=1} \tau_j^2  , \qquad \overline{\xi^2} =  \frac{1}{d}\sum^{d}_{j=1} \xi_j^2   .
 \end{eqnarray*}
 If  $\overline{\xi^2}$ is small, i.e.~a lot of  sparsity is present, then the posterior expectation of  $ \kappa^2$ will be proportional to $1/d_2$.
 Hence, smaller values of  $d_2$ encourage stronger prior shrinkage of $\theta_j$ toward  zero.

\subsection{Approximating the one-step ahead predictive density} %\label{sec:onestep}

\subsubsection{Using the Kalman filter for prediction for known parameters} \label{kalffbs}

It is well-known that the Kalman filter can be applied to derive the predictive density in a Gaussian state space model for known model parameters. We exploit this procedure  to derive the predictive density $p( y_{t}|\ym^{t-1}, \betav,\Qm,\sigma^2_t)$ for known values of the parameters $\betav$, $\Qm$,  and  $\sigma^2_t$,
given observations  $\ym^{t-1} = ({y}_1, \dots, {y}_{t-1})$.
 In the following, details  are provided for the non-centered parameterization of the TVP model, based on rewriting the observation equation
 (\ref{eq:obs}) as:
\begin{eqnarray*}
\tilde\betav_{t}&=&\tilde\betav_{t-1}+\tilde\omegav_{t},\quad \tilde\omegav_{t}  \sim \Normult{d}{0,\identy{d}}, \\
y _t &=&\xm_{t}\betav+ \Fm_{t}  \tilde \betav_{t}+\varepsilon_{t},  \quad \varepsilon_{t} \sim \Normal{0,\sigma_t^2},
\end{eqnarray*}
where $ \Fm_{t}=\xm_{t} \Diag{\sqrt{\theta_1},\dots,\sqrt{\theta_d}}$.
Starting from $\tilde \betav_0  \sim \Normult{d}{\mathbf{m}_{0},\mathbf{C}_{0} }$ with $\mathbf{m}_{0}=\bfz$ and $\mathbf{C}_{0} =\mathbf{P}_{0}=
\Diag{P_{0,11}\cdots P_{0,dd}}$, the following three	 steps are repeated for $t=1,\ldots, T-1$:\footnote{To simplify the notation, dependence on the model parameters $\betav,\Qm$, and $\sigma^2_1, \ldots, \sigma^2_T$ is not made explicit.}
\begin{itemize}
\item[(a)] A \textit{propagation step}  to determine the one-step ahead predictive density $p(\tilde \betav_t|\ym^{t-1})$:
\begin{eqnarray*}
 \tilde \betav_t|\ym^{t-1}  	&\sim& \Normult{d}{ \mathbf{m}_{t-1},\mathbf{R}_t}, \\
 %\mathbf{a}_{t} 			&=&  \mathbf{m}_{t-1}, \\
 \mathbf{R}_{t} 			&=&  \mathbf{C}_{t-1} +  \identy{d}.  \nonumber
\end{eqnarray*}
\item[(b)] A \textit{prediction step}   to determine the predictive density $p(y_{t}|\ym^{t-1})$:
\begin{eqnarray*}
&y_{t}|\ym^{t-1} 	\sim \Normal{{\hat y}_t , S_t},   &\\
& {\hat y}_{t} 		=   \xm_{t}\betav  + \Fm_t \mathbf{m}_{t-1},    \qquad
S_{t} 				= \Fm_t \mathbf{R}_{t} \mathbf{F}_t^{'} +  \sigma^2_t. &
\end{eqnarray*}
\item[(c)] A \textit{correction step} to determine the filter density $p(\tilde \betav_t |\ym^{t})$:
\begin{eqnarray*}
& \tilde{\betav} _t |\ym^{t}  	\sim \Normult{d}{\mathbf{m}_{t},\mathbf{C}_{t}},& \\
 &\mathbf{m}_{t} 		  	=   \mathbf{m}_{t-1} + \mathbf{K}_{t}( y_{t}- {\hat y}_{t}  ), \quad %& \\&
 \mathbf{K}_{t}				= \mathbf{R}_{t} \mathbf{F}_{t} ^{'}  S^{-1}_{t}, \quad % &\\&
  \mathbf{C}_{t}		  	= (\identy{d} - \mathbf{K}_{t} \mathbf{F}_{t}) \mathbf{R}_{t}.&
 \end{eqnarray*}
 \end{itemize}
Note that prediction could equally well be performed in the centered  parametrization of the TVP model.

\subsubsection{Approximations for the one-step ahead predictive density in a TVP model} \label{sec:onestep}

In Section~\ref{sec:forecasting}, the conditionally optimal Kalman mixture approximation was introduced to approximate the
one-step ahead predictive density $p(y_{t}|\ym^{t-1})$.   As  it performs \textit{exact} analytical integration with respect to the entire state process $\tilde \betav_0, \ldots,  \tilde \betav_{t}$,  not surprisingly, we found that this method outperforms alternative approximations.

\citet{bel-etal:hie-tv}, for instance,  employ a purely simulation-based approach to approximate $p(y_{t}|\ym^{t-1})$ in a univariate framework. Based on the same output of the Gibbs sampler as the  conditionally optimal Kalman mixture approximation,  they derive draws from the predictive density by using following  simulation method  suggested
  in \citet[Section 2.1]{cog-etal:bay}.
 More specifically,  for each posterior draw  $m=1,\dots,M$,     generate $   \tilde\beta_{jt} ^{(m)}$   by drawing  from the normal
 distribution $\Normal{\tilde\beta^{(m)}_{j,t-1},1}$  and transform to generate  $\beta_{jt} ^{(m)}=  \beta_{j} ^{(m)}+
 \sqrt{\theta^{(m)}_j}   \tilde\beta_{jt} ^{(m)}$ for all $j=1, \ldots, d$.
  Based on $  \betav^{(m)}_{t } =( \beta_{1t} ^{(m)}, \ldots, \beta_{dt} ^{(m)})' $,   forecast a future value $y^{(m)}_{t}$ by drawing from
 $p(y_{t}| \betav^{(m)}_{t },\sigma_{t}^{2(m)} )$:
% $$y^{(m)}_{t} = \mathbf{x}_{t} \betav^{(m)}+\mathbf{x}_{t}\text{Diag} (,\dots,\sqrt{\theta^{(m)}_d}) \tilde\betav^{(m)}_{t } 
$$y^{(m)}_{t} = \mathbf{x}_{t} \betav^{(m)}_{t }
+  \varepsilon_{t}  ^{(m)} , \quad \varepsilon_{t}  ^{(m)} \sim  \Normal{0,\sigma_{t}^{2(m)}}.$$
The same method as in Section~\ref{sec:forecasting} is applied  to forecast $\sigma_{t}^{2(m)}$ for  the SV  specification.  Finally, for each time point $t$,  \citet{bel-etal:hie-tv} perform a nonparametric kernel smoothing algorithm on $y^{(m)}_{t}$ for $m=1,\dots,M$, which is then evaluated at the observed value $y_{t}$. To this aim, they use the function \textsf{ksdensity} in Matlab, which returns a probability density estimate evaluated at the observed value $y_{t}$. This approximation method is expected to be rather inaccurate in a multivariate setting, where the scores $\LPS^*_{i,t}$ are approximated individually for each time series and aggregated  through  (\ref{cumLPD})  to obtain $ \LPS^*_{t} $ and approximation errors easily accumulate.

We also experimented with what we call   the  \emph{naive Gaussian mixture approximation}. This approximation is based on choosing $\thmod =(\betav _{t},\sigma_{t}^2)$  in the mixture approximation   (\ref{LPDSexa})  and  using the  predictive density  $y_{t}| \betav_{t}, \sigma^2_{t}   \sim \Normal{\mathbf{x}_{t} \betav_{t}, \sigma^2_{t}}$  as  component density.
The parameters $\thmod^{(m)}=(\betav^{(m)}_{t},\sigma_{t}^{2(m)})$
 are sampled from the predictive posterior distribution $p(\betav_{t}, \sigma^2_{t} |\boldsymbol{y}^{t-1})$ as in
\citet{bel-etal:hie-tv}. This yields
\begin{eqnarray} \nonumber
 p(y_{t}|\boldsymbol{y}^{t-1}) &=& \int p(y_{t}|\betav_{t}, \sigma^2_{t}) p(\betav_{t}, \sigma^2_{t} |\boldsymbol{y}^{t-1}) d(\betav_{t}, \sigma^2_{t}) \\
		  &\approx& \frac{1}{M} \sum^M_{m=1} f_N \left(y_{t}; \mathbf{x}_{t} \betav^{(m)}_{t}, (\sigma_{t}^2)^{(m)} \right). \label{naive}
\end{eqnarray}
Whereas the naive Gaussian mixture approximation (\ref{naive}) avoids kernel density estimation,  it  can give very
imprecise results, in particular for state space models with a high signal-to-noise ratio. This occurs, when   the error variance $(\sigma_{t}^2)^{(m)} $ is considerably smaller than the forecasting variance ${S}^{(m)}_{t}$ in the conditionally optimal Kalman filter approximation (\ref{Sexcat}),
see \citet{fru:app_pre} for an early discussion of this problem and Section~\ref{sec:predecb} for more empirical details.

\begin{table}[t!]
\centering
\begin{tabular}{rll}
\hline
$i$ & Symbol &  Description     \\ \hline
1 & I-1MO &one-month Euribor (Euro interbank offered rate)\\
2& I-1YR & one-year Euribor (Euro interbank offered rate)\\
3 & SENT & percentage change in economic sentiment indicator\\
4 & STOCK-1 & percentage change in equity index \\
&& (Dow Jones, Euro Stoxx, Economic sector index financial)\\
5& STOCK-2 & percentage change in equity index - Dow Jones Euro Stoxx 50 index\\
6& EXRATE & percentage change in ECB real effective exchange rate \\
 &  & (CPI deflated, broad group of currencies against euro)\\
7& IP & percentage change in industrial production index\\
8& LOANS & percentage change in loans (total maturity, all currencies combined)\\
9&  M3& annual percentage change in monetary aggregate M3\\
10& CAR& registrations of new passenger cars\\
 11& OIL& percentage change in oil price (brent crude, 1-month forward)\\
12& ORDER& change in order-book levels\\
13& UNEMP& standardized unemployment rate (all ages, male \& female)\\
  14& HICP& harmonized index of consumer prices\\
\end{tabular}
\caption{ECB data. Data description.}\label{ecb:label}
\end{table}

\section{Data description} \label{sec:data}

\subsection{ECB data} \label{sec:data_ecb}

\begin{figure}[t!]  \centering
 \includegraphics[scale=0.4]{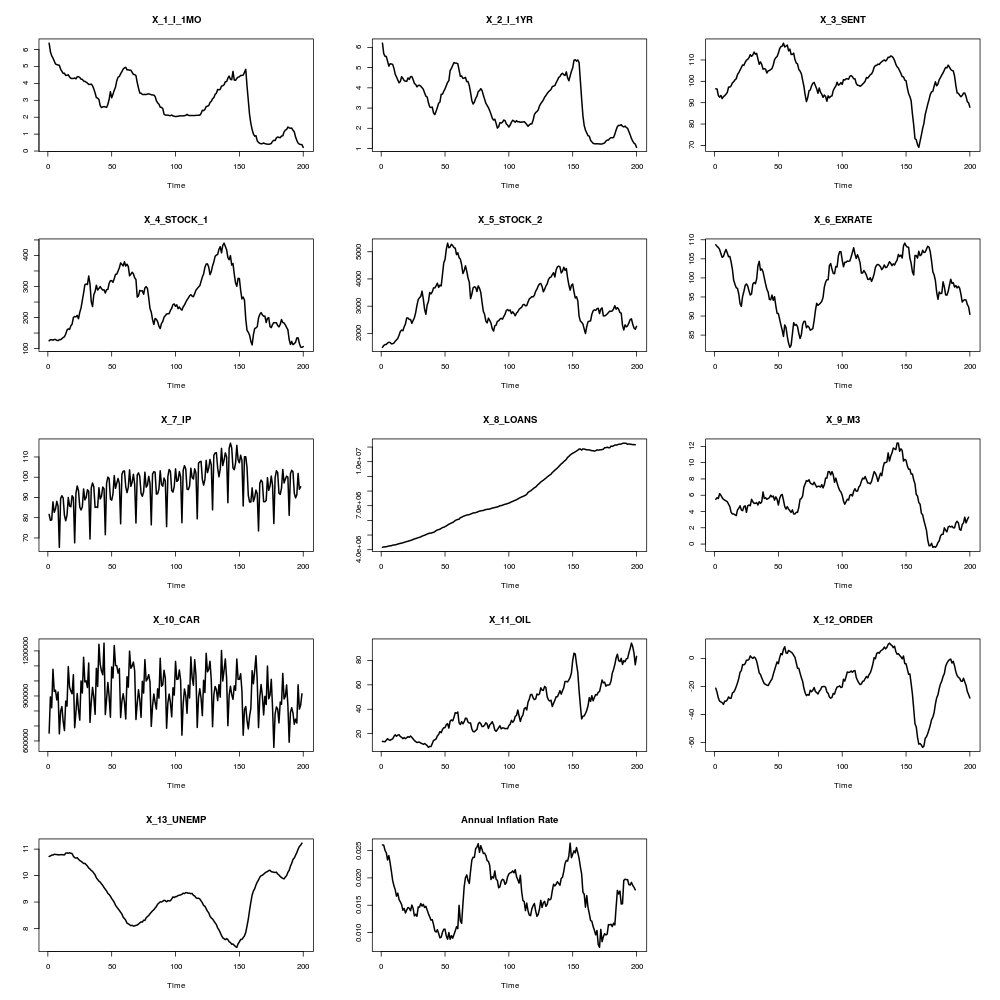}
 \caption{ECB data. Original time series.}
 \label{graph:org.pdf}
 \end{figure}

The  inflation  data  analyzed in Section~\ref{inflatiodata}
 can be retrieved freely from the ECB data warehouse.\footnote{For details and data see \url{http://sdw.ecb.europa.eu/}.}
 The data are monthly and range from February 1994 until November 2010.  The time series are listed  in Table~\ref{ecb:label} and plotted in Figure~\ref{graph:org.pdf}.
The original time series were transformed (log differences, differences) as in \citet{bel-etal:hie-tv} and each of the time series was standardized. The core inflation rate was transformed to have variance one.
 %\footcomment{Beide graphiken fehlen bei den Unterlagen vom Mai 2016.}

%\clearpage

\subsection{DAX data} \label{sec:data_dax}

The DAX - Deutscher Aktienindex (German stock index) is a blue chip
stock market index consisting of the 30 major German companies trading on the Frankfurt Stock Exchange, see Table~\ref{dax:label}.
The data set  analyzed in Section~\ref{DAXdata}
spans roughly 2500 daily stock returns from September 4th,  2001 until August 31st, 2011.  One of the companies was excluded from the
case study,  as it was not part of the DAX for the entire observation period.

\begin{table}[t!]
\centering
\begin{tabular}{rllcrll}
\hline
$i$ & Symbol &  Name  &  & $i$  &  Symbol  &  Name   \\ \hline
 1&ADS	&	Adidas 		&	&	 16&FRE &	Fresenius	         \\
 2&ALV	&	Allianz			&&	 17&FME&	Fresenius Medical Care	\\
  3&BAS	&	BASF			&&	 18&HEI	&	HeidelbergCement	\\
 4&BAYN	&	Bayer		&	&	 19&HEN3	&	Henkel	                \\
 5&BEI	&	Beiersdorf		&&	 20&IFX	&	Infineon Technologies	\\
 6&BMW	&	BMW			&&	21 &SDF	&	K + S	                \\
 7&CBK	&	Commerzbank		&& 22	 &LIN	&	Linde			\\
 8&CON	&	Continental		&&	23 &MRK	&	Merck			\\
 9&DAI	&	Daimler			&	& 24&MUV2	&	Munich Re		\\
 10&DBK	&	Deutsche Bank		&&	 25&RWE	&	RWE			\\
 11&DB1	&	Deutsche B\"orse		&&	 26&SAP	&	SAP			\\
 12&LHA	&	Deutsche Lufthansa	&	& 27&SIE	&	Siemens			\\
 13&DPW	&	Deutsche Post		&	& 28&TKA	&	ThyssenKrupp		\\
 14&DTE	&	Deutsche Telekom	&&	 29&VOW3	&	Volkswagen Group	\\
 15&EOAN	&	E.ON			&		&			&	\\ \hline
\end{tabular}
\caption{DAX data. Data description. }\label{dax:label}
\end{table}

\section{Further results} \label{sec:results}

\subsection{Detailed estimation results for the ECB data} \label{results:data_ecb}

Table~\ref{table:SV_FALSE_02_ASIS_TRUE} summarizes the posterior distributions  $p(\beta_j|\ym)$ and    $p(\sqrt{\theta_j}|\ym)$  for $j=1,\ldots, 37,$ for the EU-area  inflation  data  analyzed in Section~\ref{inflatiodata}.

 \newpage

{ \singlespacing
\small
\begin{longtable}[t!]{llrrrrr}
% \begin{longtable}[t!]{llccccc}
  \hline  \hline
   & Covariate & Mean & Stdev.& Median & 2.5\% quantil & 97.5\% quantil   \\
  \hline
  \endfirsthead
  \multicolumn{7}{l}%
{\tablename\ \thetable\ -- \textit{Continued from previous page}} \\
\hline
   & Name & Mean & Stdev.& Median & 2.5\% quantil & 97.5\% quantil \\
\hline
\endhead
\hline \multicolumn{7}{r}{\textit{Continued on next page}} \\
\endfoot
\hline
\endlastfoot
  $\beta_{1}$ &  Intercept & 2.770 & 0.655 & 2.782  & 1.450 & 4.016   \\
  $\beta_{2}$ &  $\pi_{t}$ & 1.8$\cdot 10^{-3}$ & 0.021  & 3.8$\cdot 10^{-5}$ & -0.039 & 0.049  \\
  $\beta_{3}$ &  $\pi_{t-1}$ & -3.7$\cdot 10^{-3}$ &0.012 & -3.4$\cdot 10^{-4}$ & -0.051 & 0.033   \\
  $\beta_{4}$ &  $\pi_{t-2}$ & -9.1$\cdot 10^{-3}$ & 0.021 & -2.3$\cdot 10^{-3}$ & -0.063 & 0.023   \\
  $\beta_{5}$ &  $\pi_{t-3}$ & -4.1$\cdot 10^{-4}$ & 0.018& 9.2$\cdot 10^{-8}$ & -0.040 & 0.037   \\
  $\beta_{6}$ &  $\pi_{t-4}$ & -3.6$\cdot 10^{-3}$ & 0.019 & -3.4$\cdot 10^{-4}$ & -0.050 & 0.033   \\
  $\beta_{7}$ &  $\pi_{t-5}$ & 0.019 & 0.027 & 0.010 & -0.017 & 0.084   \\
  $\beta_{8}$ &  $\pi_{t-6}$ & 1.5$\cdot 10^{-3}$ & 0.018& 6.0$\cdot 10^{-5}$ & -0.036 &0.043  \\
  $\beta_{9}$ & $\pi_{t-7}$ & -9.2$\cdot 10^{-3}$ & 0.023 & -2.0$\cdot 10^{-3}$ & -0.068 & 0.025   \\
  $\beta_{10}$ & $\pi_{t-8}$ & -1.2$\cdot 10^{-3}$ & 0.021 & -3.9$\cdot 10^{-5}$ & -0.049 & 0.043   \\
  $\beta_{11}$ & $\pi_{t-9}$ & -0.010 & 0.025 & -1.9$\cdot 10^{-3}$ & -0.079 & 0.026  \\
  $\beta_{12}$ & $\pi_{t-10}$ & -4.6$\cdot 10^{-3}$ & 0.022 & -2.5$\cdot 10^{-4}$ & -0.062 & 0.032   \\
  $\beta_{13}$ & $\pi_{t-11}$ & -0.041 & 0.047 & -0.030 & -0.150  & 0.020 \\
  $\beta_{14}$ & I-1MO & 0.595 & 0.259  & 0.586 & 0.078  & 1.130    \\
  $\beta_{15}$ & I-1YR & 0.152 & 0.245 & 0.052 & -0.172 & 0.774  \\
  $\beta_{16}$ & SENT & -6.9$\cdot 10^{-3}$ & 0.021 & -1.6$\cdot 10^{-3}$ & -0.056 &  0.029  \\
  $\beta_{17}$ & STOCK-1 &  0.024 &  0.032 &  0.015 & -0.018 & 0.104  \\
  $\beta_{18}$ & STOCK-2 & -1.2$\cdot 10^{-3}$ & 0.028 & 2.8$\cdot 10^{-6}$ & -0.068 & 0.054  \\
  $\beta_{19}$ & EXRATE & 4.5$\cdot 10^{-3}$ & 0.018 & 9.3$\cdot 10^{-4}$ & -0.028 & 0.043 \\
  $\beta_{20}$ & IP & 2.0$\cdot 10^{-3}$ & 0.012 & 2.3$\cdot 10^{-4}$ & -0.022 & 0.030  \\
  $\beta_{21}$ & LOANS & -1.0$\cdot 10^{-3}$ & 0.017 & -4.9$\cdot 10^{-5}$ & -0.038 & 0.033   \\
  $\beta_{22}$ & M3 & -0.079 & 0.251 & -0.011& -0.720 & 0.295   \\
  $\beta_{23}$ & CAR & 4.2$\cdot 10^{-3}$ & 0.028 & 1.2$\cdot 10^{-3}$ & -0.056 & 0.060   \\
  $\beta_{24}$ & OIL & -4.2$\cdot 10^{-3}$ & 0.014 & -7.1$\cdot 10^{-4}$ & -0.039 & 0.021   \\
  $\beta_{25}$ & ORDER & 3.8$\cdot 10^{-3}$ & 0.021 & 1.9$\cdot 10^{-4}$ & -0.030 & 0.054  \\
  $\beta_{26}$ & UNEMP & 0.261 & 0.428 & 0.078 & -0.255 & 1.387   \\
  $\beta_{27}$ & $D_{t,1}$  & -4.6$\cdot 10^{-3}$ & 0.044 & -4.9$\cdot 10^{-4}$ & -0.104 & 0.087  \\
  $\beta_{28}$ & $D_{t,2 }$ & -7.5$\cdot 10^{-3}$ & 0.048 & -3.7$\cdot 10^{-4}$ & -0.127 & 0.082  \\
  $\beta_{29}$ & $D_{t,3}$  & 0.022 & 0.055& 7.2$\cdot 10^{-3}$ & -0.074 & 0.153   \\
  $\beta_{30}$ & $D_{t,4}$  & 0.026 & 0.051 & 7.8$\cdot 10^{-3}$ & -0.048 & 0.155   \\
  $\beta_{31}$ & $D_{t,5}$  &0.014 & 0.044 & 2.4$\cdot 10^{-3}$ & -0.065 & 0.124  \\
  $\beta_{32}$ & $D_{t,6 }$ & -0.012 & 0.043 & -1.7$\cdot 10^{-3}$ & -0.122 & 0.067   \\
  $\beta_{33}$ & $D_{t,7 }$ & -0.026 & 0.053 & -6.5$\cdot 10^{-3}$ & -0.163 & 0.052   \\
  $\beta_{34}$ & $D_{t,8}$  & 2.3$\cdot 10^{-3}$ & 0.040 & 9.0$\cdot 10^{-5}$ & -0.086 & 0.097   \\
  $\beta_{35}$ & $D_{t,9 }$ & -4.9$\cdot 10^{-4}$ & 0.039 & 5.3$\cdot 10^{-6}$ & -0.091 & 0.086   \\
  $\beta_{36}$ & $D_{t,10}$  & -3.3$\cdot 10^{-3}$ & 0.040 & -1.0$\cdot 10^{-4}$ & -0.099 & 0.081   \\
  $\beta_{37}$ & $D_{t,11 }$ & 0.061 & 0.073 & 0.042 & -0.037 & 0.231   \\\hline
  $|\sqrt \theta_{1}|$ & Intercept & 0.166 & 0.027 & 0.168 & 0.106 & 0.213   \\
  $|\sqrt \theta_{2}|$& $\pi_{t}$ & 9.8$\cdot 10^{-4}$ & 2.1$\cdot 10^{-3}$ & 1.2$\cdot 10^{-4}$ & 2.1$\cdot 10^{-8}$ & 7.2$\cdot 10^{-3}$   \\
  $|\sqrt \theta_{3}|$& $\pi_{t-1}$ & 9.0$\cdot 10^{-4}$ & 2.0$\cdot 10^{-3}$ & 1.2$\cdot 10^{-4}$ & 2.2$\cdot 10^{-8}$ & 6.3$\cdot 10^{-3}$   \\
  $|\sqrt \theta_{4}|$& $\pi_{t-2}$ & 8.8$\cdot 10^{-4}$ & 1.8$\cdot 10^{-3}$ & 1.2$\cdot 10^{-4}$ & 2.0$\cdot 10^{-8}$ & 6.1$\cdot 10^{-3}$   \\
  $|\sqrt \theta_{5}|$& $\pi_{t-3}$ & 7.8$\cdot 10^{-4}$ & 1.6$\cdot 10^{-3}$ & 1.1$\cdot 10^{-4}$ & 2.1$\cdot 10^{-8}$ & 5.4$\cdot 10^{-3}$   \\
  $|\sqrt \theta_{6}|$ & $\pi_{t-4}$ & 9.7$\cdot 10^{-4}$ & 1.9$\cdot 10^{-3}$ & 1.4$\cdot 10^{-4}$ & 2.1$\cdot 10^{-8}$ & 6.7$\cdot 10^{-3}$   \\
  $|\sqrt \theta_{7}|$ & $\pi_{t-5}$ & 9.0$\cdot 10^{-4}$ & 1.8$\cdot 10^{-3}$ & 1.3$\cdot 10^{-4}$ & 2.2$\cdot 10^{-8}$ & 6.2$\cdot 10^{-3}$   \\
  $|\sqrt \theta_{8}|$ & $\pi_{t-6}$ & 7.6$\cdot 10^{-4}$ & 1.5$\cdot 10^{-3}$ & 1.1$\cdot 10^{-4}$ & 2.0$\cdot 10^{-8}$ & 5.2$\cdot 10^{-3}$   \\
  $|\sqrt \theta_{9}|$ & $\pi_{t-7}$ & 9.6$\cdot 10^{-4}$ & 2.1$\cdot 10^{-3}$ & 1.2$\cdot 10^{-4}$ & 2.2$\cdot 10^{-8}$ & 6.8$\cdot 10^{-3}$   \\
  $|\sqrt \theta_{10}|$ & $\pi_{t-8}$ & 1.1$\cdot 10^{-3}$ & 2.5$\cdot 10^{-3}$ & 1.3$\cdot 10^{-4}$ & 2.0$\cdot 10^{-8}$ & 7.8$\cdot 10^{-3}$   \\
  $|\sqrt \theta_{11}|$ & $\pi_{t-9}$ & 8.7$\cdot 10^{-4}$ & 1.8$\cdot 10^{-3}$ & 1.1$\cdot 10^{-4}$ & 1.8$\cdot 10^{-8}$ & 6.1$\cdot 10^{-3}$   \\
  $|\sqrt \theta_{12}|$ & $\pi_{t-10}$& 8.3$\cdot 10^{-4}$ & 1.7$\cdot 10^{-3}$ & 1.2$\cdot 10^{-4}$ & 2.1$\cdot 10^{-8}$ & 5.8$\cdot 10^{-3}$   \\
  $|\sqrt \theta_{13}|$ & $\pi_{t-11}$ & 1.6$\cdot 10^{-3}$ & 3.0$\cdot 10^{-3}$ & 2.1$\cdot 10^{-4}$ & 2.5$\cdot 10^{-8}$ & 0.011   \\
  $|\sqrt \theta_{14}|$ & I\_1MO & 0.019 & 0.027& 3.6$\cdot 10^{-3}$ & 5.2$\cdot 10^{-8}$ & 0.091 \\
  $|\sqrt \theta_{15}|$ & I\_1YR & 0.026 & 0.028& 0.018 & 1.1$\cdot 10^{-7}$ & 0.092 \\
  $|\sqrt \theta_{16}|$ & SENT & 1.2$\cdot 10^{-3}$ & 2.7$\cdot 10^{-3}$ & 1.6$\cdot 10^{-4}$ & 2.2$\cdot 10^{-8}$ & 9.0$\cdot 10^{-3}$   \\
  $|\sqrt \theta_{17}|$ & STOCK1 & 1.4$\cdot 10^{-3}$ & 2.7$\cdot 10^{-3}$ & 1.8$\cdot 10^{-4}$ & 2.3$\cdot 10^{-8}$ & 9.3$\cdot 10^{-3}$   \\
  $|\sqrt \theta_{18}|$ & STOCK2 & 1.6$\cdot 10^{-3}$ & 3.2$\cdot 10^{-3}$ & 2.0$\cdot 10^{-4}$ & 2.4$\cdot 10^{-8}$ & 0.011  \\
  $|\sqrt \theta_{19}|$ & EXRATE  & 1.2$\cdot 10^{-3}$ & 2.4$\cdot 10^{-3}$ & 1.5$\cdot 10^{-4}$ & 2.1$\cdot 10^{-8}$ & 8.1$\cdot 10^{-3}$   \\
  $|\sqrt \theta_{20}|$ & IP & 7.0$\cdot 10^{-4}$ & 1.4$\cdot 10^{-3}$ & 1.1$\cdot 10^{-4}$ & 2.2$\cdot 10^{-8}$ & 4.9$\cdot 10^{-3}$  \\
  $|\sqrt \theta_{21}|$ & LOANS & 1.2$\cdot 10^{-3}$ & 2.5$\cdot 10^{-3}$ & 1.5$\cdot 10^{-4}$ & 2.1$\cdot 10^{-8}$ & 8.5$\cdot 10^{-3}$  \\
  $|\sqrt \theta_{22}|$ & M3 & 0.027 & 0.024  & 0.023 & 3.7$\cdot 10^{-7}$ & 0.086 \\
  $|\sqrt \theta_{23}|$ & CAR & 2.9$\cdot 10^{-3}$ & 5.1$\cdot 10^{-3}$ & 5.3$\cdot 10^{-4}$ & 3.0$\cdot 10^{-8}$ & 0.018  \\
  $|\sqrt \theta_{24}|$ & OIL & 9.0$\cdot 10^{-4}$ & 1.8$\cdot 10^{-3}$ & 1.2$\cdot 10^{-4}$ & 2.0$\cdot 10^{-8}$ & 6.4$\cdot 10^{-3}$  \\
  $|\sqrt \theta_{25}|$ & ORDER & 1.3$\cdot 10^{-3}$ & 2.8$\cdot 10^{-3}$ & 1.7$\cdot 10^{-4}$ & 2.4$\cdot 10^{-8}$ & 9.5$\cdot 10^{-3}$  \\
  $|\sqrt \theta_{26}|$ &  UNEMP& 0.033 & 0.400 & 8.5$\cdot 10^{-3}$ & 5.6$\cdot 10^{-8}$ & 0.120  \\
  $|\sqrt \theta_{27}|$ & $D_{t,1}$  & 2.6$\cdot 10^{-3}$ & 4.9$\cdot 10^{-3}$ & 3.7$\cdot 10^{-4}$ & 3.0$\cdot 10^{-8}$ & 0.017  \\
  $|\sqrt \theta_{28}|$ & $D_{t,2}$  & 3.9$\cdot 10^{-3}$ & 6.8$\cdot 10^{-3}$ & 5.9$\cdot 10^{-4}$ & 3.2$\cdot 10^{-8}$ & 0.024  \\
  $|\sqrt \theta_{29}|$ & $D_{t,3 }$ & 4.0$\cdot 10^{-3}$ & 7.4$\cdot 10^{-3}$ & 5.8$\cdot 10^{-4}$ & 3.1$\cdot 10^{-8}$ & 0.026  \\
  $|\sqrt \theta_{30}|$ & $D_{t,4}$  & 2.3$\cdot 10^{-3}$ & 4.5$\cdot 10^{-3}$ & 3.0$\cdot 10^{-4}$ & 2.7$\cdot 10^{-8}$ & 0.016  \\
  $|\sqrt \theta_{31}|$ & $D_{t,5}$  & 2.4$\cdot 10^{-3}$ & 4.6$\cdot 10^{-3}$ & 3.4$\cdot 10^{-4}$ & 2.8$\cdot 10^{-8}$ & 0.016  \\
  $|\sqrt \theta_{32}|$ & $D_{t,6}$  & 1.9$\cdot 10^{-3}$ & 3.9$\cdot 10^{-3}$ & 2.5$\cdot 10^{-4}$ & 2.4$\cdot 10^{-8}$ & 0.013  \\
  $|\sqrt \theta_{33}|$ & $D_{t,7}$  & 2.6$\cdot 10^{-3}$ & 5.1$\cdot 10^{-3}$ & 3.5$\cdot 10^{-4}$ & 2.6$\cdot 10^{-8}$ & 0.018  \\
  $|\sqrt \theta_{34}|$ & $D_{t,8}$  & 2.0$\cdot 10^{-3}$ & 4.0$\cdot 10^{-3}$ & 2.2$\cdot 10^{-4}$ & 2.3$\cdot 10^{-8}$ & 0.014 \\
  $|\sqrt \theta_{35}|$ & $D_{t,9}$  & 2.3$\cdot 10^{-3}$ & 4.5$\cdot 10^{-3}$ & 3.4$\cdot 10^{-4}$ & 3.0$\cdot 10^{-8}$ & 0.016  \\
  $|\sqrt \theta_{36}|$ & $D_{t,10}$  & 2.2$\cdot 10^{-3}$ & 4.4$\cdot 10^{-3}$ & 2.6$\cdot 10^{-4}$ & 2.4$\cdot 10^{-8}$ & 0.015  \\
  $|\sqrt \theta_{37}|$ & $D_{t,11}$  & 3.2$\cdot 10^{-3}$ & 5.7$\cdot 10^{-3}$ & 5.1$\cdot 10^{-4}$ & 2.9$\cdot 10^{-8}$ & 0.020  \\\hline
  $\sigma^2$ &                                 & 0.011 & 4.2$\cdot 10^{-3}$ & 0.010 & 4.0$\cdot 10^{-3}$ & 0.020  \\
\hline  \hline
\caption{ECB data. Posterior summary statistics for $p(\beta_j|\ym)$ and    $p(\sqrt{\theta_j}|\ym)$  under the   hierarchical double gamma prior with $a^\tau \sim \Exp (10)$ and $a^\xi \sim \Exp (10)$ for $j=1,\ldots,37$.
$D_{t,k}$ is a monthly 0/1 dummy variable taking  the value 1 for month $k$ ($k=1, \ldots, 11$) and 0 otherwise.
% \comment{Version August 2017.}
}   \label{table:SV_FALSE_02_ASIS_TRUE}%
 \end{longtable}}

\clearpage
%\end{landscape}
\onehalfspacing
\normalsize

\subsection{Comparing approximations to the one-step ahead predictive density}\label{sec:predecb}

%\marginpar{\comment{LPDS of  Dax from Version October 2016 should be substituted by LPDS  for adaptive prior on $a^\tau,a^\xi$. Not feasible for first revision.}}

\begin{figure}[t!]
 \centering
\includegraphics[height=6cm,width=0.8\linewidth,page=1]{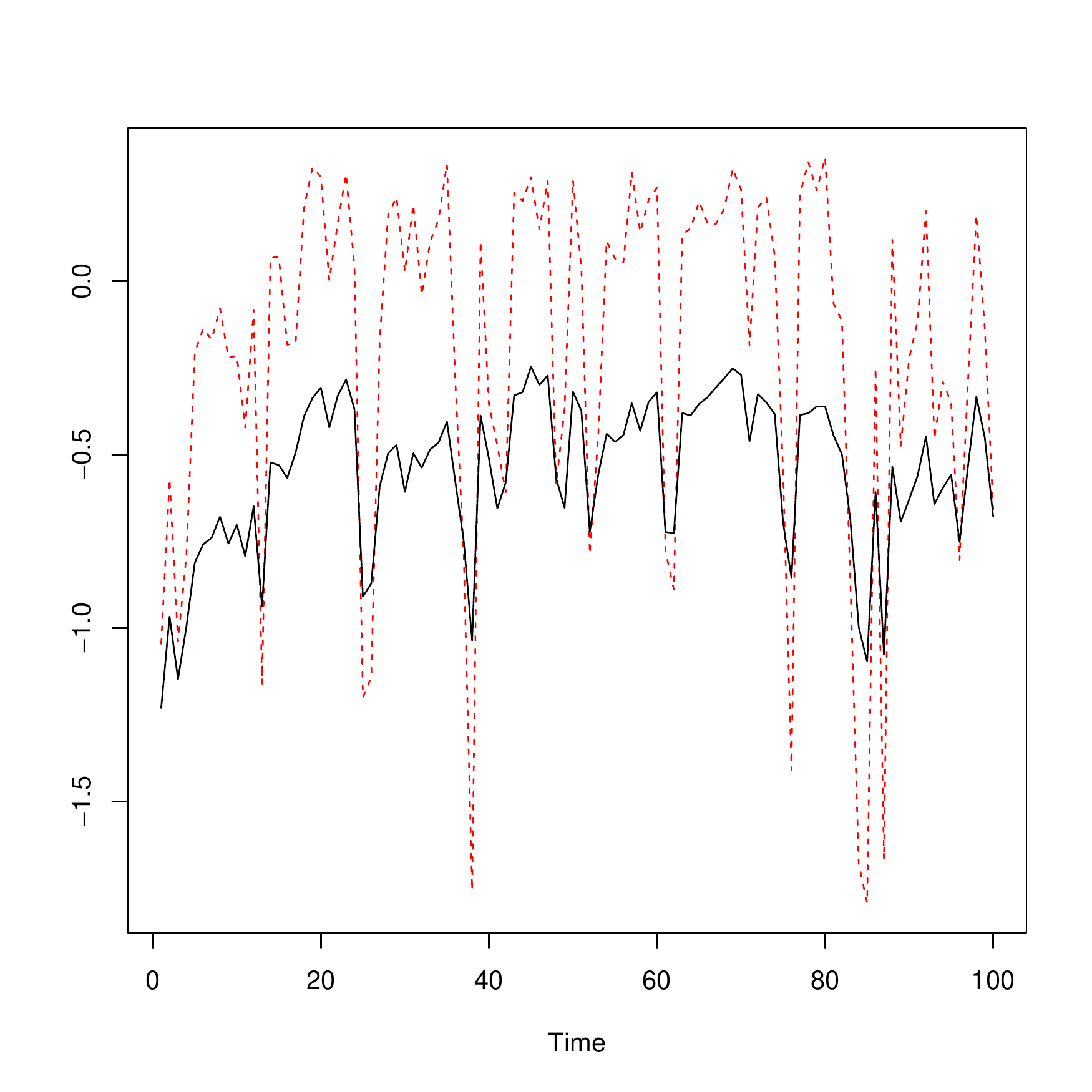}
% Based on Sylvia's MATLAB programms
\caption{ECB data.  Comparing the conditionally optimal Kalman mixture approximation (full line) and the  naive  Gaussian mixture approximation (dashed line) of the  log predictive density scores $\LPS^*_{t}$ ($t=t_0+1, \ldots, t_0+100 , t_0=90$) for the last 100 time points.} \label{fig:KFNAIecb}
\end{figure}

For the inflation data discussed in Section~\ref{inflatiodata},
 it is essential to use an accurate approximation  method  such as the conditionally optimal Kalman mixture approximation
 rather than the naive mixture approximation discussed in Section~\ref{sec:onestep}  to approximate
the one-step ahead predictive density  $p(y_{t}|\ym^{t-1})$. %, see  Appendix~\ref{sec:predecb}.
Since both approximations are finite mixtures of Gaussian distributions,  the value of each mixture component pdf need to be calculated at the actually observed value $y_{t}$   % $y_{t+1}$
in order to derive the log predictive density score $\LPS^*_{t}$. %$\LPS^*_{t+1}$.
Despite the high number of mixture components ($M=100,000$),  the resulting approximations of  $\LPS^*_{t}$  %  $\LPS^*_{t+1}$
 turned out to differ substantially,  see  Figure~\ref{fig:KFNAIecb}, where a comparison is provided  for   the   hierarchical double gamma prior with $a^\tau \sim \Exp (10)$ and $a^\xi \sim \Exp (10)$.

\begin{figure}[t!]
\centering
\begin{minipage}[b]{0.45\linewidth}
 \includegraphics[scale=0.45]{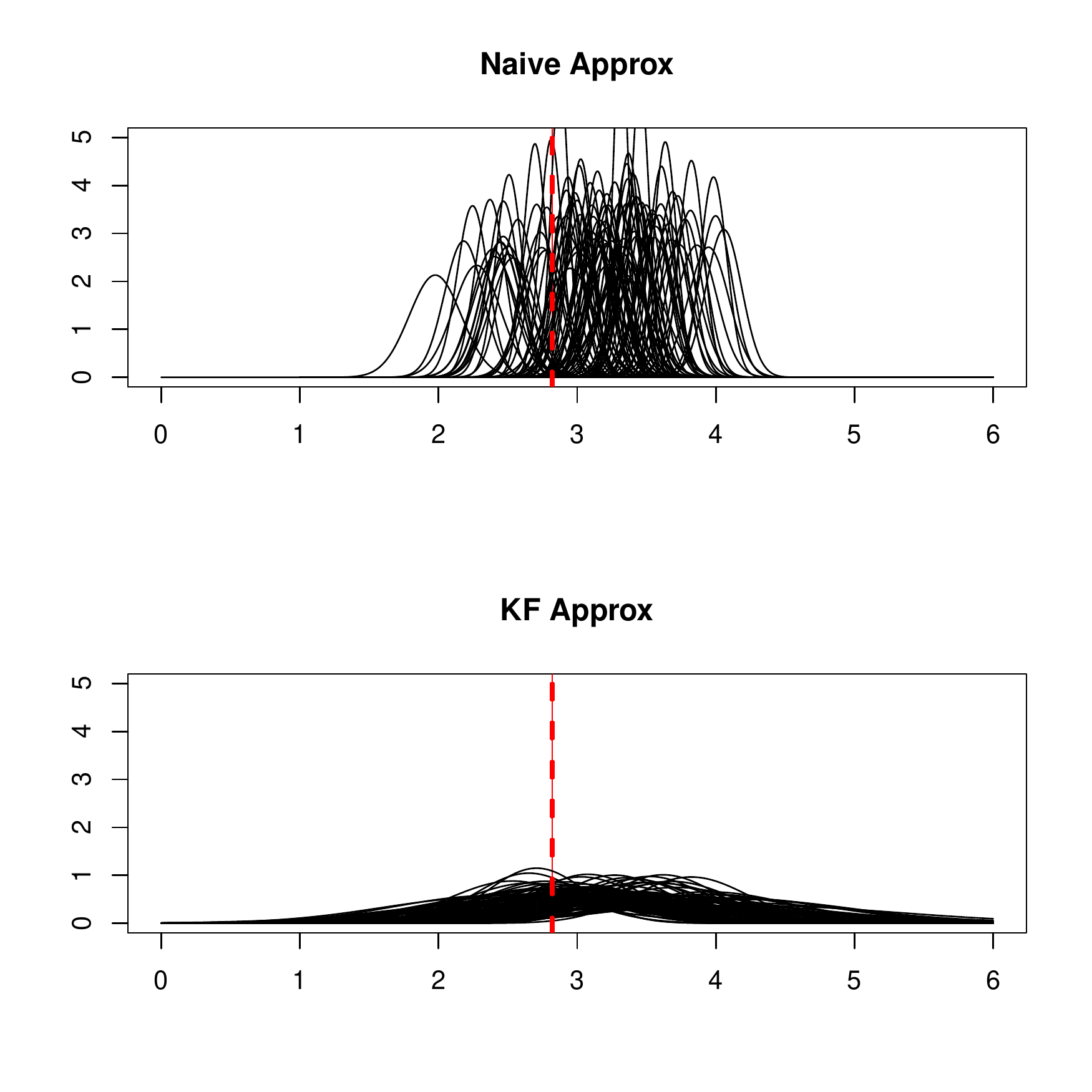}
\end{minipage}
\quad
\begin{minipage}[b]{0.45\linewidth}
\includegraphics[scale=0.45]{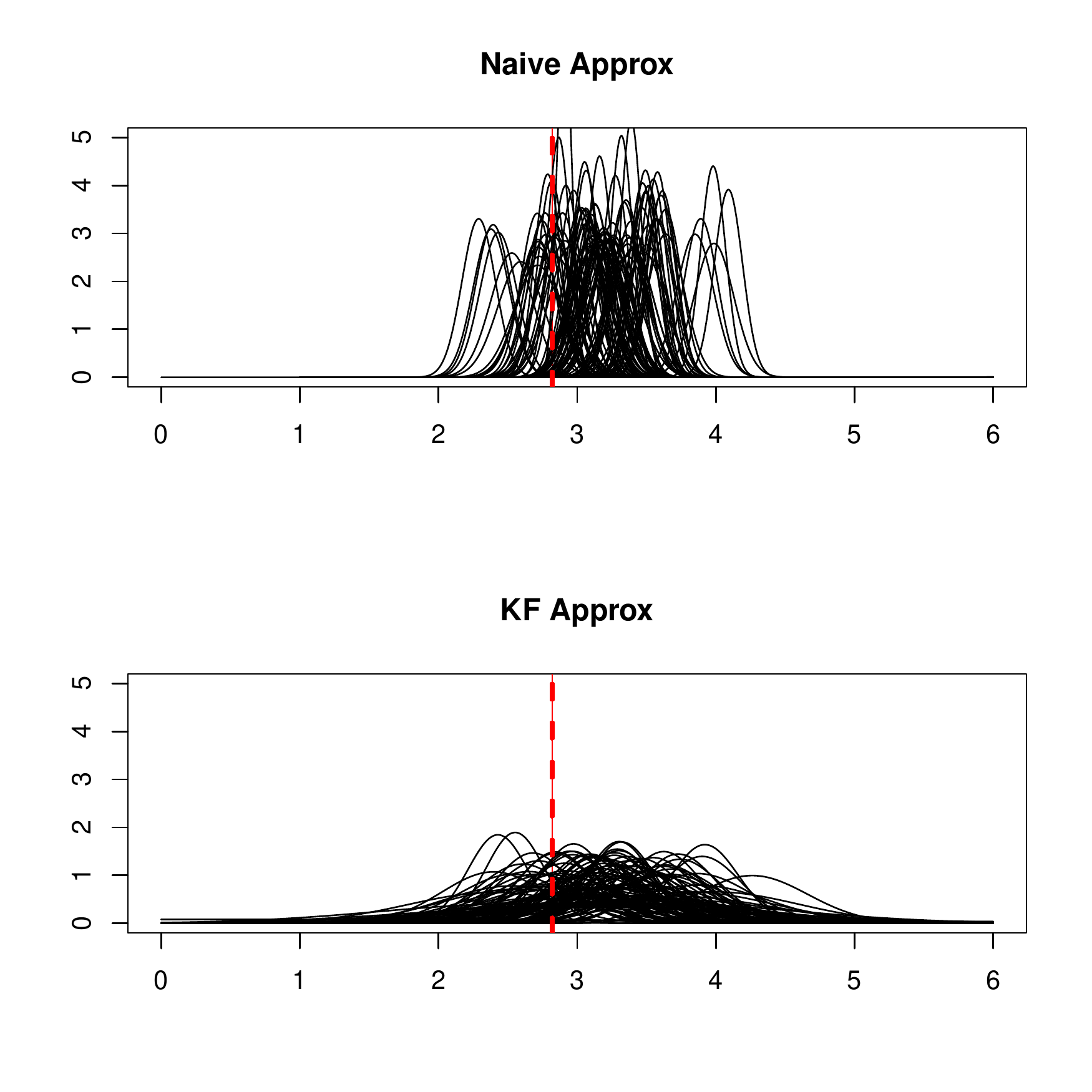}
\end{minipage} %\\
\caption{ECB data.~Illustration of the naive mixture approximation and the conditionally optimal Kalman mixture approximation for the predictive
density  $p(y_t|y^{t-1})$ at   $t=130$ % $t=t_0+40=130$
for the hierarchical double gamma prior  with  $a^\tau \sim \Exp (10)$ and $a^\xi \sim \Exp (10)$  (left-hand side) and the hierarchical Bayesian Lasso  prior (right-hand side).  The observed value  $y_{t}$ is indicated by the  vertical line.  For both mixture approximations,  40 components  densities $p(y_{t}|\ym^{t-1},\thmod \im{m})$  are plotted for randomly selected  draws $ \thmod \im{m}$ from the posterior $p(\thmod| \ym^{t-1})$.  % \comment{Version 17. J\"anner 2018}.
} \label{fig:mixture_approx}
\end{figure}

 In Figure~\ref{fig:mixture_approx} we compare a small subset of the   mixture component densities $p(y_{t}|\ym^{t-1},\thmod)$
 % $p(y_{t+1}|\ym^{t},\thmod)$
 for both approximations of  $\LPS^*_{t}$   %$\LPS^*_{t+1}$
  for a single point in time ($t=130$), where  respectively   $\thmod= (\betav_{t}, \sigma^2_{t})$    % $\thmod= (\betav_{t-1}, \sigma^2_{t+1})$
   and $\thmod = (\beta_1, \ldots,  \beta_d,  \sqrt{\theta_1} ,\ldots,\sqrt{\theta_d}, \sigma^2_{t})$.  %\sigma^2_{t+1})$.
    The  observed value  $y_{t}$ % $y_{t+1}$
  is indicated by a vertical line.
 Compared to  the conditionally optimal Kalman mixture approximation,  the mixture components of the naive approximation  have a very small spread,
  typical of a high signal-to-noise ratio.
A large number of zeros results from density evaluation and  leads to a bias of the corresponding  estimate of $\LPS^*_{t}$   % $\LPS^*_{t+1}$
toward zero  and, depending on the   true value,  the  log predictive density score   $\LPS^*_{t}$   % $\LPS^*_{t+1}$
is over- or underrated.

\begin{figure}[t!]
 \centering
\includegraphics[height=5cm,width=0.95\linewidth]{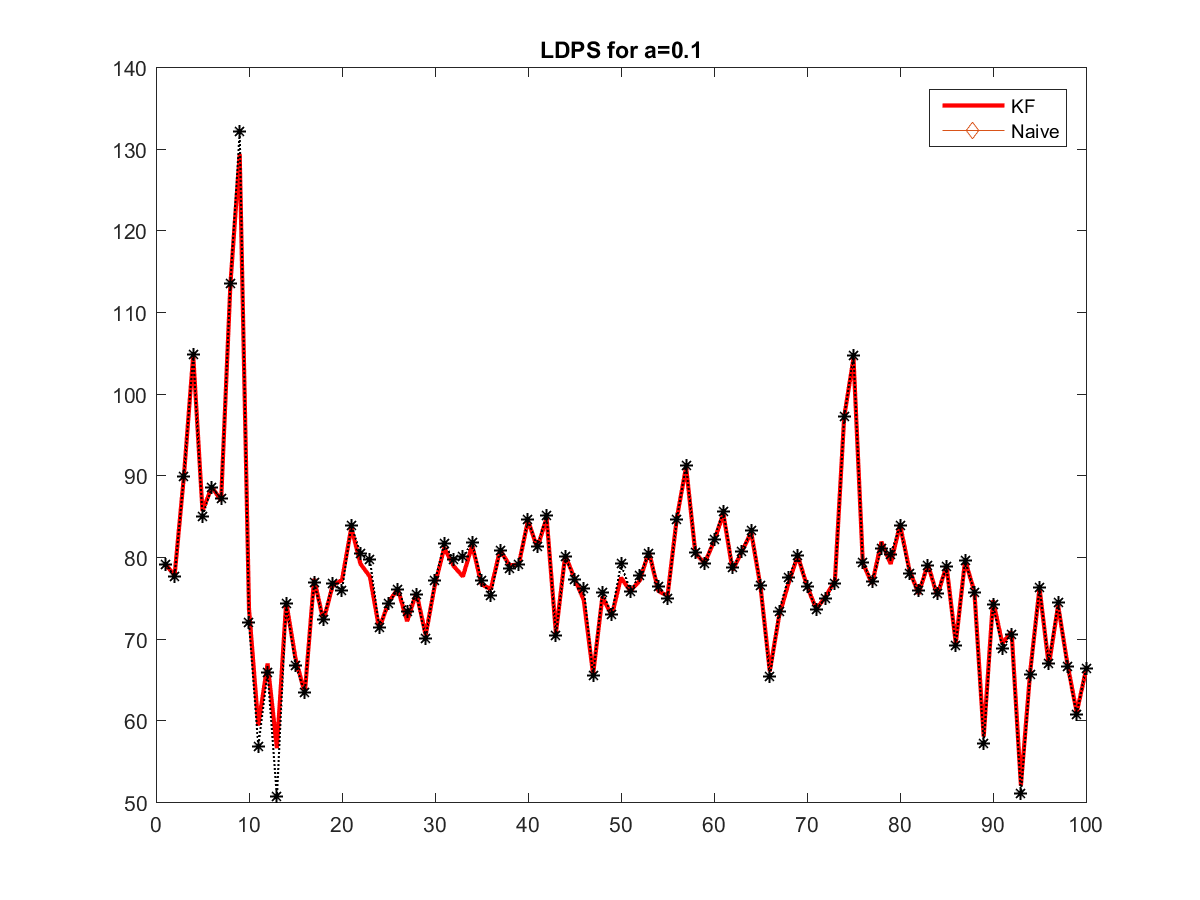}
% Based on Sylvia's MATLAB programms
\caption{DAX data.  Comparing approximations of the overall log predictive densities scores $\LPS^*_{t+1}$ for $a^\tau=a^\xi=0.1$ obtained from
the optimal Kalman mixture approximation (full line) and the  naive  Gaussian mixture approximation (dashed line with $\star$).
% \comment{TO DO ANGELA: substitute LPDS of OLD Version October 2016 by LPDS  for adaptive prior on $a^\tau,a^\xi$.}
} \label{fig:KFNAIdax}
\end{figure}

%  under   the   hierarchical double Gamma prior with $a^\tau \sim \Exp (10)$ and $a^\xi \sim \Exp (10)$ and the hierarchical Bayesian Lasso
%
 For the DAX data analyzed in Section~\ref{DAXdata}, the naive  Gaussian mixture approximation  yields identical  estimated of  $ \LPS^*_{t}$
 % $ \LPS^*_{t+1}$
for nearly all time point, see Figure~\ref{fig:KFNAIdax},
mainly because the variance of all  conditional  mixture densities $p(y_{i,t}|\ym^{t-1},\thmod)$  % $p(y_{i,t+1}|\ym^t,\thmod)$
is dominated by $\sigma_{i,t}^2$  % $\sigma_{i,t+1}^2$
for both approximations due to the low signal-to-noise ratio.

\clearpage

% \subsection{DAX Data} \label{results:dax}

%\begin{figure}
% \centering
% \begin{tabular}{c}
% \includegraphics[height=8cm,width=0.99\linewidth]{./plot_AB/plots_DAX/Xmit_text_plot_atau_01_1_08092016}\\
% \includegraphics[height=8cm,width=0.99\linewidth]{./plot_AB/plots_DAX/Xmit_text_plot_atau_1_1_08092016}
% \end{tabular}
% \caption{DAX data.~Heat plot of the posterior median of $\Bm_t$ at $t= 1150$ for $a^\tau=a^\xi =0.1$ (top) and $a^\tau=a^\xi =1$ (bottom).}
% \label{fig:heatmap_atau_all}
%\end{figure}

%\bibliographystyle{apalike}
%\bibliography{mybib}